\documentclass[iop]{emulateapj}
\usepackage{color}
\usepackage{tablefootnote}
\newcommand{\etal}{et\,al.}
\newcommand{\halpha}{H$\alpha$}
\newcommand{\kms}{km\,s$^{-1}$}
\newcommand{\lsim}{\raise0.3ex\hbox{$<$}\kern-0.75em{\lower0.65ex\hbox{$\sim$}}}

\newcommand{\HI}{H~{\sc I}}
\newcommand{\msun}{M$_{\odot}$}

\begin{document}
\slugcomment{Accepted for Publication in the Astrophysical Journal}
%{\color{red}Submission to the Astrophysical Journal planned for June 1}}

%-----------------------------------------------------------------------------%
\title{SHIELD: Comparing Gas and Star Formation in Low Mass Galaxies}
%-----------------------------------------------------------------------------%

%---------------------------
%Author list - journal style 
%---------------------------

%%%%%%%%%%%%%%%%%%%%% Tier 1:
\author{Yaron G. Teich}
\affil{Department of Physics \& Astronomy, Macalester College, 1600
  Grand Avenue, Saint Paul, MN 55105}
\affil{School of Education, Boston University, Two Silber Way, Boston, MA 02215}
\email{yateich@gmail.com}

\author{Andrew T. McNichols}
\affil{Department of Physics \& Astronomy, Macalester College, 1600
  Grand Avenue, Saint Paul, MN 55105}
\affil{NRAO Charlottesville, 520 Edgemont Road, Charlottesville, VA 
22903-2475, USA}
\email{andrew.mcnichols@macalester.edu}

\author{Elise Nims}
\affil{Department of Physics \& Astronomy, Macalester College, 1600
  Grand Avenue, Saint Paul, MN 55105}
%\email{elise.nims@gmail.com}

\author{John M. Cannon}
\affil{Department of Physics \& Astronomy, Macalester College, 1600
  Grand Avenue, Saint Paul, MN 55105}
\email{jcannon@macalester.edu}

%%%%%%%%%%%%%%%%%%%%% Tier 2:

\author{Elizabeth A. K. Adams}
\affil{ASTRON, the Netherlands Institute for Radio Astronomy, 
Postbus 2, 7990 AA, Dwingeloo, The Netherlands}
%\email{adams@astron.nl}

\author{Riccardo Giovanelli}
\affil{Center for Astrophysics and Planetary Science, Space
  Sciences Building, 122 Sciences Drive, Cornell University,
  Ithaca NY 14853 USA}
%\email{riccardo@astro.cornell.edu}

\author{Martha P. Haynes}
\affil{Center for Astrophysics and Planetary Science, Space
  Sciences Building, 122 Sciences Drive, Cornell University,
  Ithaca NY 14853 USA}
%\email{haynes@astro.cornell.edu}

\author{Kristen B.W. McQuinn}
\affil{Minnesota Institute for Astrophysics, School of Physics and Astronomy, 
116 Church Street, S.E., University of Minnesota, Minneapolis, MN 55455, USA}
\affil{University of Texas at Austin, McDonald Observatory, 2515 Speedway, 
Stop C1402, Austin, TX 78712, USA}
%\email{kmcquinn@astro.as.utexas.edu}

\author{John J. Salzer}  
\affil{Department of Astronomy, Indiana University, 727 East
  Third Street, Bloomington, IN 47405, USA}
%\email{slaz@astro.indiana.edu}

\author{Evan D. Skillman}
\affil{Minnesota Institute for Astrophysics, School of Physics and Astronomy, 
116 Church Street, S.E., University of Minnesota, Minneapolis, MN 55455, USA}
%\email{skillman@astro.umn.edu}

%%%%%%%%%%%%%%%%%%%%% Tier 3:

\author{Elijah Z. Bernstein-Cooper} 
\affil{Department of Astronomy, University of Wisconsin, 475 N
  Charter St Madison, WI 53706, USA} 
%\email{ezbc@astro.wisc.edu}

\author{Andrew Dolphin}
\affil{Raytheon Company, 1151 E. Hermans Road, Tucson, AZ 85756, USA}
%\email{adolphin@raytheon.com}

\author{E.C. Elson}
\affil{Astrophysics, Cosmology and Gravity Centre (ACGC), Department of Astronomy, University of Cape Town, Private Bag X3, Rondebosch 7701, South Africa}
%\email{ed@ast.uct.ac.za}

\author{Nathalie Haurberg} 
\affil{Physics Department, Knox College, 2 East South Street,
  Galesburg, IL 61401, USA}
%\email{nhaurber@knox.edu}

\author{Gyula I.G. J{\'o}zsa}
\affil{SKA South Africa, Radio Astronomy Research Group, 3rd Floor,
  The Park, Park Road, Pinelands, 7405, South Africa}
\affil{Rhodes University, Department of Physics and Electronics,
  Rhodes Centre for Radio Astronomy Techniques \& Technologies, PO Box
  94, Grahamstown, 6140, South Africa}
\affil{Argelander-Institut f{\"u}r Astronomie, Auf dem H{\"u}gel 71, 
53121 Bonn, Germany}
%\email{jozsa@ska.ac.za}

\author{J{\"u}rgen Ott}
\affil{National Radio Astronomy Observatory, P.O. Box O, 1003
  Lopezville Road, Socorro, NM 87801, USA}
%\email{jott@nrao.edu}

\author{Amelie Saintonge}
\affil{Department of Physics and Astronomy, University College London,
  Gower Place, London WC1E 6BT, UK}
%\email{a.saintonge@ucl.ac.uk}

\author{Steven R. Warren}
\affil{Cray, Inc.,
380 Jackson Street, Suite 210,
St. Paul, MN 55101, USA}
%\email{swarren@cray.com}

%%%%%%%%%%%%%%%%%%%%% Tier 4:

\author{Ian Cave}
\affil{Department of Physics \& Astronomy, Macalester College, 1600
  Grand Avenue, Saint Paul, MN 55105}
%\email{kylerayner8@gmail.com}

\author{Cedric Hagen}
\affil{Department of Physics \& Astronomy, Macalester College, 1600
  Grand Avenue, Saint Paul, MN 55105}
%\email{chagen2@macalester.edu}

\author{Shan Huang} 
\affil{Center for Cosmology and Particle Physics, New York University, 4 Washington Place, New York, NY 10003, USA}
%\email{shan.huang@nyu.edu}

\author{Steven Janowiecki}  
\affil{Department of Astronomy, Indiana University, 727 East
  Third Street, Bloomington, IN 47405, USA}
\affil{International Centre for Radio Astronomy Research,
University of Western Australia,
35 Stirling Highway,
Crawley, WA 6009, Australia.}
%\email{steven.janowiecki@icrar.org}

\author{Melissa V. Marshall}
\affil{Department of Physics \& Astronomy, Macalester College, 1600
  Grand Avenue, Saint Paul, MN 55105}
%\email{mveritym@gmail.com}

\author{Clara M. Thomann}
\affil{Department of Physics \& Astronomy, Macalester College, 1600
  Grand Avenue, Saint Paul, MN 55105}
%\email{claramthomann@gmail.com}

\author{Angela Van Sistine}
\affil{Department of Physics
University of Wisconsin-Milwaukee
3135 North Maryland Ave.
Milwaukee, WI 53211}
%\email{vansisti@uwm.edu}

%-----------------------------------------------------------------------------%
\begin{abstract}
%-----------------------------------------------------------------------------%

We analyze the relationships between atomic, neutral hydrogen
(\ion{H}{1}) and star formation (SF) in the 12 low-mass SHIELD
galaxies.  We compare high spectral ($\sim$0.82\,\kms\,ch$^{-1}$) and
spatial resolution (physical resolutions of 170\,pc\,--\,700\,pc)
\ion{H}{1} imaging from the VLA with \halpha\ and far-ultraviolet
imaging. We quantify the degree of co-spatiality between star forming
regions and regions of high \ion{H}{1} column densities. We calculate
the global star formation efficiencies (SFE, $\Sigma_{\rm
  SFR}$\,/\,$\Sigma_{\rm \HI}$), and examine the relationships among
the SFE and \ion{H}{1} mass, \ion{H}{1} column density, and star
formation rate (SFR). The systems are consuming their cold neutral gas
on timescales of order a few Gyr. While we derive an index for the
Kennicutt-Schmidt relation of $N$\,$\approx$\,0.68$\pm$0.04 for the
SHIELD sample as a whole, the values of $N$ vary considerably from
system to system.  By supplementing SHIELD results with those from
other surveys, we find that HI mass and UV-based SFR are strongly
correlated over five orders of magnitude.  Identification of patterns
within the SHIELD sample allows us to bin the galaxies into three
general categories: 1) mainly co-spatial \ion{H}{1} and SF regions,
found in systems with highest peak \ion{H}{1} column densities and
highest total \ion{H}{1} masses; 2) moderately correlated \ion{H}{1}
and SF regions, found in systems with moderate \ion{H}{1} column
densities; and 3) obvious offsets between \ion{H}{1} and SF peaks,
found in systems with the lowest total \ion{H}{1} masses. SF in these
galaxies is dominated by stochasticity and random fluctuations in
their ISM.

\end{abstract}						


%-----------------------------------------------------------------------------%
\section{Introduction}
\label{S1}
%-----------------------------------------------------------------------------%

%-----------------------------------------------------------------------------%
\subsection{Stars and Gas in Galaxies}
\label{S1.1}
%-----------------------------------------------------------------------------%

The conversion of gas into stars is one of the most fundamental
processes in astronomy. Yet, despite decades of effort, a simple
prescription of star formation (SF) that successfully describes all
observations of galaxies across a range of halo masses has remained
elusive.  In broad terms, more massive star-forming galaxies will have
larger gas reservoirs (both atomic and molecular) and higher global
star formation rates (SFR) than less massive systems \citep[see, e.g.,
][]{kennicutt1998a}. However, the gas mass fractions in star forming
galaxies tend to increase with decreasing mass
\citep[e.g.,][]{fishertully1975}.

Empirical correlations between gas properties and various tracers of
instantaneous (using H$\alpha$ emission, with a characteristic
timescale of $\lsim$ 10 Myr) or ongoing (using FUV emission, with a
characteristic timescale of $\lsim$ 100 Myr) SF have been numerous in
the literature.  The most common parameterization relates a SFR
surface density to a gas surface density:

\begin{equation}
\Sigma_{\rm SFR} \propto (\Sigma_{\rm gas})^{N}
\end{equation}

\noindent with the SFR surface density ($\Sigma_{\rm SFR}$) in units
of \msun\,yr$^{-1}$\,kpc$^{-2}$, the gas surface density ($\Sigma_{\rm
  gas}$) in units of \msun\,pc$^{-2}$, and $N$ being a positive
number. \citet{schmidt1959} found that $N \approx 2$, and similar
indices have been derived numerous times over the last half-century
(see Elmegreen 2011\nocite{elmegreen2011} for a recent review). For
example, \citet{kennicutt1998b} found $N =$ 1.4\,$\pm$\,0.15 for 61
large spiral galaxies when relating the \halpha-based SFR to the total
gas surface density (via both \ion{H}{1} and CO observations, where
the CO is used as a tracer for molecular gas).  

In a study of this relation on sub-kpc scales, \citet{bigiel2008}
found that the relationship between the total gas surface density and
the SFR surface density varied dramatically among and within
individual spiral galaxies, and that most of the sample showed little
or no correlation between $\Sigma_{\rm \HI}$ and $\Sigma_{\rm
  SFR}$. In an associated paper, \citet{leroy2008} found a molecular
Schmidt power-law slope of $N = 1.0 \pm 0.2$ in 18 nearby spiral
galaxies; similarly, \citet{momose2013} found $N =$ 1.3 - 1.8 for 10
nearby spiral galaxies. In a subsequent study, \citet{bigiel2010} find
no clear evidence for SF thresholds and emphasize that it may not be
realistic to expect them.

Molecular gas in low mass galaxies remains largely undetectable via
traditional CO tracers, thus studying the relationship between atomic
gas and SF is especially important in these metal-poor low-mass
systems \citep[e.g., ][]{bolatto2013}. In fact, very few detections of
CO gas exist at metallicities less than $\sim$10\% Z$_{\odot}$, even
in systems that are actively forming stars (see, e.g., {Taylor
  \etal\ 1998}\nocite{taylor1998}, {Schruba
  \etal\ 2012}\nocite{schruba2012}, {Warren
  \etal\ 2015}\nocite{warren2015}). For low mass galaxies, this
implies that studying the relationship between $\Sigma_{\rm SFR}$ and
$\Sigma_{\rm gas}$ as a function of galaxy mass is necessarily
constrained to only the atomic gas component.

It is thus interesting to note that a correlation between $\Sigma_{\rm
  \HI}$ and $\Sigma_{\rm SFR}$ appears to hold in some galaxies and
not in others.  For example, \citet{bigiel2008} find that $\Sigma_{\rm
  \HI}$ and $\Sigma_{\rm SFR}$ are not related in individual spiral
disks. This can be compared to the results in \citet{skillman1987},
which show that 10$^{21}$ cm$^{-2}$ (corresponding to 7.9
\msun\ pc$^{-2}$ or 10.6 \msun\ pc$^{-2}$ when accounting for helium)
represents a requisite threshold \ion{H}{1} surface density for
massive SF. Similarly, \citet{wyder2009} examined 19 low
surface-brightness galaxies and found an apparent threshold in the
\ion{H}{1} gas surface density in the range 3 $-$ 10 \msun\ pc$^{-2}$
below which very little SF (traced by \halpha) is observed.  Extending
to even lower masses, \citet{roychowdhury2009} and
\citet{roychowdhury2011} find that all \ion{H}{1} gas in their
galaxies with $\Sigma_{\rm \HI}$ $\gtrsim$ 10 \msun\ pc$^{-2}$
($\approx$ 1.2$\times$10$^{21}$ cm$^{-2}$) have associated SF, but
there is no threshold below which SF is not observed (that is, SF is
observed in regions with \ion{H}{1} columns $<$ 10$^{21}$ cm$^{-2}$).
Most recently, \citet{roychowdhury2014} found that the $\Sigma_{\rm
  \HI}$ $-$ $\Sigma_{\rm SFR}$ relation for a set of dwarf irregular
galaxies was nearly linear.  \citet{roychowdhury2015} find consistency
across a range of scales (400pc and 1kpc) and galaxy types, including
both low-mass galaxies and more massive spiral disks.

Based on the above results, it is not trivial to anticipate where in a
galaxy one will observe ongoing SF - regardless of how massive that
galaxy is.  Knowledge of the \ion{H}{1} properties alone is often
insufficient to predict where in a given galaxy the conditions are
ripe for SF.  For example, \citet{krumholz2012} suggests that
molecular gas is a better predictor of SF than the neutral ISM. We
examine this issue from the \ion{H}{1} perspective in
Section~\ref{S4.2}.

In addition to the $\Sigma_{\rm gas}$ vs. $\Sigma_{\rm SFR}$ analysis,
we also study the star formation efficiency (SFE), which is a useful
metric when discussing where in a galaxy SF is occurring
\citep{leroy2008}. Several ways of describing the SFE exist, but for
consistency with other surveys we use SFE = $\Sigma_{\rm SFR} /
\Sigma_{\rm gas}$ with units of yr$^{-1}$. The SFE is more useful than
SFR alone to identify where conditions are conducive to SF because it
is normalized by the gas mass surface density. Thus, it quantifies the
local physical properties in regions where the gas is being turned
into stars efficiently: regions of elevated gas surface density which
have no young stars associated with them are inefficient, while
regions of elevated gas surface density that show co-spatiality with
young stars are efficient. Conveniently, the inverse of the SFE is the
gas depletion time, which is the time required for SF to consume the
gas reservoir at the present-day SFR. For a sample of low-mass
galaxies, \citet{roychowdhury2014} found gas depletion timescales of
$\sim$10$^{10}$ yr, an order of magnitude lower than is estimated for
the outer regions of large spiral galaxies \citep{bigiel2010}. We
calculate gas consumption times for our sample of galaxies in
Section~\ref{S4.2}.

%-----------------------------------------------------------------------------%
\subsection{Low-Mass Galaxies from ALFALFA}
\label{S1.2}
%-----------------------------------------------------------------------------%

The ALFALFA survey \citep{giovanelli2005} has produced one of the
largest and most statistically useful catalogs of nearby, gas-rich
galaxies to date.  The final ALFALFA database will include source
parameters for more than 30,000 systems.  With the acquisition of data
for ALFALFA now complete, a unique database exists to facilitate the
study of fundamental galaxy properties across an unprecedented range
of physical parameters.  One particularly rich area of exploration
that has been enabled by ALFALFA is a robustly-populated faint end of
the \ion{H}{1} mass function \citep{martin2010}.  Specifically, it is
possible to identify a complete sample of gas-bearing, low-mass
galaxies by matching the ALFALFA database to existing optical survey
data.

As introduced in \citet{cannon2011}, the ``Survey of \ion{H}{1} in
Extremely Low-mass Dwarfs'' (hereafter, SHIELD) is a multi-wavelength,
detailed study of the properties of ALFALFA-discovered or cataloged
systems with extremely small \ion{H}{1} mass reservoirs (see
Section~\ref{S2} for detailed discussion of the sample selection).
Subsequent works have established the distances \citep{mcquinn2014},
the nebular abundances \citep{haurberg2015}, and the qualities of SF
based on spatially resolved Hubble Space Telescope (HST) imaging
\citep{mcquinn2015a}.  The primary goals of SHIELD are to 1)
characterize the nature of SF in very low-mass galaxies and to 2)
determine what fraction of the mass in these low-mass galaxies is
baryonic.  In this work, we undertake a comparative study of the
\ion{H}{1} gas and various tracers of recent SF in order to address
goal \#1.  A companion paper by \citet{mcnichols16} explores the
dynamical properties of our sample galaxies in order to address goal
\#2.

SHIELD is one of a number of recent \ion{H}{1} surveys of dwarf
galaxies using interferometric data. This list includes WHISP (The
Westerbork \ion{H}{1} Survey of Irregular and Spiral Galaxies;
{Swaters 2002}\nocite{swaters2002}), FIGGS (Faint Irregular Galaxies
GMRT Survey; {Begum \etal\ 2008}\nocite{begum2008}), VLA-ANGST (Very
Large Array Survey of ACS Nearby Galaxy Survey Treasury Galaxies; {Ott
  \etal\ 2012}\nocite{ott2012}), LITTLE-THINGS (Local Irregulars That
Trace Luminosity Extremes in The \ion{H}{1} Nearby Galaxy Survey;
{Hunter \etal\ 2012}\nocite{hunter2012}), and LVHIS (The Local Volume
\ion{H}{1} Survey; {Kirby 2012}\nocite{kirby2012}). These studies have
yielded valuable insights into a total of nearly two dozen systems
with M$_{\rm \HI}$ $\lsim$ $10^{7}$ \msun. SHIELD adds to this
relatively understudied region of parameter space by significantly
increasing the number of sources with resolved \ion{H}{1} imaging.

%-----------------------------------------------------------------------------%
\section{Galaxy Sample, Observations, and Data Handling}
\label{S2}
\subsection{Sample Selection}
\label{S2.1}
%-----------------------------------------------------------------------------%

SHIELD is a multi-wavelength survey of 12 low-mass galaxies in the
Local Volume.  Sample members were selected from the first $\sim$10\%
of the ALFALFA-detected galaxies on the basis of estimated \ion{H}{1}
mass (M$_{\rm \HI}$ $<$ 10$^{7.2}$, based on flow model distances
using the prescription of {Masters 2005}\nocite{masters2005}). The
W$_{\rm 50}$ condition (\ion{H}{1} line width at 50\% of peak $<$ 65
\kms, with no correction for inclination which would increase
rotational velocity) discriminates against massive but \ion{H}{1}-poor
galaxies and identifies truly low-mass galaxies.  Following the
presentation of early SHIELD results in \citet{cannon2011}, the
sources were observed with HST to derive their distances via the tip
of the red giant branch (TRGB) method. The details are given in
\citet{mcquinn2014}; all sources moved to somewhat higher distances
than the flow model predictions and so the \ion{H}{1} masses of the
systems are slightly increased. Even with the updated distances, all
but one of the galaxies has M$_{\rm \HI}$ $<$ 10$^{7.3}$. The median
distance, \ion{H}{1} mass, and \ion{H}{1} line width for the sample
are 7.86 Mpc, 10$^{7.06}$ \msun, and 25 \kms,
respectively. Table~\ref{t1} provides a summary of physical
characteristics of the SHIELD galaxies.

%-----------------------------------------------------------------------------%
\subsection{Data Products}
\label{S2.2}
\subsubsection{Karl G. Janksy Very Large Array \ion{H}{1} Data}
\label{S2.2.1}
%-----------------------------------------------------------------------------%

The \ion{H}{1} data for the survey were obtained using the Karl
G. Janksy VLA\footnote{The National Radio Astronomy
  Observatory is a facility of the National Science Foundation
  operated under cooperative agreement by Associated Universities,
  Inc.} in multiple array configurations for programs
VLA/10B-187 (legacy code AC\,990; P.I. Cannon) and VLA/13A-027 (legacy
code AC\,1115; P.I. Cannon). Our observational strategy (9, 4, and 2
hours of observation per source in the B, C, and D arrays of the VLA,
respectively, with typical calibration overheads of 25\%) achieves
high spatial ($\sim$6\arcsec\ synthesized beam at full resolution) and
spectral resolution ($\sim$0.824 \kms\ ch$^{-1}$), while retaining
sensitivity to extended structure. The WIDAR correlator is used to
provide a single 1 MHz sub-band with 2 polarization products and 256
channels each, covering 211 \kms\ of frequency space at 3.906 kHz
ch$^{-1}$, which is the setup for the B- and C-configuration
observations. For the D-configuration observations, the frequency
coverage was widened to 4 MHz (corresponding to 1024 channels and 844
\kms\ at 3.906 kHz ch$^{-1}$). VLA data acquisition for SHIELD began
in 2010 October, and was completed in 2013. All of the sample members
were observed in the B-configuration except for AGC\,111164,
AGC\,111977, and AGC\,112521.

The VLA \ion{H}{1} data reduction techniques are standard and were
done using the Common Astronomy Software Application (CASA; McMullin
\etal\ 2007\nocite{mcmullin2007})\footnote{https://casaguides.nrao.edu/}. Radio
Frequency Interference was excised by hand. Calibrations were then
derived for antenna position, antenna-based phase delays, atmospheric
opacity, and bandpass frequency response. After continuum subtraction
in the $uv$-plane, the B, C, and D-configuration measurement sets were
gridded together using CASA's \textsc{clean} algorithm to produce data
cubes. We generated a Boolean mask based on the dirty cube for use in
our deep cleans (threshold set at 2.5$\sigma$). We also performed
residual flux rescaling on our data cubes.  Deconvolution of the dirty
map induces a difference area between the `dirty beam' of the residual
map and the `clean beam' of the deconvolved map and thus the resultant
flux density in the final `combined' image; in order to increase the
accuracy of the flux scaling in the final image (which is the linear
sum of the residual map and clean map) the residual must be rescaling
by some factor proportional to the difference in the (frequency
dependent) `dirty' and `clean beam' areas. This is a higher-order
correction, as the error on the flux scaling of these observations is
primarily limited by the calibration model uncertainties. The details
and motivation for this procedure are discussed in
\citet{jorsater1995}.  Following these corrections, we then
implemented a correction for the primary beam attenuation. The robust
data parameters are summarized in Table~\ref{t2}.

To produce two-dimensional images from the three-dimensional cubes, we
implemented a threshold mask followed by a manual inspection and
hand-blanking of the cubes. Our final data products include 4 cubes
and 4 moment-0 (integrated intensity) maps for each of the 12
galaxies. The cubes are either natural-weighted or robust-weighted
(explicitly, weighting$=$``briggs'' and robust$=$0.5), and at either
native spectral resolution (0.82 \kms\ ch$^{-1}$) or spectrally
smoothed by a factor of 3 (2.46 \kms\ ch$^{-1}$). We estimate the
uncertainty on our \ion{H}{1} flux densities to be a minimum of 10\%,
and propagate this throughout the calculation of \ion{H}{1} masses,
column densities, and $\Sigma_{\rm gas}$. This estimate accounts for
random noise in our maps as well as errors in flux calibration. The
flux densities of the \ion{H}{1} maps are converted to \ion{H}{1}
masses using the standard transformation for optically thin \ion{H}{1}
emission:

\begin{equation}
M_{\rm \HI} \approx 2.36 \times 10^{5}\ (D)^{2} \int S(v)dv
\end{equation}

\noindent where M$_{\rm \HI}$ is the \ion{H}{1} mass in M$_{\odot}$, D
is the distance in Mpc, and the integral $\int$$S(v)dv$ is the line
flux of the source in Jy km s$^{-1}$ \citep{giovanellihaynes1988}.

One of the SHIELD sources, AGC\,111164, presented unique challenges in
the \ion{H}{1} data processing.  AGC\,111164 is located
$\sim$15\arcmin\ east of the more massive, gas-rich galaxy
NGC\,784. Both sources overlap in velocity space; NGC\,784 lies almost
exactly one half-power beam-width away from AGC\,111164, and thus the
sidelobes from NGC\,784 are co-spatial with the SHIELD source and are
challenging to clean properly. We used the \textit{outlierfile}
parameter for CASA's \textsc{clean} task, which proved to be helpful
in extracting the larger source from our field. We also shifted the
phasecenter of the \textsc{clean} to lie directly between the two
sources.

%-----------------------------------------------------------------------------
\subsubsection{GALEX Archival Data Products}
\label{S2.2.2}
%-----------------------------------------------------------------------------%

GALEX is a 50\,cm diameter UV telescope that images the sky
simultaneously in both a far ultraviolet (FUV) and a near ultraviolet
(NUV) band, with effective wavelengths of 1528 and 2271 \AA,
respectively \citep{martin2005}. The field of view of GALEX is
approximately circular with a diameter of 1.25 degrees, with an
intrinsic angular resolution of 4.2\arcsec\ and 5.3\arcsec\ FWHM in
the FUV and NUV, respectively. The ultraviolet data presented here
are derived from three GALEX programs: the Guest Investigator Program
(GI), the Medium Imaging Survey (MIS), and the All-sky Imaging Survey
(AIS). All sources had exposure times between $\sim$1600 and
$\sim$2800 seconds except for AGC\,174605, which only has AIS-depth
imaging (120 seconds of integration). AGC\,749237 has no FUV data
because its 2010 observation followed the 2009 suspension of FUV
operations due to an electrical overcurrent; all the other sources
have FUV data.  All 12 of the sources in the survey have NUV data. The
GALEX data were processed with a pipeline which performed calibration
and background subtraction. Details of the GALEX detectors, pipeline,
calibration, and source extraction can be found in
\citet{morrissey2007}. GALEX pipeline products have an assumed 10\%
flux calibration error.

The FUV and NUV images were cropped to a region centered on the galaxy
and excluding most foreground and background sources.  By comparison
with the HST images, remaining contaminants were excised by hand.
Photometry was then extracted using standard techniques, and a
conversion from raw counts per second in the GALEX images to
magnitudes was performed. The GALEX
website\footnote{http://asd.gsfc.nasa.gov/archive/galex/FAQ/counts\_background.html}
and associated papers \citep{morrissey2007} provide the standard
prescriptions to convert from background-subtracted images to AB
magnitudes.

As \citet{kennicutt1998a} and others have shown, accounting for severe
dust attenuation and extinction of light from extragalactic sources is
a difficult problem to address. \citet{buat2005} and
\citet{burgarella2005} found that in nearby star-forming galaxies,
dust attenuation in the UV regime can vary from zero to several
magnitudes. The galaxies in our sample are low-metallicity dust-poor
dwarfs (see Haurberg \etal\ 2015\nocite{haurberg2015} for details),
and so we expect and assume neglible internal dust attenuation. While
we would prefer to use an energy balance approach with FUV and total
infrared (TIR) emission to probe the dust-free luminosity (e.g.,
McQuinn \etal\ 2015b\nocite{mcquinn2015b}), our sources were not
imaged at 24 $\mu$m or longer wavelengths by \textit{Spitzer}. Even if
we did have \textit{Spitzer} MIPS imaging and were able to use this
TIR+FUV flux method, it is likely that the extinction corrections
derived in this way would be small; for example, the median A$_{\rm
  FUV}$ found by \citet{mcquinn2015b} for a sample of low-metallicity
dwarf galaxies using this method was 0.76 mag.

Galactic extinction along the line-of-sight can be significant in the
FUV and NUV.  We applied the method of using the dust maps from
\citet{schlafly2011} to account for the effects of Galactic
extinciton\footnote{http://irsa.ipac.caltech.edu/applications/DUST/}. The
extinction at FUV wavelengths is calculated using the following
prescription from \citet{wyder2007}:

\begin{equation}
\rm{A_{FUV} = 8.24 \times E(B-V)} 
\end{equation}

\noindent The ultraviolet extinction values for the sample members
were all below 1 mag and are included in Table~\ref{t1}. After
correcting the measured FUV magnitudes for Galactic extinction, we
converted to FUV luminosities using standard prescriptions and the TRGB
distances in Table~\ref{t1}.

Other issues affecting the data were minor. AGC\,111946 resides less
than 200\arcsec\ from the edge of the frame in both the FUV and NUV so
vignetting is significant. Although the background subtraction
performed on this image via the GALEX pipeline is adequate, our
background subtraction uncertainty has still increased to account for
the vignetting. For the FUV observation of AGC\,174585, the pipeline
background subtraction was unsatisfactory due to a bright foreground
source in the Northwest corner of the image, so a manual background
subtraction was performed using an average sky value from a region in
the image which was unaffected by the contaminating source.

%-----------------------------------------------------------------------------
\subsubsection{WIYN 3.5m Data Products}
\label{S2.2.3}
%-----------------------------------------------------------------------------%

Ground-based optical images were obtained using the Mini-Mosaic Imager
on the WIYN 3.5m telescope at Kitt Peak National Observatory
(KPNO). The observations were performed over four nights: 2010 October
7-8 and 2011 March 29-30. The Fall 2010 nights had good seeing
($\approx$ 0.5\arcsec) and were done under photometric conditions; the
Spring 2011 nights had degraded seeing ($\approx$ 1.2\arcsec) and the
conditions were not photometric. All sources were observed in four
separate filters: broadband Johnson B, V, and R, and narrowband W036
\halpha\ (FWHM = 60\AA). Exposure times for the Fall images were 900, 720, 600, and
900 s for the B, V, R, and \halpha. The Spring images were 1200, 720,
900, and 1200 s exposures. Our treatment of the data used standard
prescriptions in IRAF\footnote{IRAF is distributed by the National
  Optical Astronomy Observatory, which is operated by the Association
  of Universities for Research in Astronomy (AURA) under a cooperative
  agreement with the National Science Foundation.}. A complete
description of the WIYN 3.5m datasets and imaging procedures can be
found in \citet{haurberg2013} and Haurberg et al. (in prep). Note that
AGC\,748778 and AGC\,749241 are non-detections in \halpha, and the
latter is a non-detection in the \textit{Spitzer} images as well. In
Table~\ref{t3} we include the final \halpha\ luminosities for all
sample members; related quantities are presented in Table~\ref{t4}.
\halpha\ SFRs have uncertainties which include both the distance and
photometric errors.

%-----------------------------------------------------------------------------
\subsubsection{Hubble Space Telescope Data Products}
\label{S2.2.4}
%-----------------------------------------------------------------------------%

\begin{figure}
\centering
% old location = \plotone{/Users/research/data/vla/ELLIPSES.ar}
\plotone{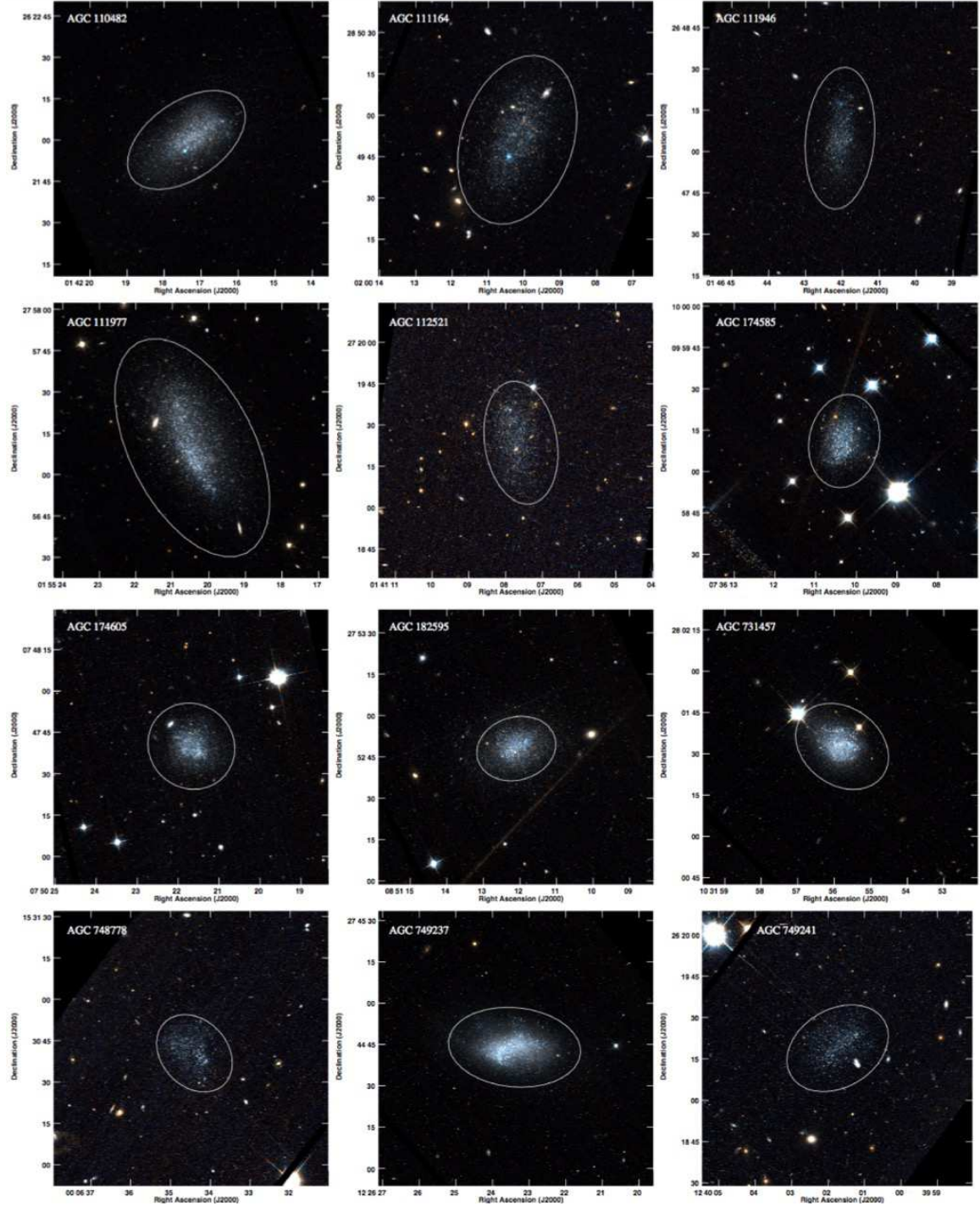}
\caption{HST 3-color images of the SHIELD galaxies; these images are 
reproductions of those in \citet{mcquinn2014}.  Overlaid on each panel 
is a single elliptical aperture to demonstrate the position and orientation of 
the elliptical apertures that were used in surface brightness analysis.}
\label{ELLIPSES}
\end{figure}

Imaging of the 12 SHIELD galaxies was obtained for program GO-12658
(P.I. Cannon). The HST observations were conducted with the Advanced
Camera for Surveys (ACS). The F606W and F814W filters were used, with
average total exposure times of 1000 s and 1200 s, respectively. Final
3-color images were produced by creating a third ``green'' image,
which is the linear average of the blue and red images;
Figure~\ref{ELLIPSES} contains these final images. A complete
description of the data handling and analysis is given in
\citet{mcquinn2014}; that manuscript also derives the distances of the
SHIELD galaxies shown in Table~\ref{t1}. Further analysis of the
recent SF histories of the SHIELD galaxies, as well as a
discussion of the birthrate parameter ($b = SFR_{recent} /
SFR_{lifetime}$) in the context of distance to neighboring systems,
are given in \citet{mcquinn2015a}. A discussion of the uncertainty in
the TRGB distances, which are involved in many calculations and are
propagated through in quadrature, may be found in \citet{mcquinn2014}.

%-----------------------------------------------------------------------------
\subsubsection{\textit{Spitzer} Space Telescope Data Products}
\label{S2.2.5}
%-----------------------------------------------------------------------------%

Observations of the SHIELD galaxies with \textit{Spitzer} were
acquired in 2011-2012 for program GO-80222 (P.I. Cannon).
Near-infrared images at 3.6 and 4.5 $\mu$m were acquired with the
InfraRed Array Camera (IRAC) operating during the ``warm phase''
\citep{werner2004,fazio2004}. The average exposure time was 96 seconds
before mosaicking. Data were acquired from the \textit{Spitzer}
Heritage
Archive\footnote{http://sha.ipac.caltech.edu/applications/Spitzer/SHA}
and corrected for the effects of column pull-down from bright,
saturated foreground objects using the IRAC instrument team
software\footnote{Written by M. Ashby and J. Hora of the IRAC
  instrument team and available at
  http://irsa.ipac.caltech.edu/data/SPITZER/
  docs/dataanalysistools/tools/contributed/ irac/fixpulldown/}.  The
Mosaicker and Point Source Extractor
(MOPEX\footnote{http://irsa.ipac.caltech.edu/data/SPITZER/
  docs/dataanalysistools/tools/mopex/}) was then employed to produce
flux calibrated, artifact-corrected, mosaicked science images for each
galaxy in each filter \citep{marshall2012}. In this paper, the
\textit{Spitzer} images are included both for completeness and for
straightforward visual comparison of the locations of the old stellar
populations of the systems (seen in 3.6 $\mu$m in
Figures~\ref{fig110482} - \ref{fig749241}) to the younger gaseous and
stellar regions imaged with other telescopes. \citet{haurberg2013}
used the \textit{Spitzer} data to derive stellar masses of the SHIELD
systems; the companion paper to this work, \citep{mcnichols16}, shows
the 4.5 $\mu$m images.

%-----------------------------------------------------------------------------
\subsection{Photometric Measurements}
\label{S2.3}
%-----------------------------------------------------------------------------%

Since the SHIELD galaxies have highly irrgular morphologies,
determinations of basic galaxy parameters such as inclination are
non-trivial. Thus, we derived parameters for photometric analysis
using the custom IDL\footnote{Interactive Data Language,
  http://idlastro.gsfc.nasa.gov} program \textsc{CleanGalaxy}
\citep{hagen2014}, which fits surface brightness contours as a
function of galactocentric radius to the HST F606W images of each
source. The position angles, ellipticities, semi-major axes, and
inclinations derived for the ellipses are included in Table
\ref{t5}. A representative ellipse is overlaid on each of the HST
3-color images in Figure~\ref{ELLIPSES}. Dependent values in the plots
and tables of this paper are corrected for inclination effects
following the prescriptions described in \citet{haurberg2013}.

\begin{figure}
\centering
% old location = \plotone{/Users/research/data/vla/110482.ar.eps}
\epsscale{0.8}
\plotone{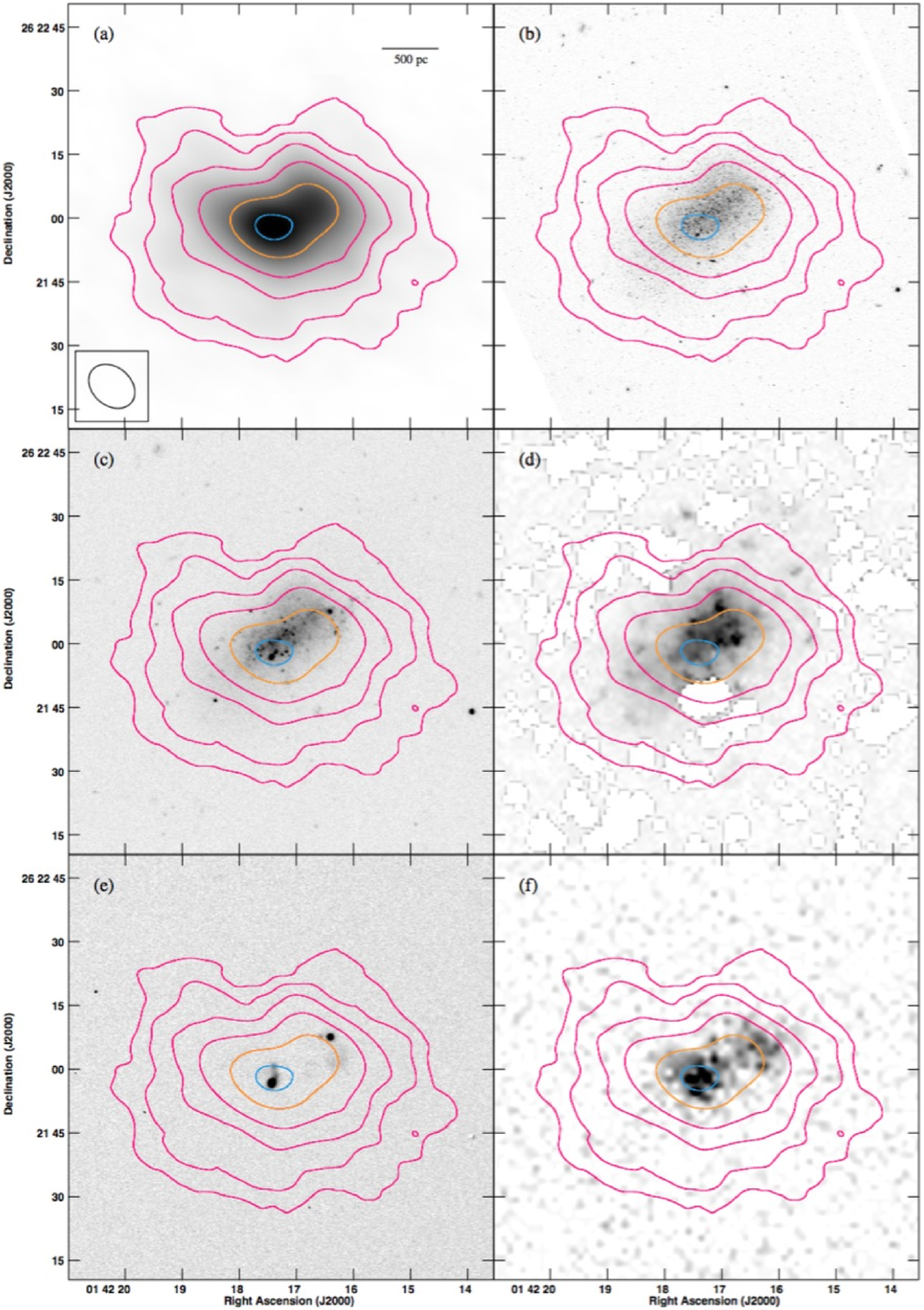}
\caption{AGC\,110482 in VLA \ion{H}{1} (a), HST F606W (b), KPNO WIYN 3.5m
  B-band (c), \textit{Spitzer} 3.6 $\mu$m (d), KPNO WIYN 3.5m
  continuum-subtracted \halpha\ (e), and GALEX FUV (f). The \ion{H}{1} column
  density contours, in units of 10$^{20}$ cm$^{-2}$, are overlaid at
  levels of (0.6125, 1.25, 2.5, 5, 10, 16). The highest two contour
  levels are highlighted in orange (10$\times$10$^{20}$ cm$^{-2}$) and
  blue (16$\times$10$^{20}$ cm$^{-2}$).  The beam size of
  11.98\arcsec$\times$9.04\arcsec\ is shown in panel (a); the \ion{H}{1} images are created using the 
robust-weighted, spectrally averaged data as discussed in Section \ref{S2.2.1}.}
\label{fig110482}
\end{figure}

\begin{figure}
\centering
% old location = \plotone{/Users/research/data/vla/111164.ar.eps}
\epsscale{0.8}
\plotone{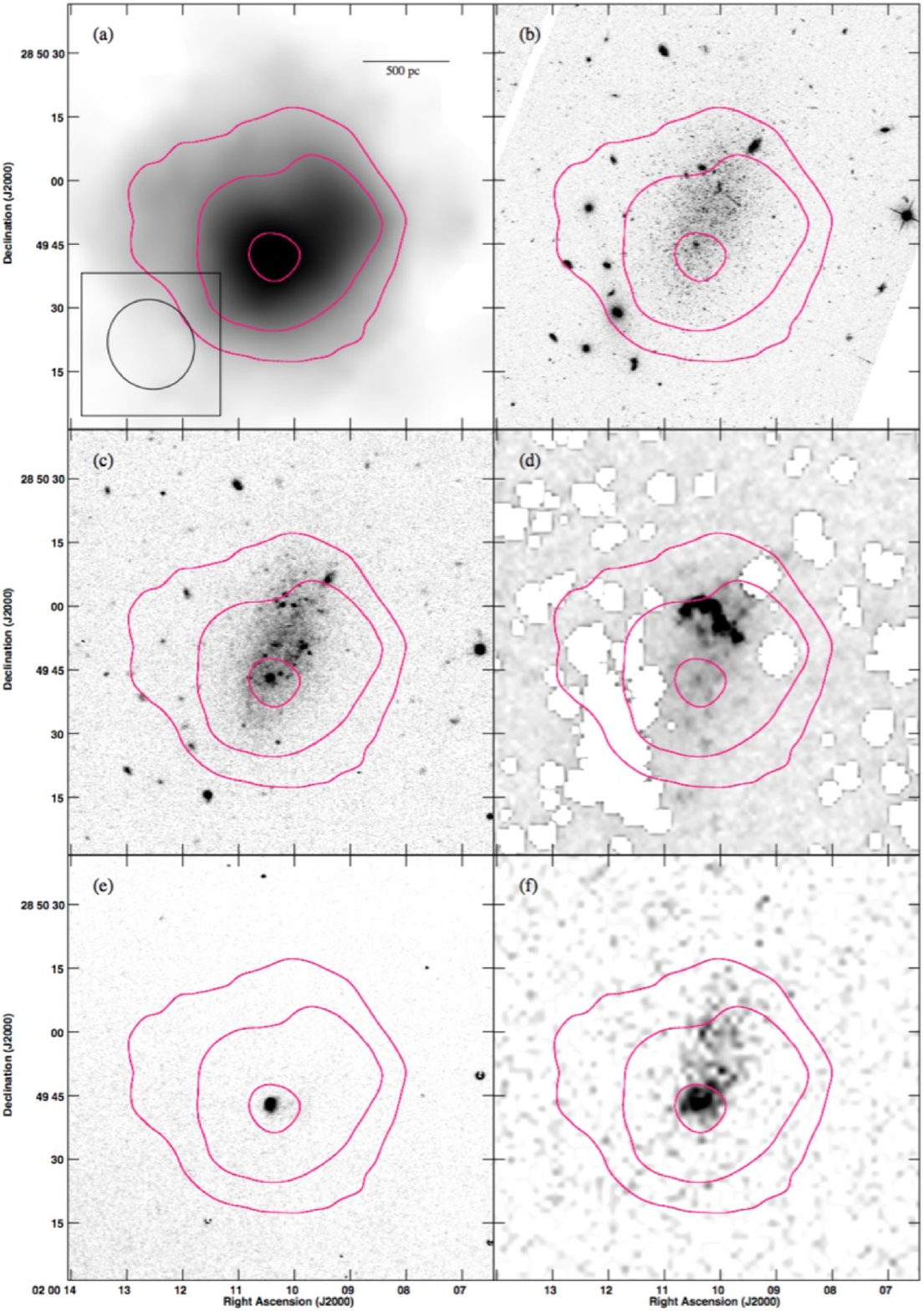}
\caption{AGC\,111164 in VLA \ion{H}{1} (a), HST F606W (b), KPNO WIYN 3.5m
  B-band (c), \textit{Spitzer} 3.6 $\mu$m (d), KPNO WIYN 3.5m continuum-subtracted \halpha\ (e), and
  GALEX FUV (f). The \ion{H}{1} column density contours, in units of 10$^{20}$
  cm$^{-2}$, are overlaid at levels of (0.7, 1.4, 2.8).
  The beam size of 21.56\arcsec$\times$20.24\arcsec\ is shown in panel
  (a); the \ion{H}{1} images are created using the 
robust-weighted, spectrally averaged data as discussed in Section \ref{S2.2.1}.}
\label{fig111164}
\end{figure}

\begin{figure}
\epsscale{1.0}
\centering
% old location = \plotone{/Users/research/data/vla/111946.ar.eps}
\epsscale{0.8}
\plotone{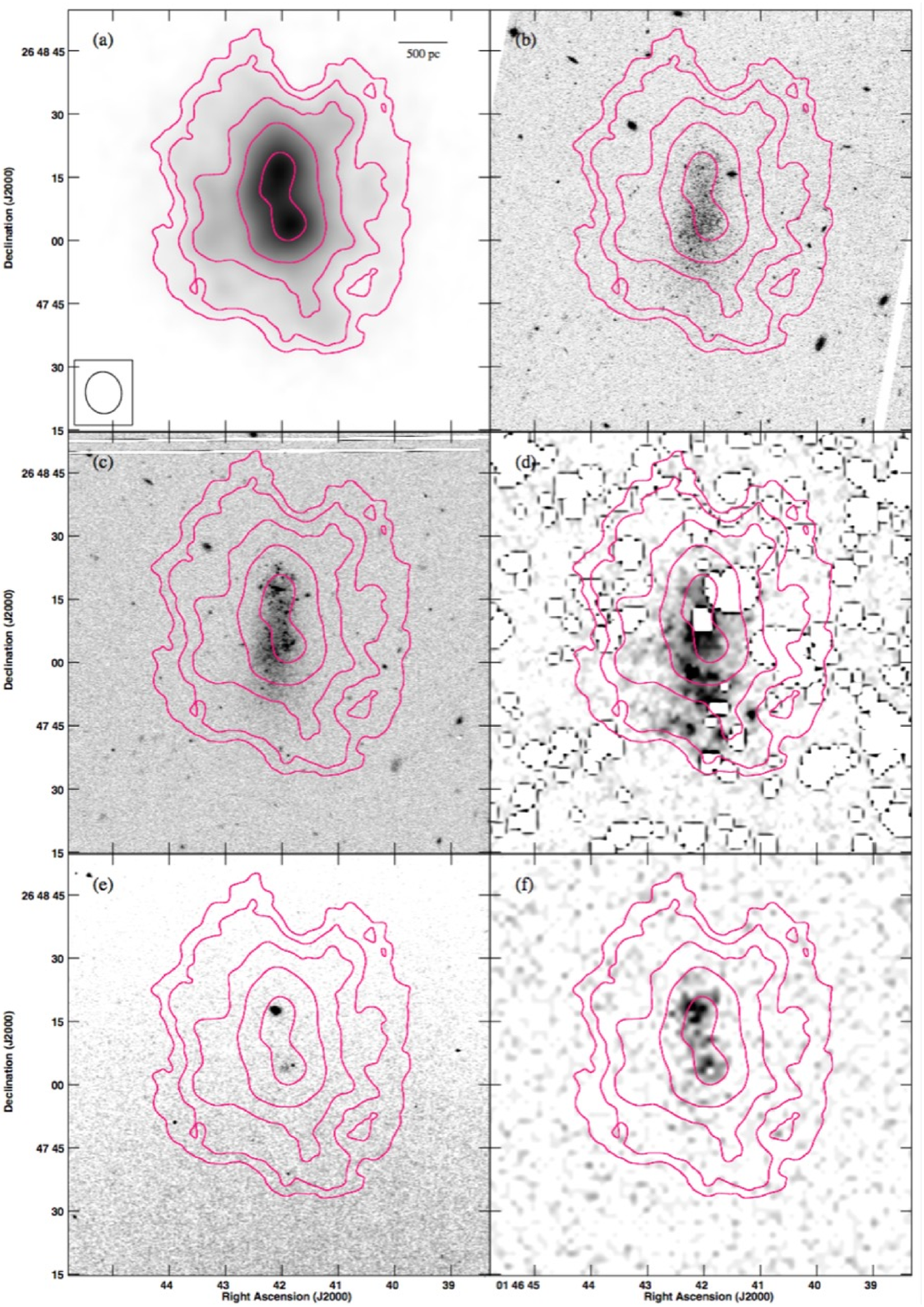}
\caption{AGC\,111946 in VLA \ion{H}{1} (a), HST F606W (b), KPNO WIYN
  3.5m B-band (c), \textit{Spitzer} 3.6 $\mu$m (d), KPNO WIYN 3.5m
  continuum-subtracted \halpha\ (e), and GALEX FUV (f). The \ion{H}{1}
  column density contours, in units of 10$^{20}$ cm$^{-2}$, are
  overlaid at levels of (0.5, 1, 2, 4, 8). The beam size of
  10.30\arcsec$\times$8.86\arcsec\ is shown in panel (a); the \ion{H}{1} images are created using the 
robust-weighted, spectrally averaged data as discussed in Section \ref{S2.2.1}.}
\label{fig111946}
\end{figure}

\begin{figure}
\epsscale{1.0}
\centering
% old location = \plotone{/Users/research/data/vla/111977.ar.eps}
\epsscale{0.8}
\plotone{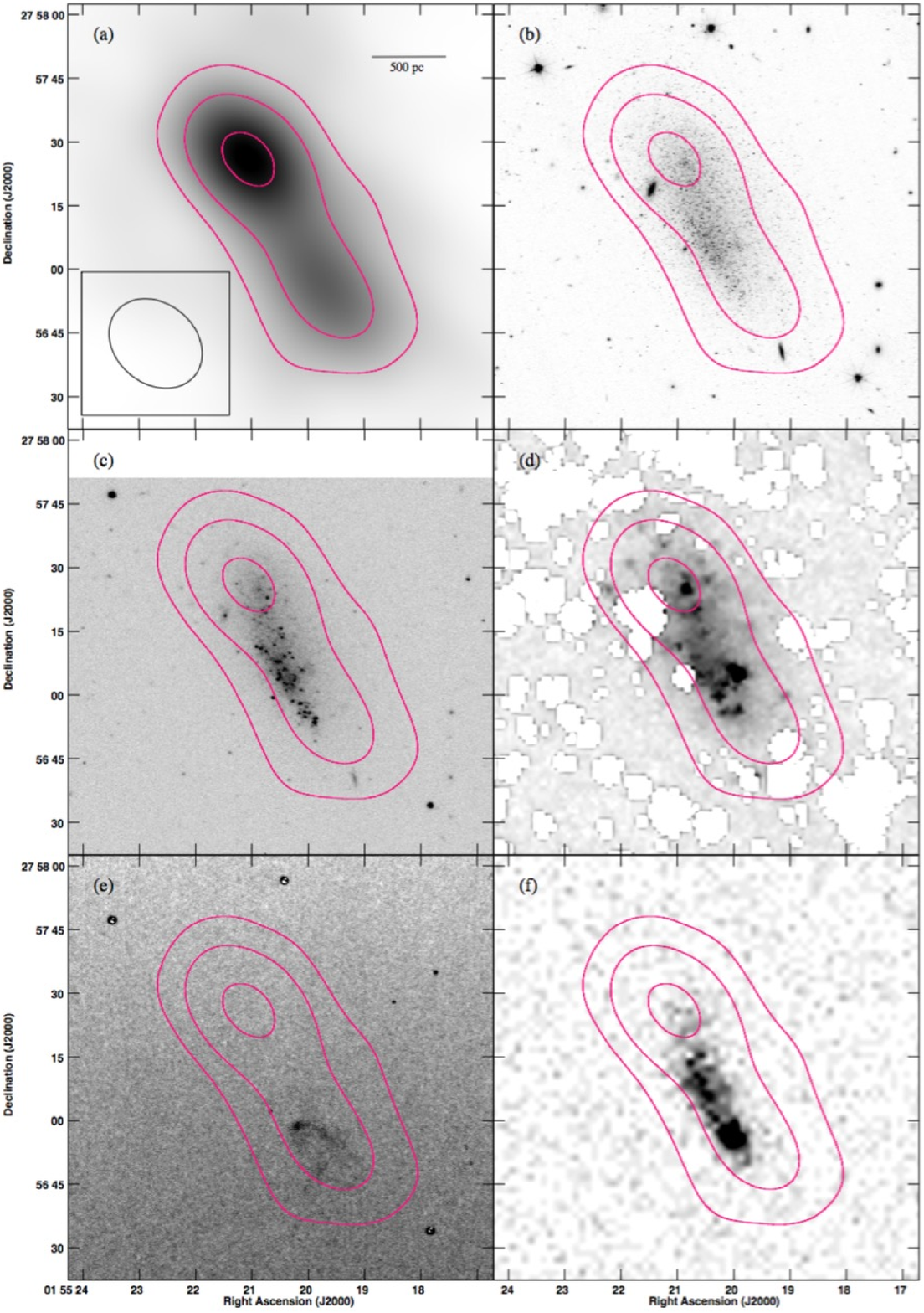}
\caption{AGC\,111977 in VLA \ion{H}{1} (a), HST F606W (b), KPNO WIYN 3.5m
  B-band (c), \textit{Spitzer} 3.6 $\mu$m (d), KPNO WIYN 3.5m continuum-subtracted \halpha\ (e), and
  GALEX FUV (f). The \ion{H}{1} column density contours, in units of 10$^{20}$
  cm$^{-2}$, are overlaid at levels of (0.6125, 1.25, 2.5).
  The beam size of 24.01\arcsec$\times$19.95\arcsec\ is shown in panel
  (a); the \ion{H}{1} images are created using the 
robust-weighted, spectrally averaged data as discussed in Section \ref{S2.2.1}.}
\label{fig111977}
\end{figure}

\begin{figure}
\epsscale{1.0}
\centering
% old location = \plotone{/Users/research/data/vla/112521.ar.eps}
\epsscale{0.8}
\plotone{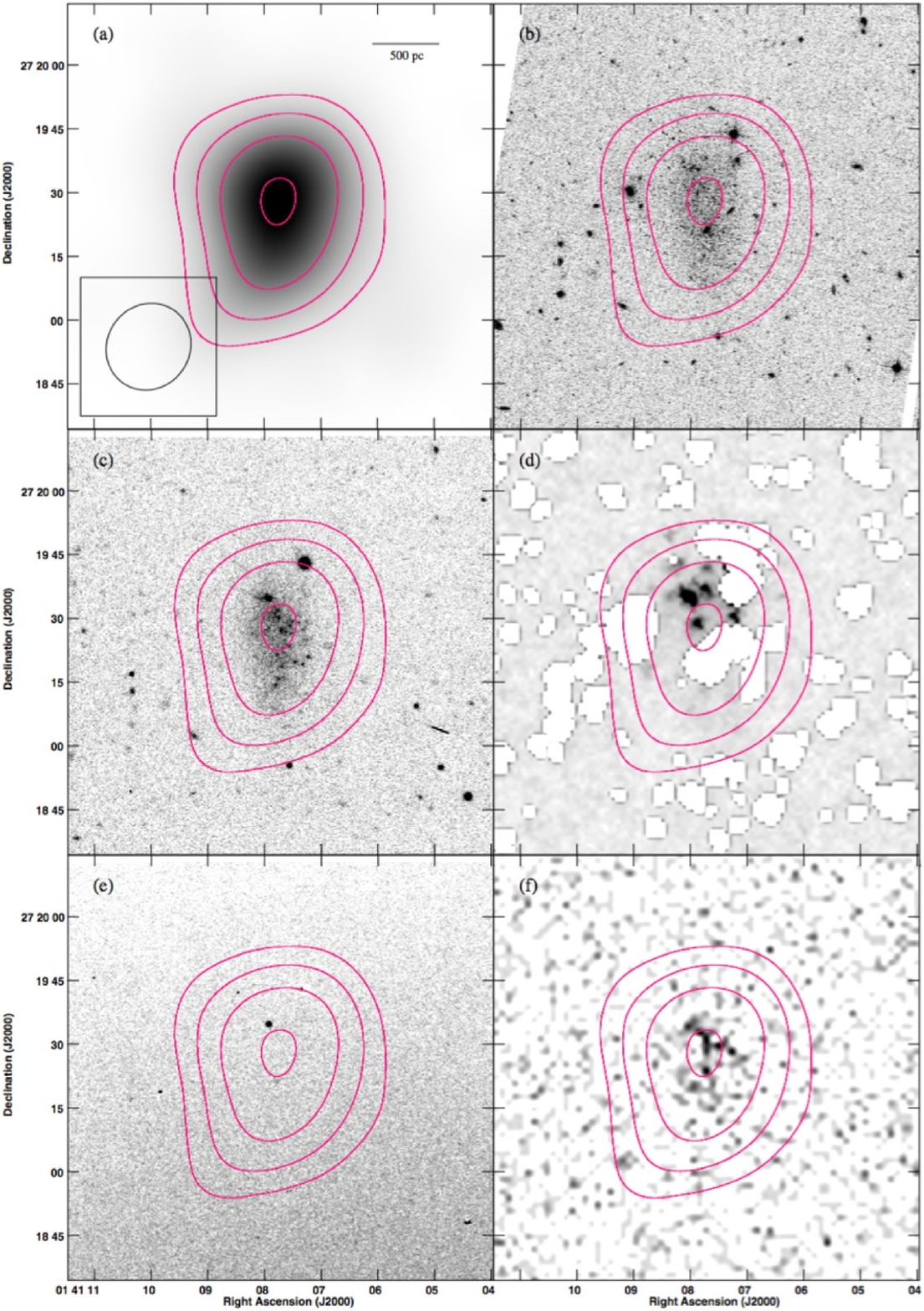}
\caption{AGC\,112521 in VLA \ion{H}{1} (a), HST F606W (b), KPNO WIYN 3.5m
  B-band (c), \textit{Spitzer} 3.6 $\mu$m (d), KPNO WIYN 3.5m continuum-subtracted \halpha\ (e), and
  GALEX FUV (f). The \ion{H}{1} column density contours, in units of 10$^{20}$
  cm$^{-2}$, are overlaid at levels of (0.5625, 1.125, 2.25, 4.5).
  The beam size of 22.03\arcsec$\times$19.51\arcsec\ is shown in panel
  (a); the \ion{H}{1} images are created using the 
robust-weighted, spectrally averaged data as discussed in Section \ref{S2.2.1}.}
\label{fig112521}
\end{figure}

\begin{figure}
\epsscale{1.0}
\centering
% old location = \plotone{/Users/research/data/vla/174585.ar.eps}
\epsscale{0.8}
\plotone{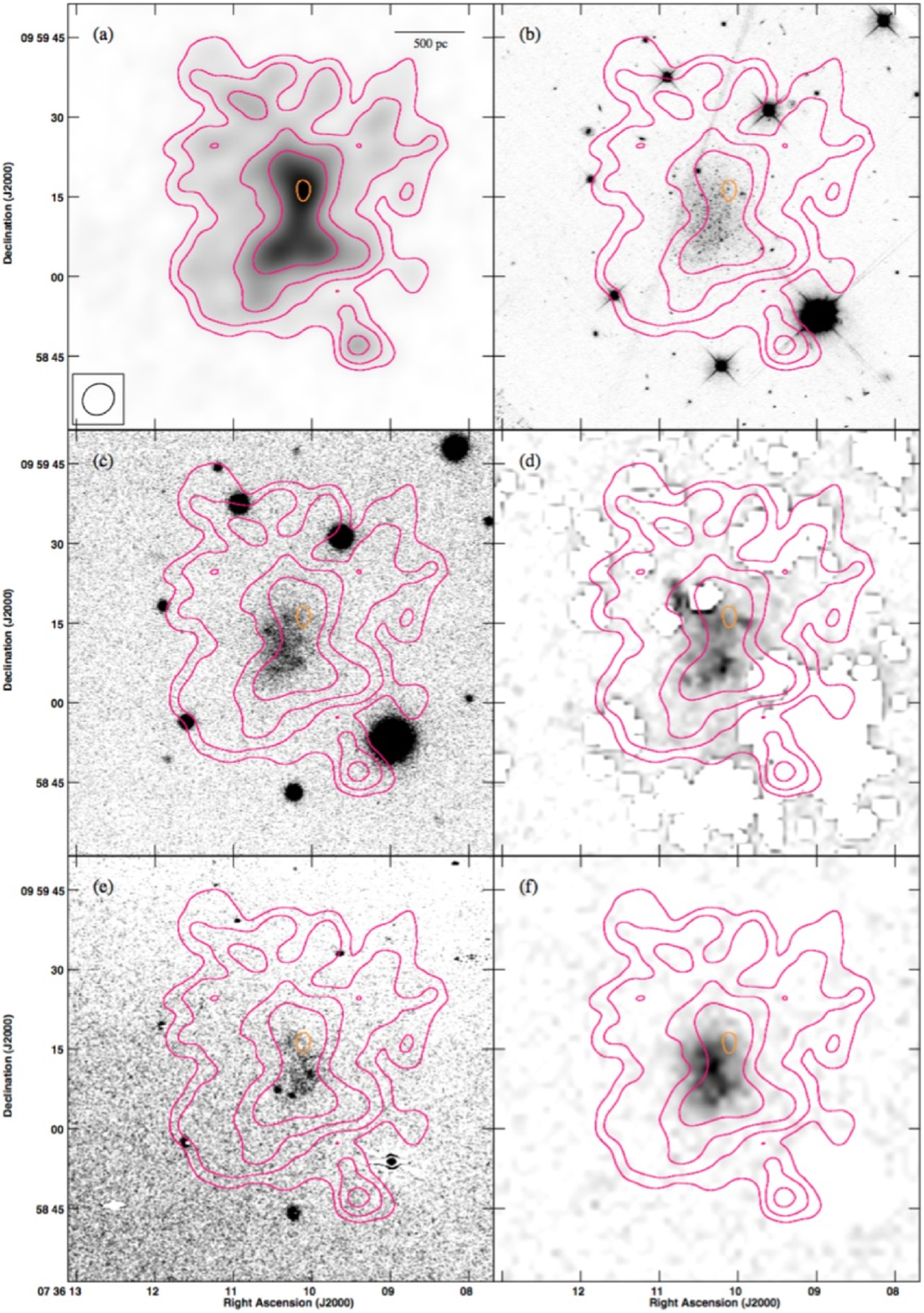}
\caption{AGC\,174585 in VLA \ion{H}{1} (a), HST F606W (b), KPNO WIYN 3.5m
  B-band (c), \textit{Spitzer} 3.6 $\mu$m (d), KPNO WIYN 3.5m continuum-subtracted \halpha\ (e), and
  GALEX FUV (f). The \ion{H}{1} column density contours, in units of 10$^{20}$
  cm$^{-2}$, are overlaid at levels of (0.6125, 1.25, 2.5, 5, 10). The
  highest contour level is highlighted in orange (10$\times$10$^{20}$
  cm$^{-2}$). The beam size of 6.19\arcsec$\times$5.52\arcsec\ is
  shown in panel (a); the \ion{H}{1} images are created using the 
robust-weighted, spectrally averaged data as discussed in Section \ref{S2.2.1}.}
\label{fig174585}
\end{figure}

\begin{figure}
\epsscale{1.0}
\centering
% old location = \plotone{/Users/research/data/vla/174605.ar.eps}
\epsscale{0.8}
\plotone{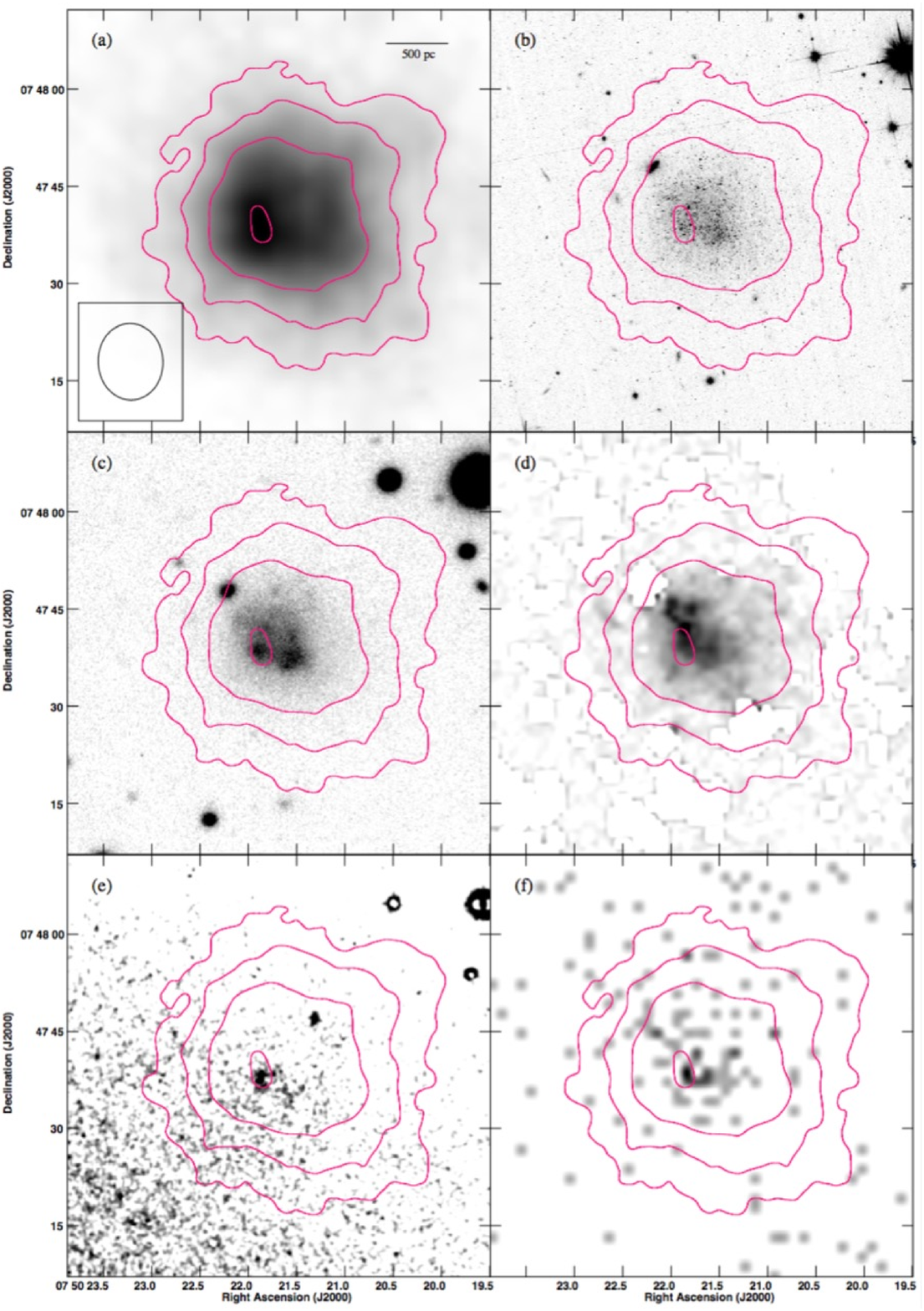}
\caption{AGC\,174605 in VLA \ion{H}{1} (a), HST F606W (b), KPNO WIYN 3.5m
  B-band (c), \textit{Spitzer} 3.6 $\mu$m (d), KPNO WIYN 3.5m continuum-subtracted \halpha\ (e), and
  GALEX FUV (f). The \ion{H}{1} column density contours, in units of 10$^{20}$
  cm$^{-2}$, are overlaid at levels of (0.55, 1.1, 2.2, 4.4).  The beam
  size of 11.81\arcsec$\times$9.99\arcsec\ is shown in panel (a); the \ion{H}{1} images are created using the 
robust-weighted, spectrally averaged data as discussed in Section \ref{S2.2.1}.}
\label{fig174605}
\end{figure}

\begin{figure}
\epsscale{1.0}
\centering
% old location = \plotone{/Users/research/data/vla/182595.ar.eps}
\epsscale{0.8}
\plotone{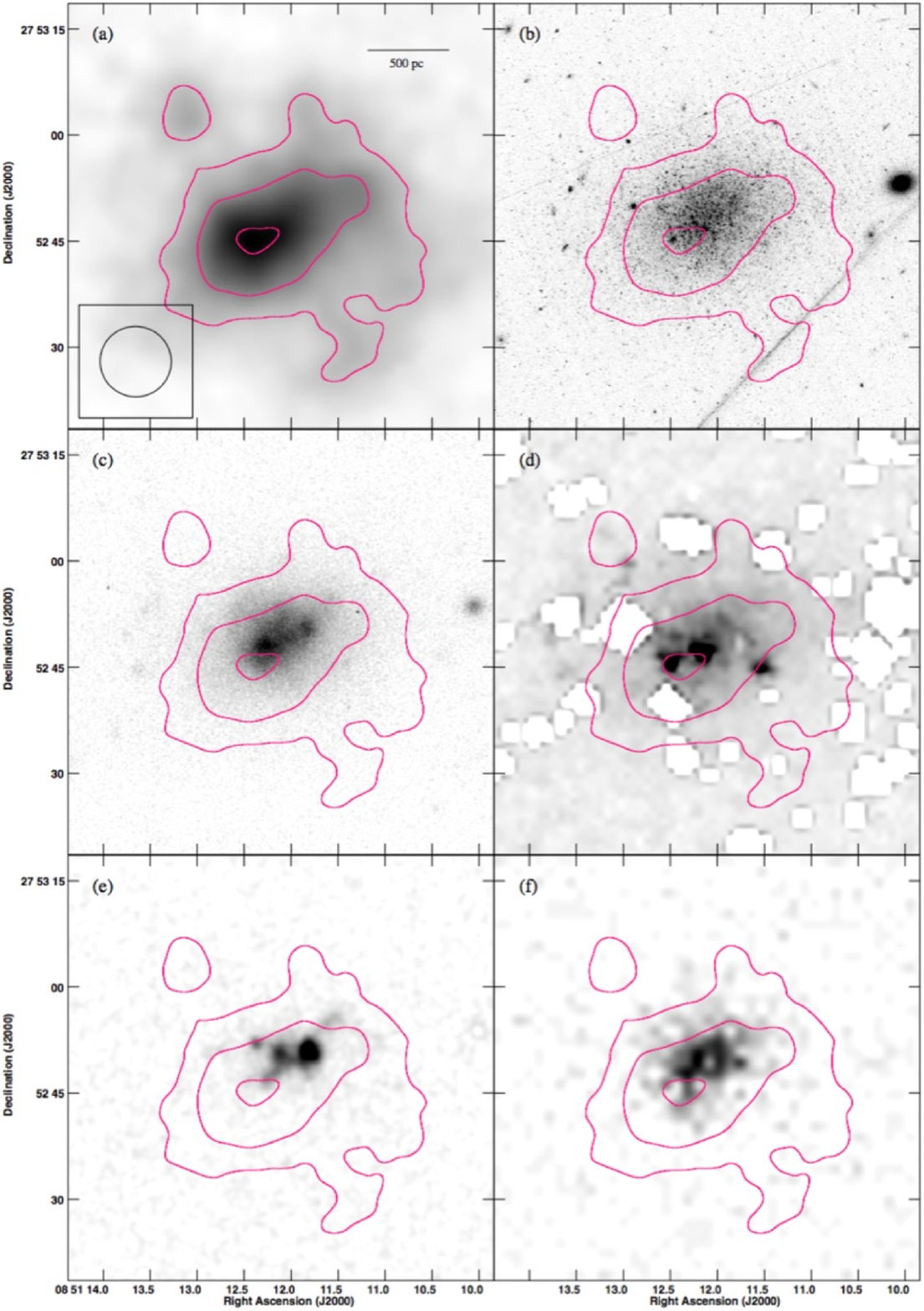}
\caption{AGC\,182595 in VLA \ion{H}{1} (a), HST F606W (b), KPNO WIYN 3.5m
  B-band (c), \textit{Spitzer} 3.6 $\mu$m (d), KPNO WIYN 3.5m continuum-subtracted \halpha\ (e), and
  GALEX FUV (f). The \ion{H}{1} column density contours, in units of 10$^{20}$
  cm$^{-2}$, are overlaid at levels of (0.75, 1.5, 3).  The beam size of
  10.05\arcsec$\times$9.93\arcsec\ is shown in panel (a); the \ion{H}{1} images are created using the 
robust-weighted, spectrally averaged data as discussed in Section \ref{S2.2.1}.}
\label{fig182595}
\end{figure}

\begin{figure}
\epsscale{1.0}
\centering
% old location = \plotone{/Users/research/data/vla/731457.ar.eps}
\epsscale{0.8}
\plotone{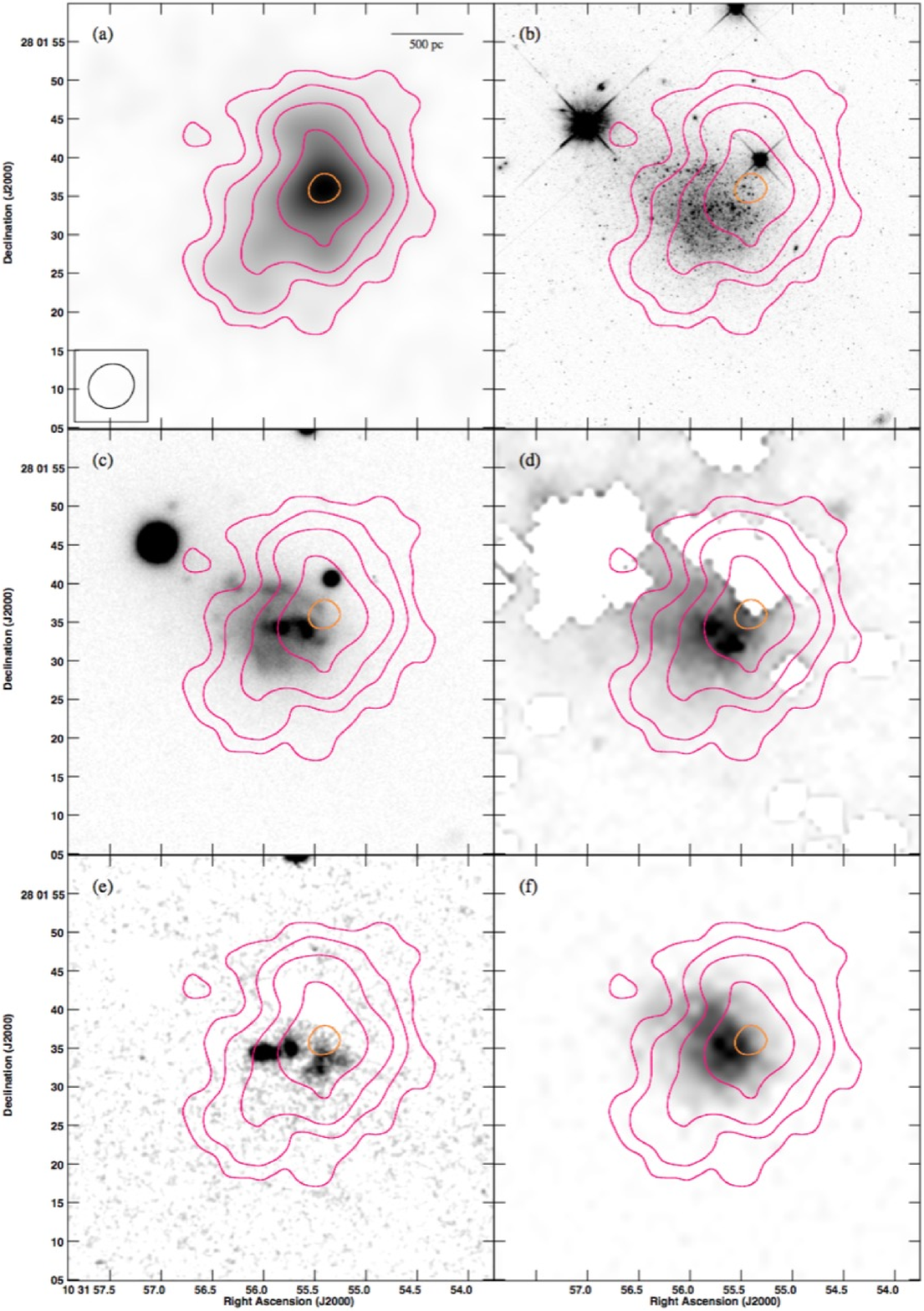}
\caption{AGC\,731457 in VLA \ion{H}{1} (a), HST F606W (b), KPNO WIYN 3.5m
  B-band (c), \textit{Spitzer} 3.6 $\mu$m (d), KPNO WIYN 3.5m continuum-subtracted \halpha\ (e), and
  GALEX FUV (f). The \ion{H}{1} column density contours, in units of 10$^{20}$
  cm$^{-2}$, are overlaid at levels of (0.6125, 1.25, 2.5, 5, 10). The
  highest contour level is highlighted in orange (10$\times$10$^{20}$
  cm$^{-2}$). The beam size of 6.04\arcsec$\times$5.53\arcsec\ is
  shown in panel (a); the \ion{H}{1} images are created using the 
robust-weighted, spectrally averaged data as discussed in Section \ref{S2.2.1}.}
\label{fig731457}
\end{figure}

\begin{figure}
\epsscale{1.0}
\centering
% old location = \plotone{/Users/research/data/vla/748778.ar.eps}
\epsscale{0.8}
\plotone{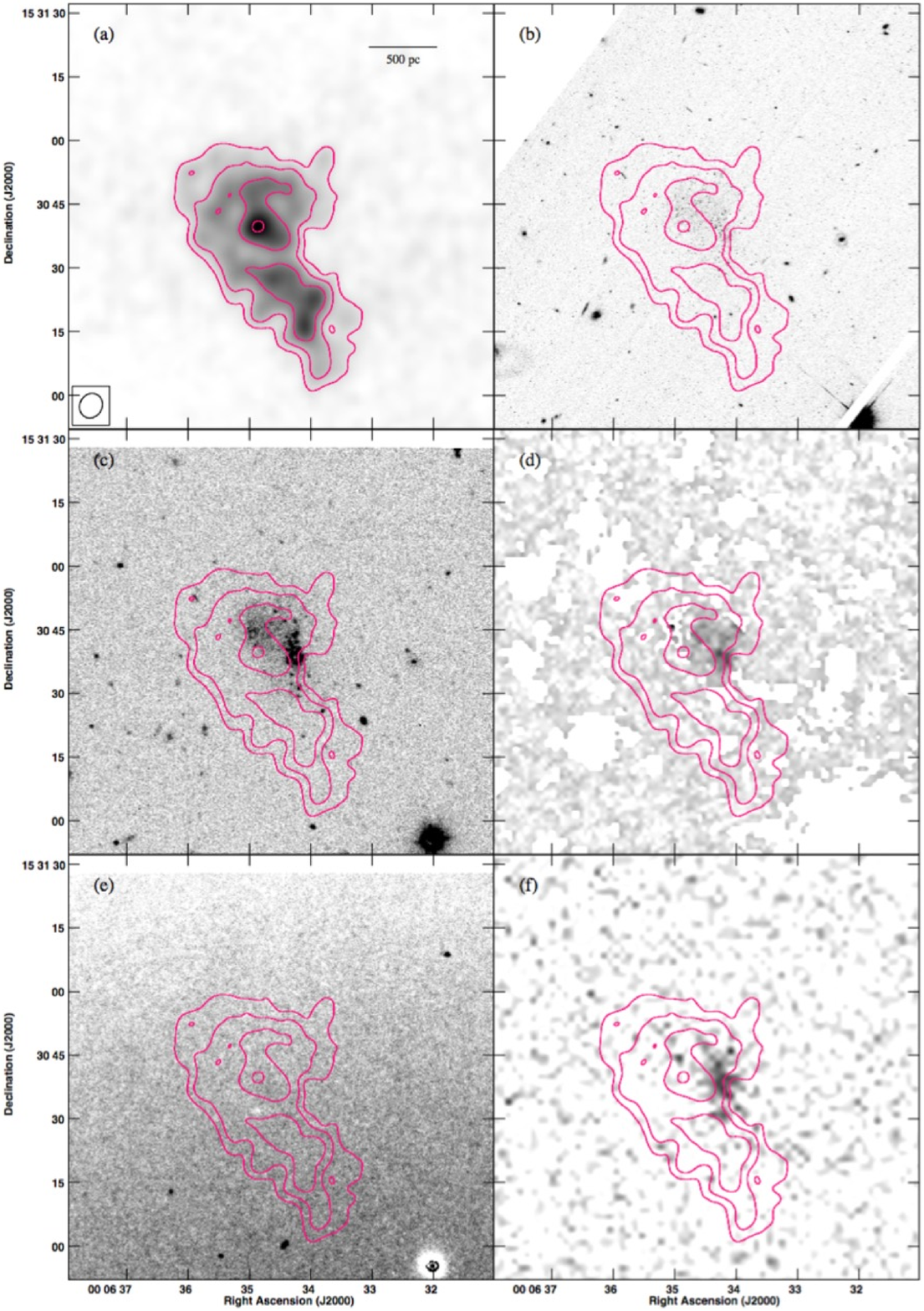}
\caption{AGC\,748778 in VLA \ion{H}{1} (a), HST F606W (b), KPNO WIYN 3.5m
  B-band (c), \textit{Spitzer} 3.6 $\mu$m (d), KPNO WIYN 3.5m continuum-subtracted \halpha\ (e), and
  GALEX FUV (f). The \ion{H}{1} column density contours, in units of 10$^{20}$
  cm$^{-2}$, are overlaid at levels of (0.6875, 1.375, 2.75, 5.5).  The
  beam size of 5.91\arcsec$\times$5.23\arcsec\ is shown in panel (a); the \ion{H}{1} images are created using the 
robust-weighted, spectrally averaged data as discussed in Section \ref{S2.2.1}.}
\label{fig748778}
\end{figure}

In Figures~\ref{fig110482} through \ref{fig749241}, we show a
six-panel mosaic for each SHIELD galaxy.  Panel (a) shows the
\ion{H}{1} image used in our SF analysis in greyscale (created using
the robust-weighted, spectrally-smoothed data products described in
Section \ref{S2.2.1}) and with column density contours overlaid to
demonstrate dynamic range.  The same contours are shown on all
remaining panels: (b) shows a greyscale representation of the HST
F606W image; (c) shows the WIYN 3.5\,m B-band image; (d) shows the
Spizter 3.6\,$\mu$m image; (e) shows the WIYN 3.5\,m
continuum-subtracted \halpha\ image; (f) shows the GALEX FUV image.
All images are regridded to the HST image pixel scale and orientation.
Note that all \textit{Spitzer} images are shown with identified
foreground and background objects removed, except for AGC\,749241,
which is a non-detection. These mosaics facilitate an important visual
assessment of the degree to which the gas in these galaxies correlates
with regions of ongoing SF. The FUV and \halpha\ regions are sometimes
co-spatial with the \ion{H}{1} knots (e.g., AGC\,110482, AGC 111946),
but some systems show the opposite: SF regions appear where \ion{H}{1}
minima occur (e.g., AGC\,111977, AGC\,749241). An in-depth discussion
of the situation for each individual galaxy follows in
Section~\ref{S4.1}.

In order to compare photometric measurements at multiple wavelengths,
the images must be registered to the same coordinate grid. For the
surface brightness profile analysis, we preserved the original pixel
scale of the individual images in order to faithfully represent the
fluxes contained within the elliptical annuli. The final images
(Figures~\ref{fig110482} - \ref{fig749241}) were regridded to the HST
fields using the MIRIAD\footnote{http://bima.astro.umd.edu/miriad/}
task \textsc{regrid}. For the pixel correlation procedures discussed
in Section~\ref{S4}, the FUV data were smoothed with a Gaussian kernel
equal to the size of the restoring beam in the \ion{H}{1} maps and
then regridded to those \ion{H}{1} maps. Radial profiles were produced
by integrating over concentric elliptical annuli in the resulting
\ion{H}{1}, \halpha, and FUV images.

%-----------------------------------------------------------------------------%
\section{Star Formation Rates}
\label{S3}
%-----------------------------------------------------------------------------%

Observations of the SHIELD sources in the far-ultraviolet (FUV)
continuum and in the \halpha\ emission line provide constraints on the
SFRs of the galaxies. Both of these tracers are attributable to the
formation of young stars \citep{kennicutt1998a}. FUV radiation has
been used as a tracer of spatially and temporally extended SF,
revealing populations of relatively young stars (lifetimes
$\sim$10$^{8}$ yr, masses $>$ 6 \msun; e.g., {Salim
  \etal\ 2007}\nocite{salim2007}). Note that usage of any FUV scaling
relation to infer a SFR assumes that SF has been constant over a
$\sim$100 Myr timescale. Near-ultraviolet (NUV) emission is also
occasionally converted to a SFR, although FUV is generally preferred
over NUV: 1) NUV images are more contaminated by foreground stellar
sources than FUV images; 2) the NUV emission is generally less
reliable for tracing the recent SF since the flux at the redder NUV
wavelengths will have a greater contribution from stars with lifetimes
$>$ 10$^{8}$ yr \citep{lee2009}.

While the ultraviolet continuum probes SF over the most recent
$\sim$100 Myr, \halpha\ line emission provides an almost instantaneous
snapshot of the formation of massive stars.  Since only stars more
massive than $\sim$17 \msun\ can produce significant numbers of
photons capable of ionizing neutral hydrogen \citep{lee2009}, and
these stars have very short main sequence lifetimes, the presence of
significant recombination line emission requires the presence of such
short-lived stars.  Due to this direct coupling between the nebular
emission and massive SF, various works have used \halpha\ emission to
probe the ongoing SF in spiral and irregular galaxies (e.g.,
{Kennicutt 1983}\nocite{kennicutt1983}, {Kennicutt
  1998}\nocite{kennicutt1998a}).

Even though FUV and H$\alpha$ emission have different characteristic
timescales, both SF indicators are expected to agree in systems with
fully populated initial mass functions (IMF) and constant
SFRs. Indeed, \citet{lee2009} find a constant ratio of \halpha\ to FUV
emission in systems with global SFRs larger than $\sim$0.1
\msun\,yr$^{-1}$.  However, as the integrated SFR falls, the
\halpha-based SFR no longer tracks the FUV-based SFR. Possible reasons
are numerous, including non-constant SF histories, stochasticity in
the upper IMF, and leakage of ionizing photons, among others
\citep{lee2016}. The SHIELD galaxies provide a unique perspective on
this issue; as Figures~\ref{fig110482} - \ref{fig749241} demonstrate,
and as discussed below, some of the SHIELD galaxies harbor only single
\ion{H}{2} regions. The effects of stochasticity are expected to be
significant.

%-----------------------------------------------------------------------------%
\subsection{FUV SFRs}
\label{S3.1}
%-----------------------------------------------------------------------------%

Many conversions from FUV luminosity to SFR exist in the literature.
Most assume a fully populated IMF and/or a Solar metallicity
\citep{kennicutt1998a,salim2007,hao2011,kennicuttevans2012}.  As
discussed in \citet{haurberg2015} and \citet{mcquinn2015a}, both of
these assumptions fail for the SHIELD galaxies.  A preferable approach
takes into account a varying and stochastically populated IMF.  We
thus apply the recent empirical calibration of \citet{mcquinn2015b},
which is based on a randomly populated Salpeter IMF with mass limits
between 0.1 and 120 \msun\ for low-metallicity dwarf galaxies with
SFR$_{\rm FUV}$ between 10$^{-3}$ and 10$^{-1}$ \msun\ yr$^{-1}$:

\begin{equation}
\rm{SFR_{FUV} = 2.04 \pm 0.81 \times 10^{-28} \times L^{0}_{FUV}}
\end{equation}

\noindent where SFR$_{\rm FUV}$ is in \msun\ yr$^{-1}$ and L$^{0}_{\rm
  FUV}$ is in erg s$^{-1}$ Hz$^{-1}$. This prescription yields
somewhat higher SFR$_{\rm FUV}$ values (by $\sim53\%$) than the other
prescriptions in the works cited above.  The final global FUV-derived
SFRs are included in Table~\ref{t3}.

Note that no FUV data are available for AGC\,749237, and so we have
calculated an approximate value for the SFR$_{\rm FUV}$ based on its
NUV emission. We empirically determined the correlation between the
FUV and NUV counts per second for each galaxy and use the resulting
line-of-best fit equation to estimate what the FUV counts for
AGC\,749237 would be, given its observed NUV counts per second. We
then correct for extinction and carry this through to a SFR$_{\rm
  FUV}$.

In Figure~\ref{SFRvsHImass} we plot the total \ion{H}{1} mass versus
the FUV-based SFR for the SHIELD galaxies and for the members of other
nearby galaxy surveys. Note that we did not revise the methods used to
calculate SFR$_{\rm FUV}$ for the galaxies in the other surveys; the
scatter in SFR$_{\rm FUV}$ may be larger as a result. This figure
demonstrates that the observed SFR$_{\rm FUV}$ of our sample members
is comparable to members of some of the other low-mass dwarf galaxy
surveys. As expected, there is a correlation between the total $M_{\rm
  \HI}$ of the systems and the SFR$_{\rm FUV}$. Treating each survey's
sample of galaxies independently, the plot also fits a linear
regression to each group; the SHIELD and FIGGS galaxies, selected in
similar ways (i.e., had to meet \ion{H}{1} line width criteria), show
the closest agreement in slope. Note that except for a small fraction
of the LITTLE-THINGS galaxies, all systems which have log(M$_{\rm
  \HI}$) $\lsim$ 7.5 \msun\ have log(SFR$_{\rm FUV}$) $\lsim$ $-$2.0
\msun\ yr$^{-1}$.

In Figure~\ref{SFRvsHImass-composite} we treat all of the galaxy
surveys uniformly and fit the composite sample with a single linear
regression.  When no errorbars for archival data were available, we
assumed a 10\% uncertainty on either variable.  The slope of the
resulting line is 0.88\,$\pm$\,0.01; if the more massive THINGS
galaxies \citep{walter2008} are excluded the slope becomes slightly
more shallow (0.85\,$\pm$\,0.01). We note that these formal
uncertainties are likely underestimates, due to the different SFR
metrics used by different comparison studies.  With this caveat in
mind, from Figure~\ref{SFRvsHImass-composite} we conclude that the HI
mass and UV-based SFR are strongly correlated; the SFE is essentially
constant over five orders of magnitude.

%-----------------------------------------------------------------------------%
\subsection{\halpha\ SFRs}
\label{S3.2}
%-----------------------------------------------------------------------------%

The most widely-used conversion from \halpha\ luminosity to an
instantaneous SFR is given by \citet{kennicutt1998a}:

\begin{equation}
\rm{SFR_{H\alpha} = L_{H\alpha}/(1.26 \times 10^{41})}
\end{equation}

\noindent where the SFR$_{\rm H\alpha}$ is the \halpha-based SFR in
units of \msun\ yr$^{-1}$ and L$_{\rm H\alpha}$ is the
\halpha\ luminosity in units of erg\,s$^{-1}$. This prescription
assumes a fully populated Salpeter IMF (i.e., there is a sufficient number of
stars throughout the entire mass range, in this case 0 - 100
\msun). Based on the very low FUV-based SFRs, stochasticity in the
upper IMF (the high-mass star regime) is expected to be significant in
the SHIELD galaxies. However, in the absence of an alternative
calibration of \halpha\ luminosities into SFRs in the extreme low SFR
regime, we adopt the \citet{kennicutt1998a} calibration for purposes
of direct comparison. The global SFR$_{\rm H\alpha}$ for the SHIELD
galaxies are included in Table~\ref{t3}. Note that AGC\,748778 and
AGC\,749241 are \halpha\ non-detections, and so we provide upper
limits for the \halpha-based SFR. The range of SFR$_{\rm H\alpha}$ of
the sample is 10$^{-2.34}$ \msun\ yr$^{-1}$ (AGC\,749237) to
10$^{-4.10}$ \msun\ yr$^{-1}$ (AGC\,112521).

%-----------------------------------------------------------------------------%
\subsection{Comparison of \halpha\ and FUV SFRs}
\label{S3.3}
%-----------------------------------------------------------------------------%

In Figure~\ref{FUVvsHAsfrs} we plot SFR$_{\rm H\alpha}$ as a function
of SFR$_{\rm FUV}$ for the SHIELD galaxies and for comparison samples
drawn from the 11HUGS \citep{lee2009}, LITTLE-THINGS
\citep{hunter2012}, and FIGGS \citep{roychowdhury2014} samples. The
overall trend of increasing scatter at decreasing SFR is apparent in
all survey samples. The SHIELD galaxies are some of the most
extremely low-SFR galaxies in these surveys.

In addition to increased scatter at the faint end of
Figure~\ref{FUVvsHAsfrs}, the SFR$_{\rm H\alpha}$/SFR$_{\rm FUV}$
ratio becomes progressively smaller as a function of decreasing SFR.
This trend is demonstrated clearly in Table~\ref{t3} and
Figure~\ref{SFRratios}, where the same samples of galaxies are shown
in the SFR$_{\rm H\alpha}$/SFR$_{\rm FUV}$ vs. SFR$_{\rm H\alpha}$
plane. The most quiescent systems detected in \halpha\ are also the
most quiescent systems detected in the FUV. As seen in multiple
previous works, the decreased \halpha\ luminosity relative to FUV is
an expected effect of stochasticity at the upper end of the IMF in the
low SF regime \citep{lee2009, salim2007,
  bellkennicutt2001}. \citet{meurer2009, lee2009, boselli2009} all
noted discrepancies between UV and \halpha\ SFRs in the sense that
\halpha\ SFRs are systematically lower with decreasing galaxy
luminosity. This was originally interpreted as potentially due to a
change in the upper IMF with decreasing galaxy mass/metallicity.
However, {Fumagalli \etal\ (2011)}\nocite{fumagalli2011}, {da Silva
  \etal\ (2012)}\nocite{dasilva2012}, and {Eldridge
  \etal\ (2012)}\nocite{eldridge2012} all showed that the observed
trend could be explained by a stochastically sampled cluster and
stellar mass function scenario. \citet{weisz2012} demonstrated that
the observed trend was a natural result of fluctuating SF rates $-$
exactly what is seen in low mass galaxies.

%-----------------------------------------------------------------------------%
\section{Star Formation vs. \ion{H}{1} in SHIELD}
\label{S4}
%-----------------------------------------------------------------------------%

We now examine the complex relationships between the spatial
distribution of \ion{H}{1} gas and the spatial distribution of the SF
tracers. We perform this analysis using several different
strategies, each of which is discussed in detail below.  

First, and most simplistically, we visually inspect the images of
each galaxy (Figures~\ref{fig110482} - \ref{fig749241}) at each of the
wavelengths available. Analyzing the morphology of the galaxies allows
us to identify parts of a system in which the HI gas and SF tracers
are co-spatial, as well as areas in a system which have elevated SF
but are devoid of HI gas. It is also interesting to explore the
reasons why this co-spatiality does or does not occur; the visual
arrangement of the SF regions and gas within a galaxy can inform us
about how much the SF events disrupt the gas and how long it takes for
SF to clear out clumps of atomic gas.

Second, we calculate and examine radial profiles of the \ion{H}{1} gas
surface density ($\Sigma_{\rm \HI}$, units of \msun\,pc$^{-2}$), the
\halpha-based SFR surface density ($\Sigma_{\rm H\alpha\ SFR}$, units
of \msun\,yr$^{-1}$\,kpc$^{-2}$), and the FUV-based SFR surface
density ($\Sigma_{\rm FUV\ SFR}$, units of
\msun\,yr$^{-1}$\,kpc$^{-2}$) as functions of radius within each
galaxy.  In order to be systematic and uniform, all profiles are
centered on the fitted center of the HST F606W images as determined by
\textsc{CleanGalaxy} \citep{hagen2014}; these positions are the
centers of the white ellipses shown in Figure~\ref{ELLIPSES}. They are
presented as surface densities in concentric annuli, so the axes in
these plots are radius (in kpc) and $\Sigma_{\rm SFR}$ or $\Sigma_{\rm
  \HI}$. Figure~\ref{RadProfs} shows these profiles for all 12 SHIELD
galaxies. In general, the \ion{H}{1} profiles show smooth curves while
the \halpha\ and FUV profiles are more choppy and reach the level of
noise at smaller radii. In some cases the variations seen in the
\halpha\ curves are reflected in the FUV curves (e.g., AGC\,111164,
AGC\,174585), but in other cases the profiles do not mimic each
other's shape (e.g., AGC\,110482, AGC\,174605). The reason for these
matchups or discrepancies could be associated with the temporal nature
of the \halpha\ emission compared to FUV: a recently-depleted
\halpha\ region could certainly still be bright in FUV.

Third, we plot $\Sigma_{\rm SFR}$ against $\Sigma_{\rm \HI}$ in a
variety of ways. One method uses the FUV, \halpha, and \ion{H}{1}
emssion within each annulus described above; the surface densities
within the concentric annuli are plotted and a slope is derived from
the points with values $>$ 5$\sigma$. We present plots of $\Sigma_{\rm
  FUV\ SFR}$ vs. $\Sigma_{\rm \HI}$ (using the FUV SFR prescription
from {McQuinn \etal\ 2015b}\nocite{mcquinn2015b}) and $\Sigma_{\rm
  H\alpha\ SFR}$ vs. $\Sigma_{\rm \HI}$ (using the {Kennicutt
  1998}\nocite{kennicutt1998a} \halpha\ SFR prescription) in Figures~
\ref{FUVSFE} and \ref{K98SFE}, respectively. Because the \halpha\ data
is often clumpy, not radially extended, and in some cases does not
have a good S/N ratio, Figure~\ref{K98SFE} is difficult to interpret
with confidence; thus, we do not consider this plot in our final
assessment of the average K-S slope for the
sample. Figure~\ref{FUVSFE} includes data for every source in the
sample, and because the emission is more smoothly distributed than the
\halpha, we have more confidence in using elliptical annuli to find
the slope value. The slopes derived in each of these figures and the
sample average are included in Table~\ref{t6}; note that the
uncertainties quoted are formal uncertainties only.

Fourth, we examine $\Sigma_{\rm FUV\ SFR}$ vs. $\Sigma_{\rm \HI}$ on a
per-pixel basis.  For this analysis, we use a Gaussian kernel to
smooth the FUV images to the FWHM of the respective \ion{H}{1} images.
The FUV images are regridded to the \ion{H}{1} fields of view and then
both are cropped to a 64$\times$64 pixel grid (sufficient to include
all detected \ion{H}{1} emission in all galaxies). The resulting plots
are shown in Figure~\ref{SFEpixcorr}; the best-fit regression line to
all pixels is shown in red for each galaxy, and the slope $N$ of the
$\Sigma_{\rm SFR}$ $\propto$ $\Sigma_{\rm gas}^{N}$ relation is shown.
If $\Sigma_{\rm FUV\ SFR}$ and $\Sigma_{\rm \HI}$ are both high in the
same pixels $-$ that is, the FUV and \ion{H}{1} emission are
co-spatial $-$ and they are both relatively low in the same pixels,
then we expect a positive slope of the best-fit line. If the
$\Sigma_{\rm FUV\ SFR}$ falls off more quickly, then the slope is
steeper, but if the $\Sigma_{\rm \HI}$ falls off more quickly, then
the slope is shallower. An inspection of the images shown in
Figures~\ref{fig110482} through \ref{fig749241} reveals that both the
\ion{H}{1} gas and the FUV emission have resolved structure; further,
some sources have significant offsets of \ion{H}{1} versus FUV. Thus,
the slopes of the lines are sometimes negative (e.g., AGC\,748778 and
AGC\,749241), indicating an anti-correlation between the atomic gas
and the areas of elevated SF activity. The slopes of the K-S relations
seen in Figure~\ref{SFEpixcorr} and the average slope for the sample
are included in Table~\ref{t6}.

Finally, we show a series diagnostic plots which serve two functions:
comparing the \ion{H}{1} and SF properties, and demonstrating
potential effects of the physical resolution on a system's peak HI
column density and K-S slope. The first of these,
Figure~\ref{coldentrend}, compares the peak \ion{H}{1} column density
of each galaxy to its SFR (derived from both FUV and from
\halpha). This allows us to search for trends related to the maximum
\ion{H}{1} surface density detected in a given source. The second,
Figure~\ref{globalsigmas}, compares the global \ion{H}{1} and SFR
surface densities of the SHIELD sample against those in other relevant
studies. For the SHIELD galaxies, the global values are calculated by
summing all FUV emission and all \ion{H}{1} emission in concentric
annuli that encompass all of the FUV flux; regions of extended
\ion{H}{1} gas are not included in the calculation of the areas. This
plot allows us to contextualize the global SF properties of the SHIELD
sample with respect to sources with significantly higher and lower SFR
surface densities; however, we note that the derivation of the SFR is
different in each of the included surveys ({Kennicutt
  1998a}\nocite{kennicutt1998a}, {Wyder
  \etal\ 2009}\nocite{wyder2009}, {Roychowdhury
  \etal\ 2014}\nocite{roychowdhury2014}). Figures~\ref{resolutionvscolden}
and \ref{slopevalsvsresolution} compare the physical resolution sizes
of the SHIELD sources against their observed peak HI column density
and derived K-S $N$ value.

Using the above diagnostics, we examine causes for the appearances and
trends and compare our composite sample to other low-mass galaxy
surveys. In Section~\ref{S4.1} we discuss each galaxy individually,
focusing on the morphology, physical reasons for the derived K-S
slopes, and local environment surrounding each source. In
Section~\ref{S4.2} we describe the results for the sample as a whole
and contextualize our results with regard to other relevant
studies. Because we have several different metrics for determining the
slope for the Kennicutt-Schmidt (herefafter, K-S) relation, we will
discuss the relative strengths and weaknesses of each method.  Note
that error bars are shown in the figures, but are often unseen because
they are smaller than the points themselves. All lines-of-best-fit
that are fitted to data in the plots have been weighted by their
uncertainties. Average values for the sample, such as the average
consumption time, are weighted averages.

%-----------------------------------------------------------------------------%
\subsection{Discussion of Individual Galaxies}
\label{S4.1}
%-----------------------------------------------------------------------------%

\begin{figure}
\epsscale{1.0}
\centering
% old location = \plotone{/Users/research/data/vla/749237.ar.eps}
\epsscale{0.8}
\plotone{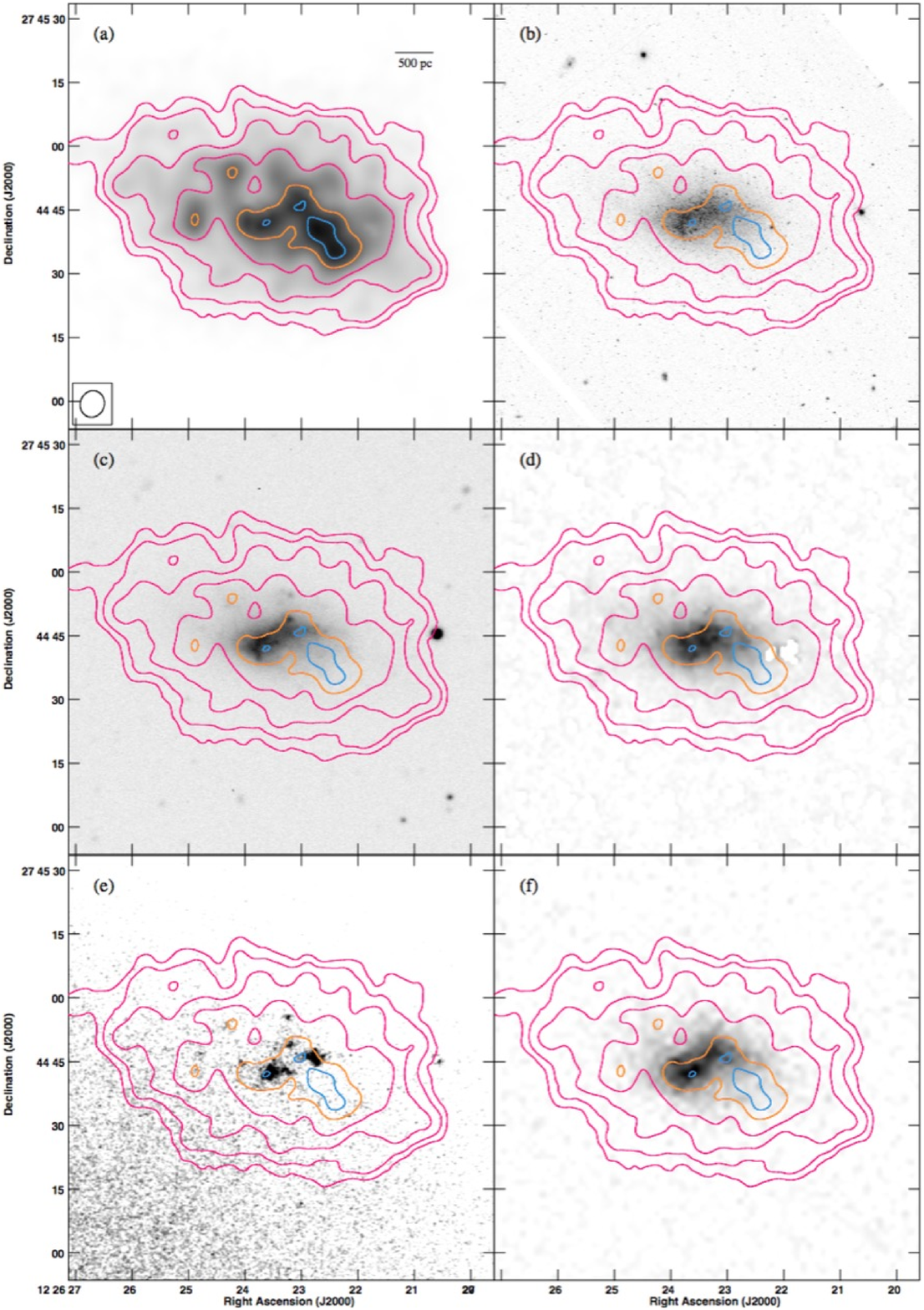}
\caption{AGC\,749237 in VLA \ion{H}{1} (a), HST F606W (b), KPNO WIYN 3.5m
  B-band (c), \textit{Spitzer} 3.6 $\mu$m (d), KPNO WIYN 3.5m continuum-subtracted \halpha\ (e), and
  GALEX FUV (f). The \ion{H}{1} column density contours, in units of 10$^{20}$
  cm$^{-2}$, are overlaid at levels of (0.6125, 1.25, 2.5, 5, 10,
  14). The highest two contour levels are highlighted in orange
  (10$\times$10$^{20}$ cm$^{-2}$) and blue (14$\times$10$^{20}$ cm$^{-2}$).
  The beam size of 6.21\arcsec$\times$5.59\arcsec\ is shown in panel
  (a); the \ion{H}{1} images are created using the 
robust-weighted, spectrally averaged data as discussed in Section \ref{S2.2.1}.}
\label{fig749237}
\end{figure}

\begin{figure}
\epsscale{1.0}
\centering
% old location = \plotone{/Users/research/data/vla/749241.ar.eps}
\epsscale{0.8}
\plotone{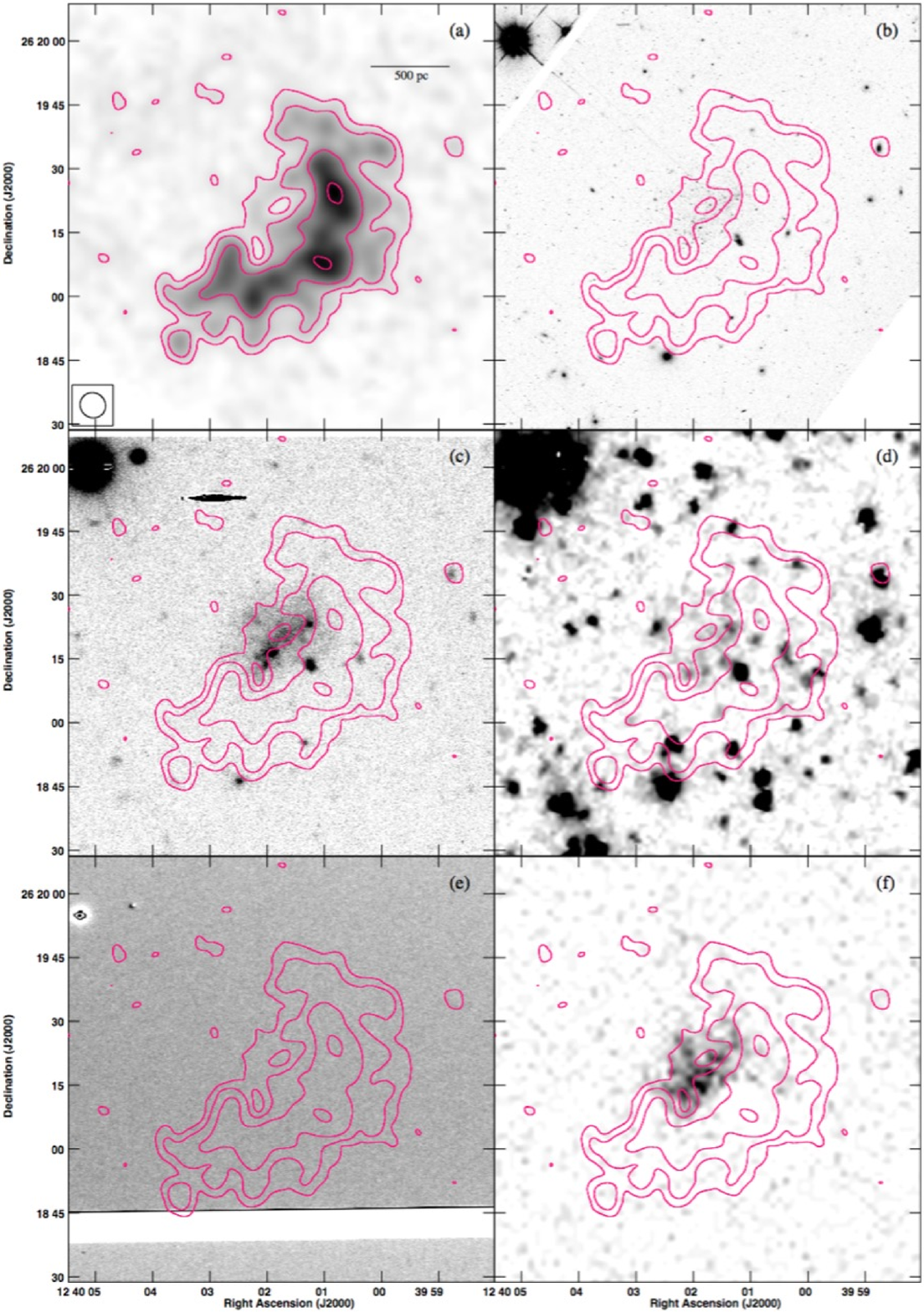}
\caption{AGC\,749241 in VLA \ion{H}{1} (a), HST F606W (b), KPNO WIYN 3.5m
  B-band (c), \textit{Spitzer} 3.6 $\mu$m (d), KPNO WIYN 3.5m continuum-subtracted \halpha\ (e), and
  GALEX FUV (f). The \ion{H}{1} column density contours, in units of 10$^{20}$
  cm$^{-2}$, are overlaid at levels of (0.6125, 1.25, 2.5, 5).  The
  beam size of 6.06\arcsec$\times$5.82\arcsec\ is shown in panel (a); the \ion{H}{1} images are created using the 
robust-weighted, spectrally averaged data as discussed in Section \ref{S2.2.1}.}
\label{fig749241}
\end{figure}

\begin{figure}
\epsscale{1.00}
% old location = \plotone{/Users/research/Desktop/PLOTSFORPAPER/SFRvsHImasscomparison.eps}
\plotone{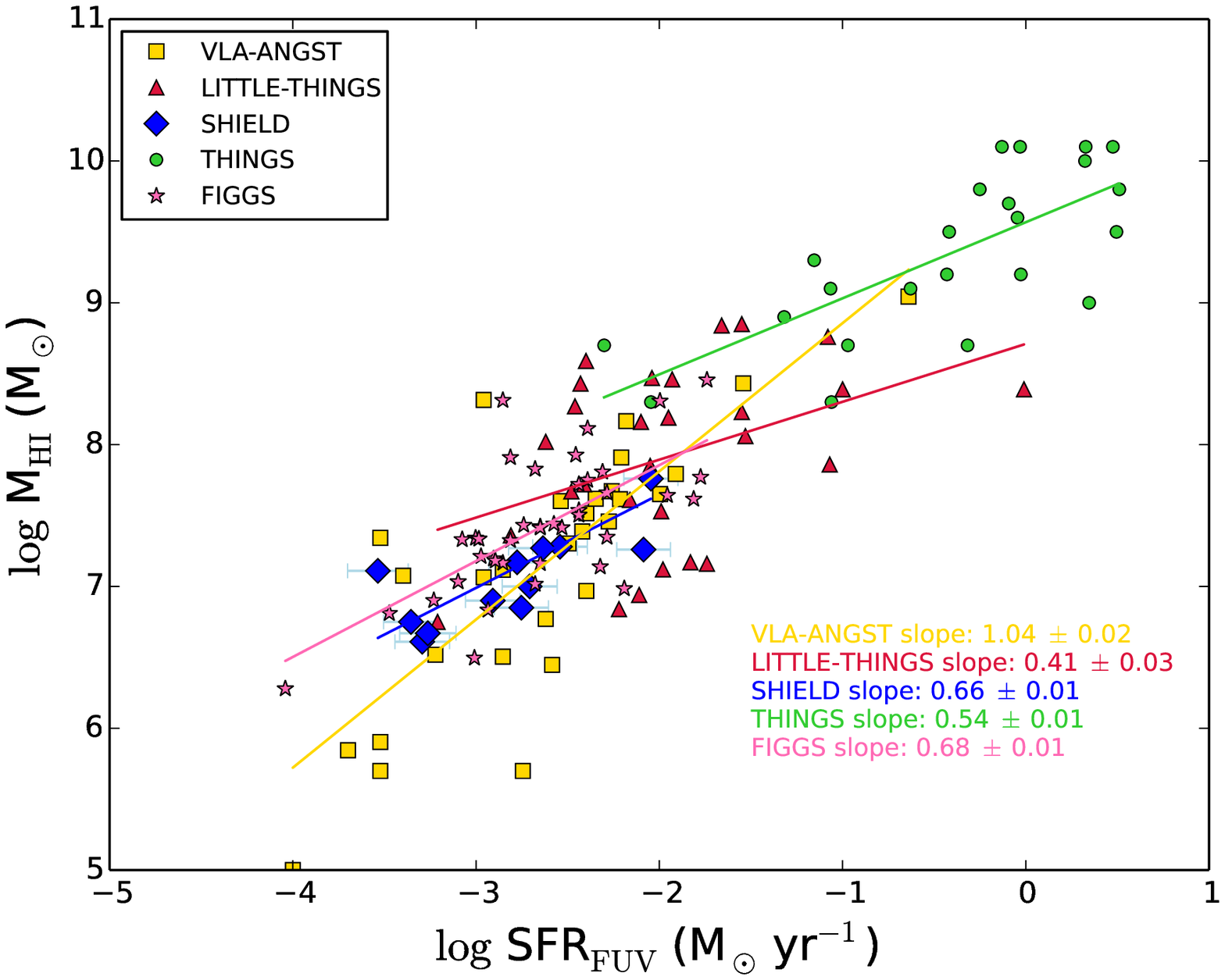}
\caption{M$_{\rm \HI}$ vs. SFR$_{\rm FUV}$ for the 12 SHIELD
  galaxies. Data from four other nearby galaxy surveys are included
  for comparison: VLA-ANGST \citep{ott2012}, LITTLE-THINGS
  \citep{hunter2012}, THINGS \citep{walter2008}, and FIGGS
  \citep{roychowdhury2014}.  For each galaxy sample, we fit a linear
  regression for the slope and show these values at lower right.  The
  most similar trends can be seen in the SHIELD and FIGGS datasets;
  this is not surprising, given that both surveys were selected to
  study low-mass, gas-rich systems.  Note that while a significant
  portion of the sources from the low-mass surveys lie in the space
  where log(SFR$_{\rm FUV}$) $\lesssim$ -2.5, very few of these
  sources have log(M$_{\rm \HI}$) $\gtrsim$ 7.5; galaxies which are
  forming fewer stars have smaller \ion{H}{1} reservoirs.}
%VLA-ANGST (Ott et al. 2012): converted from GALEX FUV asymptotic magnitudes given by Lee et al. (2011) and using SFR = 1.4 × 10−28 Lν (erg s−1 Hz−1) (Kennicutt 1998)
%THINGS (Leroy et al. 2008): SFRTot = 0.68 × 10−28Lν (FUV) + 2.14+0.82 −0.49 × 10−42L(24 μm) --- combined FUV and 24micron intensities to get total SFR; they used Salim et al. 2007 for the FUV part and their own derived factor for the 24micron part based on the ratio of FUV/24micron emission 
%LITTLE-THINGS (Hunter et al. 2012): SFR-FUV is the star formation rate determined from GALEX FUV fluxes (Hunter et al. 2010, with an update of the GALEX FUV photometry to the GR4/GR5 pipeline reduction).
%FIGGS (Roychowdhury et al. 2014): The calibration we use for converting FUV luminosity to SFR is taken from Kennicutt & Evans II (2012); Hao et al. (2011): log SFR M⊙ yr−1 = log ν Hz Lν ergs s−1 Hz−1 − 43.35
\label{SFRvsHImass}
\end{figure}

\begin{figure}
\epsscale{1.00}
% old location = \plotone{/Users/research/Desktop/PLOTSFORPAPER/SFRvsHImasscomparison.eps}
\plotone{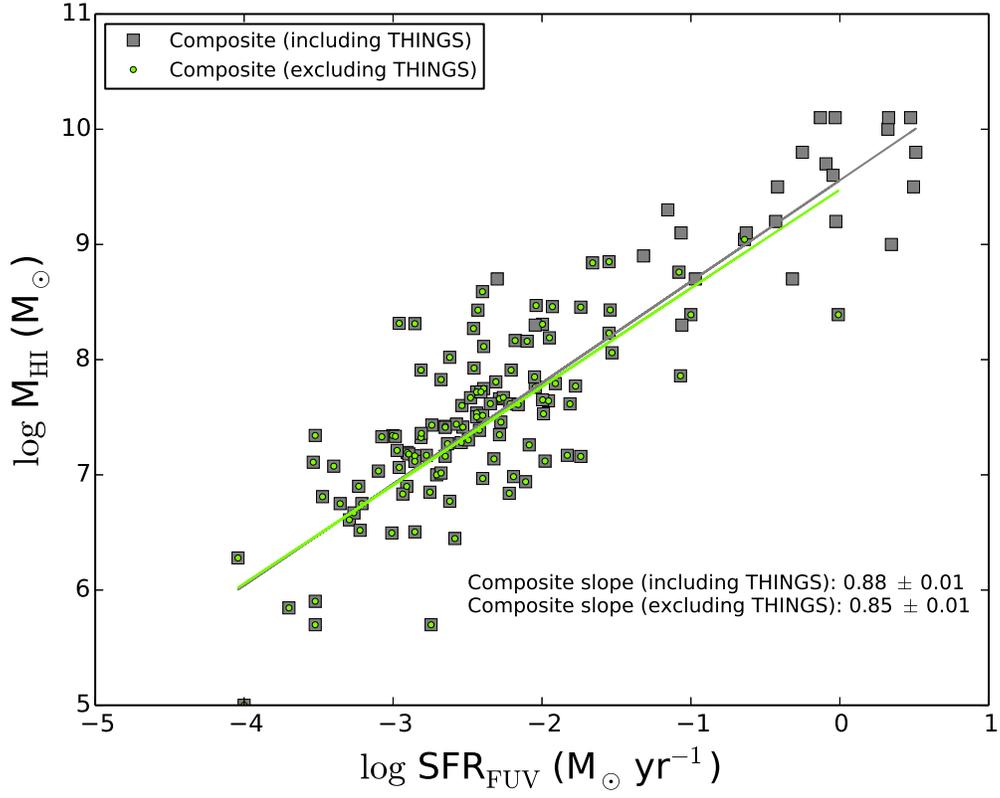}
\caption{M$_{\rm \HI}$ vs. SFR$_{\rm FUV}$ for all surveys shown in
  Figure~\ref{SFRvsHImass}.  Two fits are shown: one to the composite
  sample of all galaxies (black line) and another to all galaxies
  excluding the THINGS sources.  The slopes of the two lines are
  essentially indistinguishable.}
\label{SFRvsHImass-composite}
\end{figure}

\begin{figure}
\epsscale{1.0}
% old location = \plotone{/Users/research/Desktop/PLOTSFORPAPER/FUVvsHAsfrscomparison.eps}
\plotone{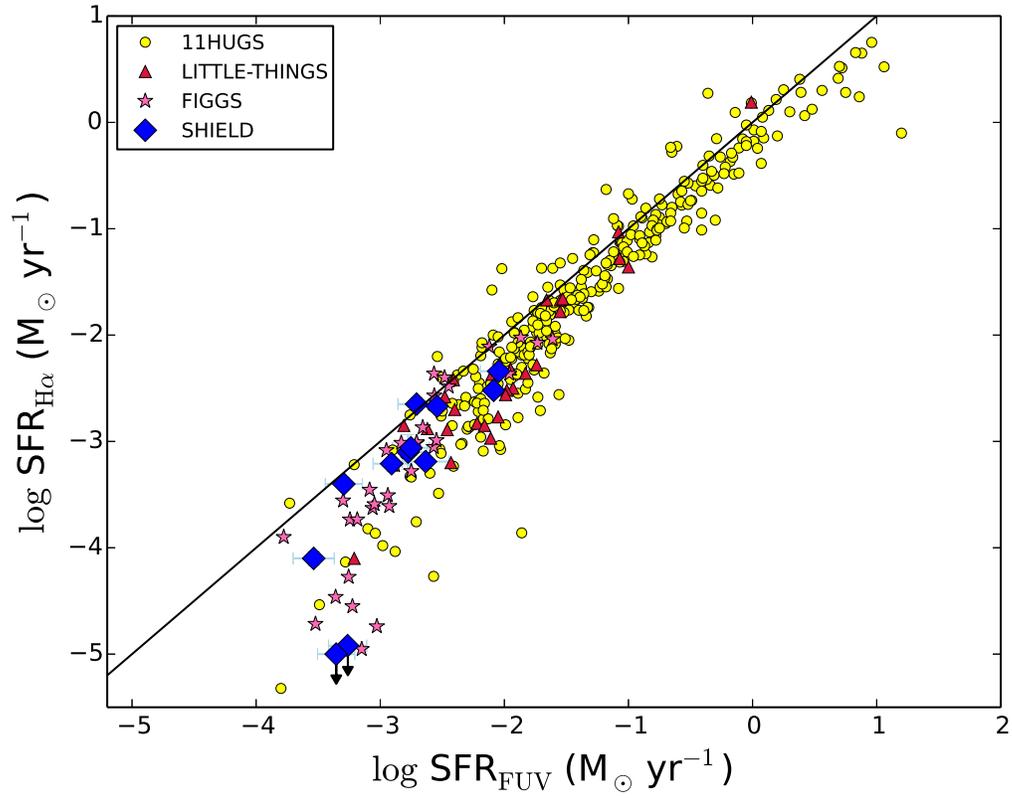}
\caption{SFR$_{\rm H\alpha}$ vs. SFR$_{\rm FUV}$ for the SHIELD
  sample.  The line plotted shows equality. The bottom-most points for
  SHIELD represent the \halpha\ SFRs for AGC\,748778 and AGC\,749241,
  which are both formally upper limits.}
\label{FUVvsHAsfrs}
\end{figure}

\begin{figure}
\epsscale{1.0}
% old location = \plotone{/Users/research/Desktop/PLOTSFORPAPER/SFRratios.eps}
\plotone{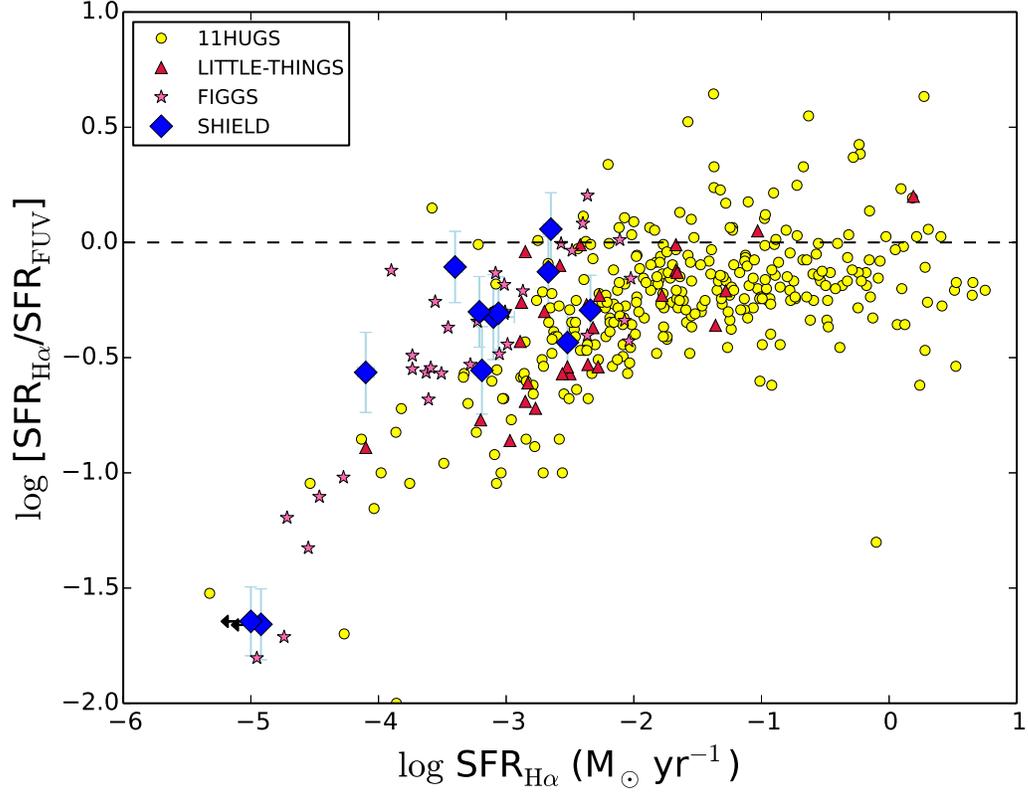}
\caption{SFR ratio [\halpha/FUV] vs. SFR$_{\rm H\alpha}$ for the
  SHIELD sample. Data from three other nearby galaxy surveys are
  included for comparison: 11HUGS \citep{lee2009}, LITTLE-THINGS
  \citep{hunter2012}, and FIGGS \citep{roychowdhury2014}.}
\label{SFRratios}
\end{figure}

As Figures~\ref{fig110482} through \ref{fig749241} demonstrate, some
galaxies in the SHIELD sample have ongoing SF in regions of relatively
high \ion{H}{1} column density while others do not. The strength of
the correlation between the SF tracers and the locations of elevated
\ion{H}{1} column densities varies dramatically among the galaxies.
As shown in Table~\ref{t2}, only 4 of galaxies (AGC\,110482,
AGC\,174585, AGC\,731457, and AGC\,749237) have \ion{H}{1} column
densities that exceed 10$^{21}$ cm$^{-2}$ (shown by an orange contour
in those respective figures), yet several of the other sources with
``sub-critical'' peak \ion{H}{1} column densities have co-spatial
\ion{H}{1} knots and SF regions as traced by both FUV and \halpha.  We
now briefly discuss each of the 12 galaxies in turn.

\begin{figure}
\epsscale{1.0}
\centering
% old location = \plotone{/Users/research/Desktop/PLOTSFORPAPER/RadProfsall12.eps}
\plotone{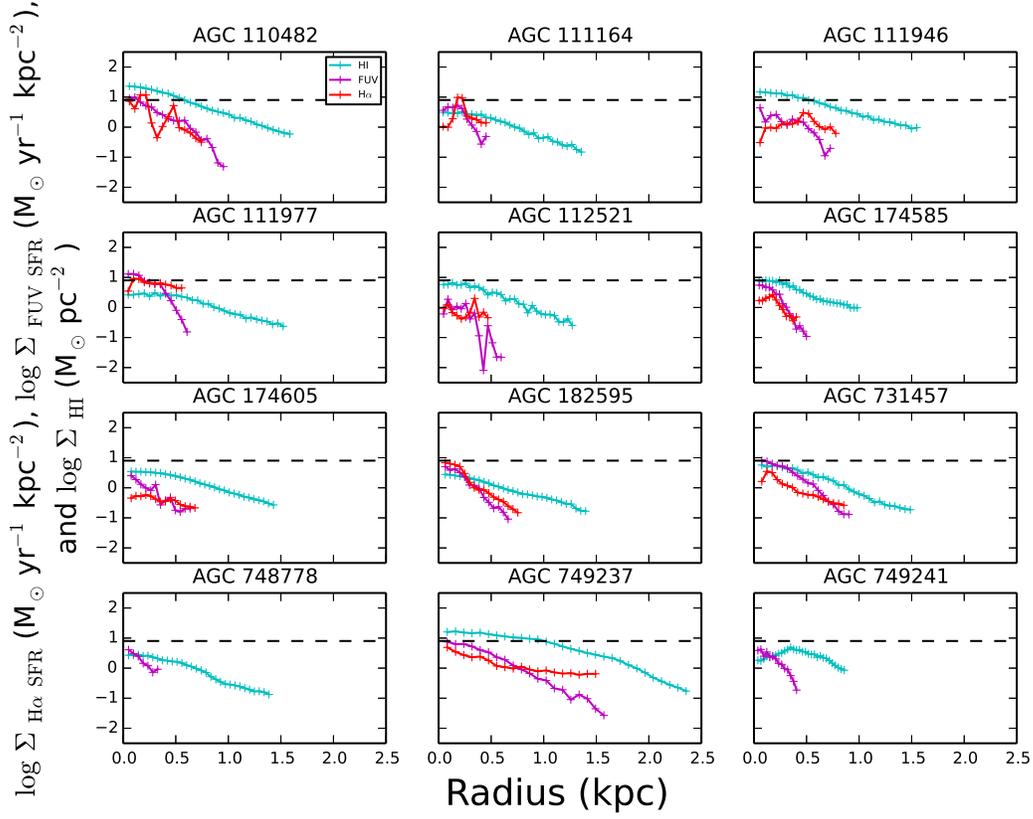}
\caption{$\Sigma_{\rm \HI}$ (cyan), $\Sigma_{\rm H\alpha\ SFR}$ (red,
  from the Kennicutt 1998 prescription), and $\Sigma_{\rm FUV\ SFR}$
  (magenta) vs. radius in kpc for each of the sample members.  Only
  points measured at high confidence (greater than 3\,$\sigma$) are
  shown. These radial profiles demonstrate the variety of
  distributions of star-forming regions and gas in the galaxies. FUV
  data for AGC\,749237 is based on a conversion from the associated
  NUV counts. AGC\,748778 and AGC\,749241 are non-detections in
  \halpha.  The horizontal axis is set by the physically largest
  system, AGC\,749237.  The horizontal dashed line represents the HI
  column density threshold of 10$^{21}$ cm$^{-2}$.}
\label{RadProfs}
\end{figure}

\begin{figure}
\epsscale{1.0}
\centering
% old location = \plotone{/Users/research/Desktop/PLOTSFORPAPER/fullFUVSFEplotsall12.eps}
\plotone{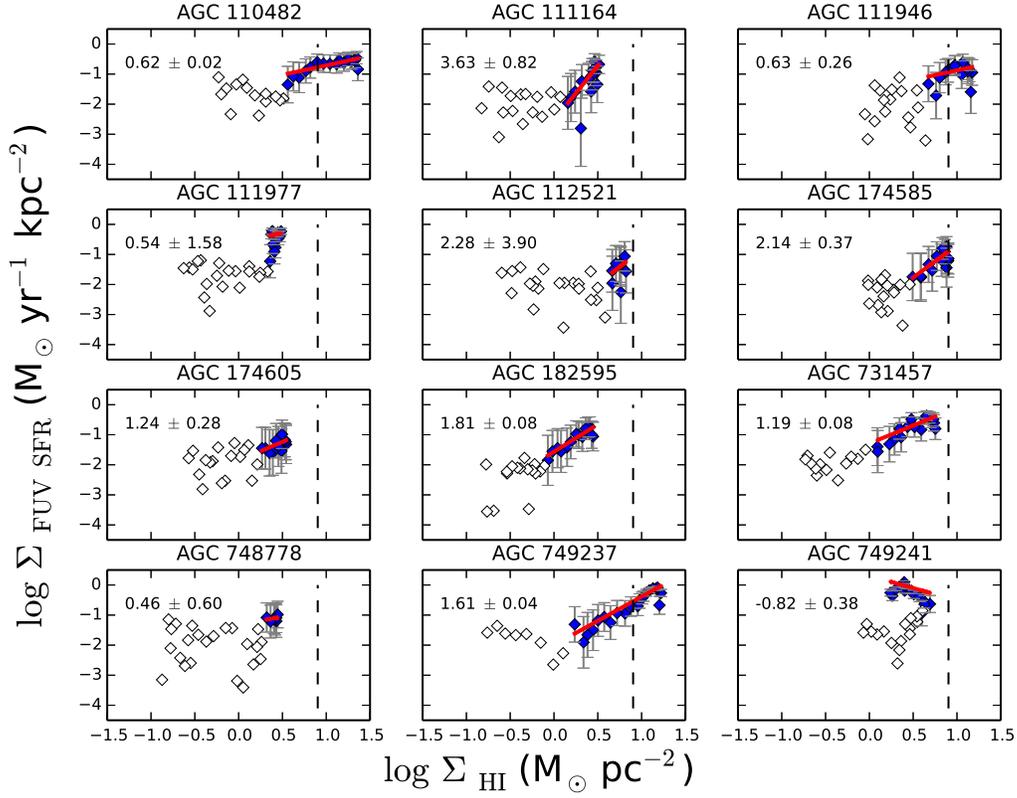}
\caption{$\Sigma_{\rm FUV\ SFR}$ vs. $\Sigma_{\rm \HI}$ for each of
  the sample members. The slope of the line is the N index in the
  Kennicutt-Schmidt relation and is included in Table \ref{t6}. The
  white points are below our 5$\sigma$ noise threshold and thus are
  not included in the calculation of the trendline of the data. The
  average value for the slope is N $\approx$ 1.50$\pm$0.02. The dashed
  vertical line represents an \ion{H}{1} column density of 10$^{21}$
  atoms cm$^{-2}$.}
\label{FUVSFE}
\end{figure}

\begin{figure}
\epsscale{1.0}
\centering
% old location = \plotone{/Users/research/Desktop/PLOTSFORPAPER/fullK98SFEplotsall12.eps}
\plotone{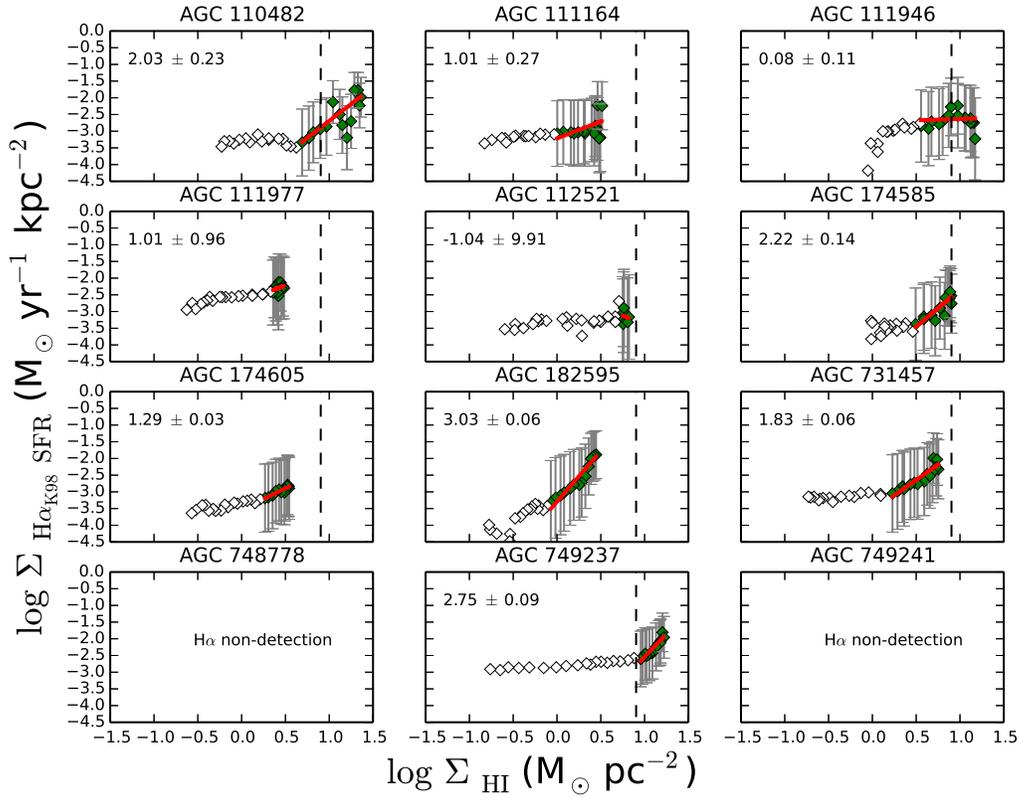}
\caption{$\Sigma_{\rm H\alpha\ SFR}$ vs. $\Sigma_{\rm \HI}$ for each
  of the sample members, using the SFR$_{\rm H\alpha}$ derived from
  the \citet{kennicutt1998a} prescription. The slope of the line is
  the N index in the Kennicutt-Schmidt relation and is included in
  Table \ref{t6}. The white points are below our 5$\sigma$ noise
  threshold and thus are not included in the calculation of the
  trendline of the data. The average value for the slope is N
  $\approx$ 0.34$\pm$0.02. The dashed vertical line represents an
  \ion{H}{1} column density of 10$^{21}$ atoms cm$^{-2}$.}
\label{K98SFE}
\end{figure}

\begin{figure}
\epsscale{1.0}
% old location = \plotone{/Users/research/Desktop/PLOTSFORPAPER/SFEpixcorr.eps}
\plotone{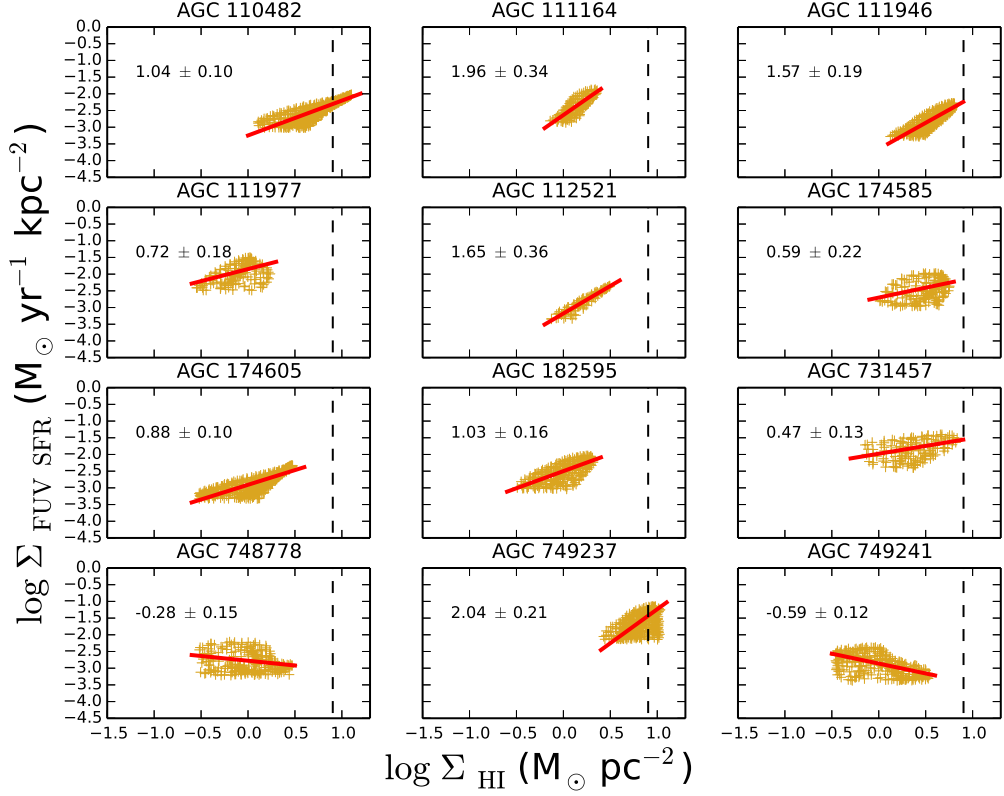}
\caption{$\Sigma_{\rm FUV\ SFR}$ vs. $\Sigma_{\rm \HI}$ for each of
  the sample members. The points plotted are from the pixel-by-pixel
  correlation process described in Section \ref{S3}. All FUV pixel
  values are above the 10\% contour level for each source (i.e., all
  pixels plotted are greater than 10\% of the peak flux), and the same
  is true of the \ion{H}{1} values relative to the peak \ion{H}{1}
  column density. The slope of the line is the N index in the
  Kennicutt-Schmidt relation and is included in Table \ref{t6}. The
  average value for the slope is $N$ $\approx$ 0.68$\pm$0.04. A positive
  slope indicates high FUV and \ion{H}{1} emission in the same
  pixels. The dashed vertical line represents the column density
  threshold of 1$\times$10$^{21}$ atoms cm$^{-2}$.}
\label{SFEpixcorr}
\end{figure}

\begin{figure}
\epsscale{1.0}
% old location = \plotone{/Users/research/Desktop/PLOTSFORPAPER/coldentrend.eps}
\plotone{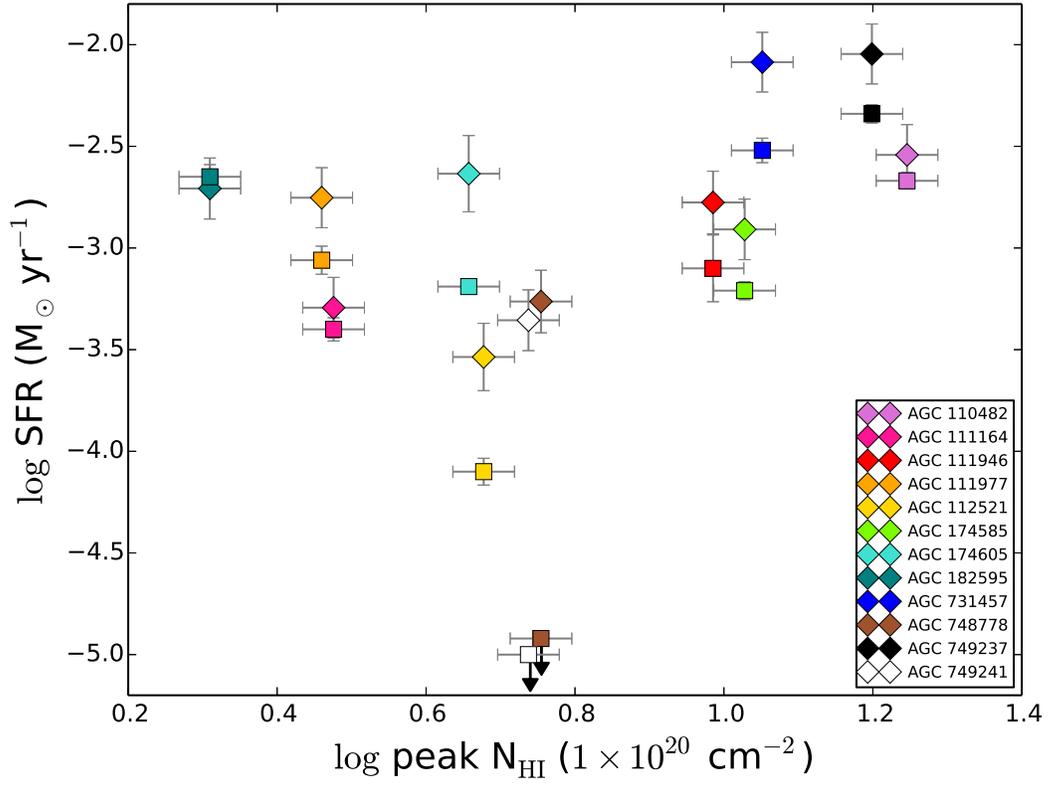}
\caption{UV and \halpha-based SFR vs. peak \ion{H}{1} column density
  for each of the sample members. FUV values are represented by
  diamonds and \halpha\ values are represented by squares. In some
  cases the error bars are smaller than the points themselves. The
  bottom-most two points represent upper-limits on the \halpha\ SFRs
  for AGC\,748778 and AGC\,749241 (formally non-detections).}
\label{coldentrend}
\end{figure}

\begin{figure}
\epsscale{1.0}
% old location = \plotone{/Users/research/Desktop/PLOTSFORPAPER/globalsigmas.eps}
\plotone{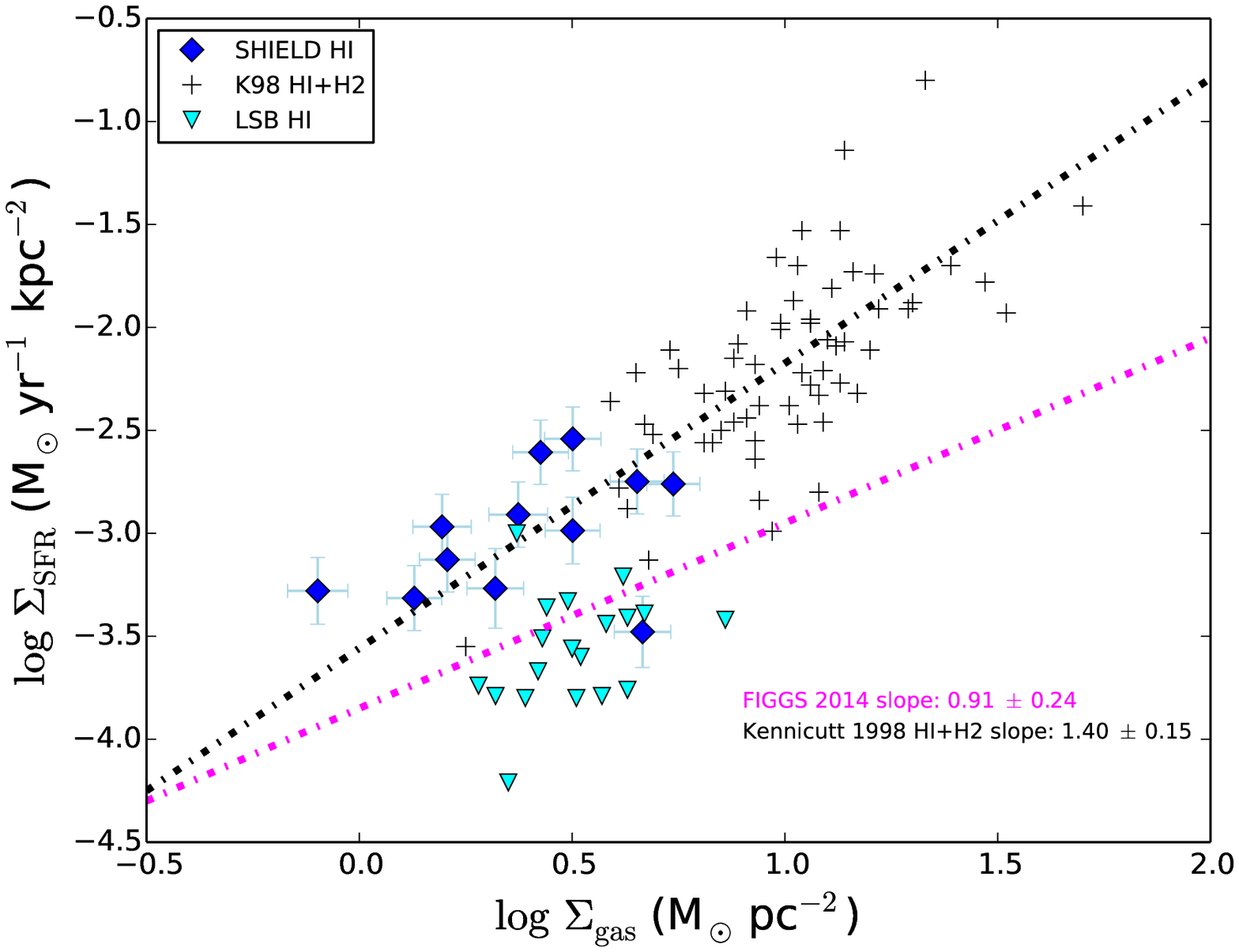}
\caption{$\Sigma_{\rm SFR_{\rm FUV}}$ vs. $\Sigma_{\rm \HI}$ for the
  SHIELD sample. Data from a study of LSB galaxies \citep{wyder2009}
  and slopes from studies of normal spiral galaxies
  \citep{kennicutt1998b} and FIGGS dwarf irregular galaxies
  \citep{roychowdhury2014} are shown for comparison; note that
  different SFR metrics are used from one survey to the next. While
  the LSB galaxies lie below the canonical power-law slope, both
  studies of low-mass dwarfs appear to have shallower K-S relations
  than the higher-mass spiral galaxies. Note that the data from
  \citep{kennicutt1998b} is for HI+H2, while the other studies use
  just HI but assume the molecular contribution is negligible.}
\label{globalsigmas}
\end{figure}

\begin{figure}
\epsscale{1.0}
\plotone{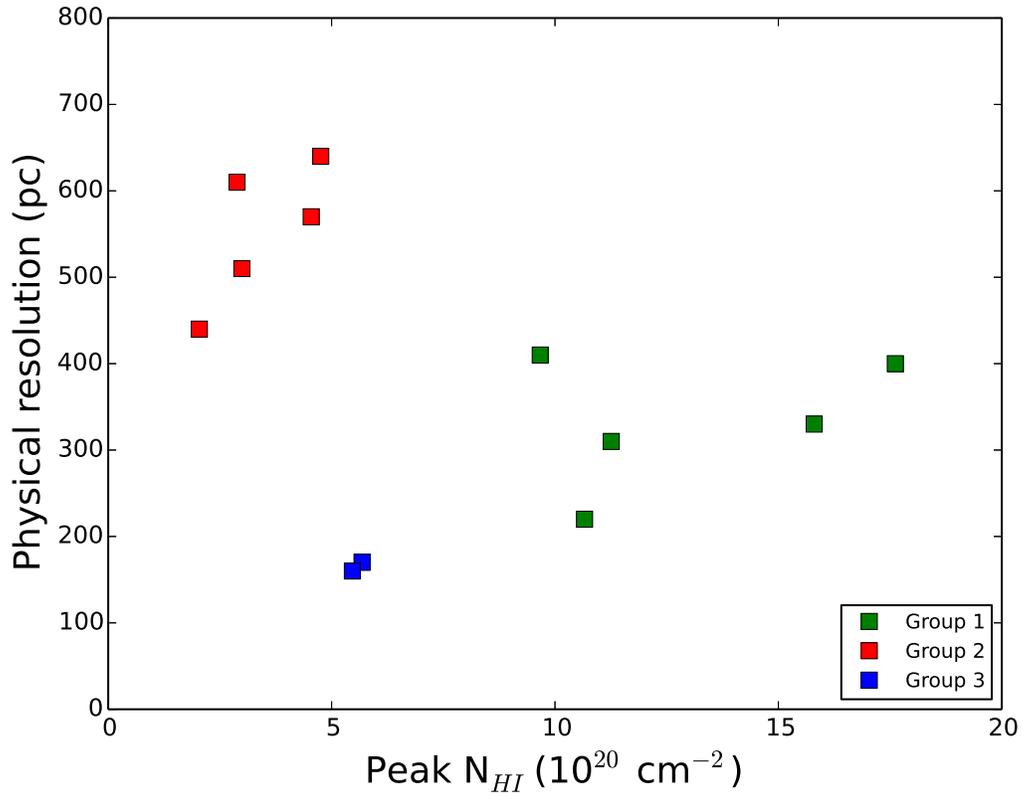}
\caption{Physical resolution in pc vs. peak \ion{H}{1} column density
  for the SHIELD sample. The galaxies are grouped in the manner
  described in Section \ref{S4.2} according to their similar
  properties. While 2 of the 5 galaxies in Group 1 have the largest
  distances in the sample, the presence of B-configuration data
  results in smaller beam dimensions and fine resolution elements,
  allowing us to see the highest column density HI gas. In Group 2, 3
  of the 5 galaxies lack B-configuration data, which contributes to
  their coarser resolution elements and inability to detect higher
  column density HI gas. The Group 3 galaxies are the outliers: the
  inclusion of B-configuration data provides fine resolution elements,
  but the galaxies are low-mass and devoid of higher column density HI
  gas.}
\label{resolutionvscolden}
\end{figure}

\begin{figure}
\epsscale{1.0}
\plotone{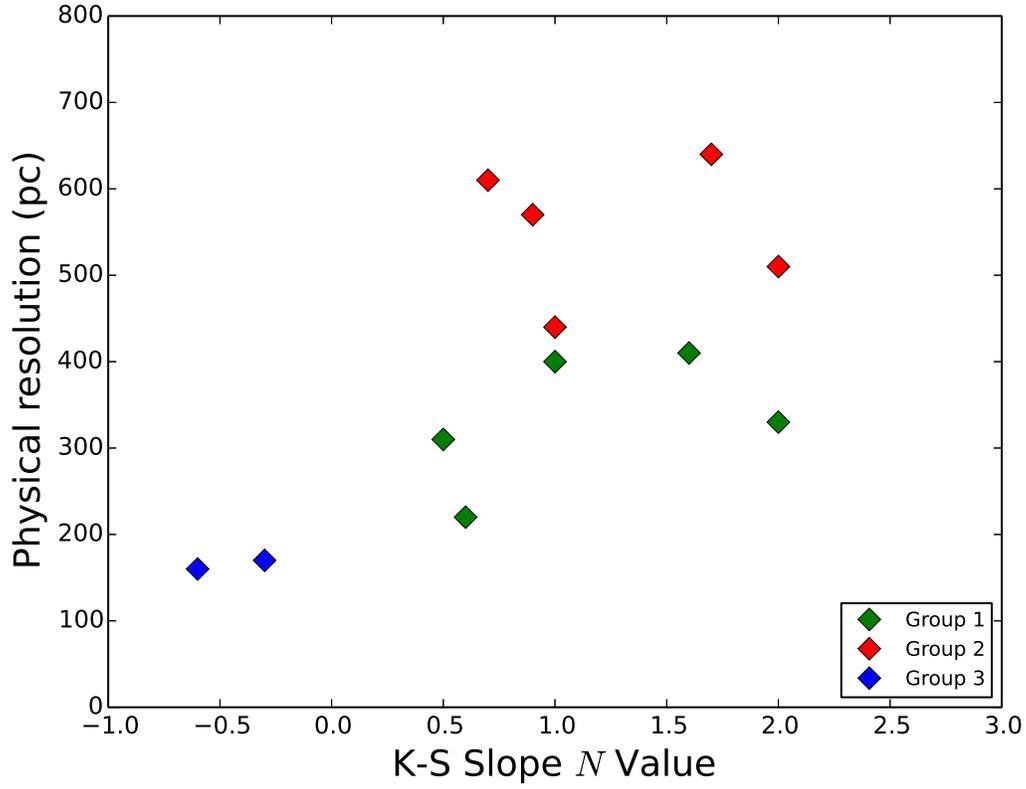}
\caption{Physical resolution in pc vs. $N$ value (the derived slope
  for the Kennicutt-Schmidt relation) for the SHIELD sample. The
  galaxies are grouped in the manner described in Section \ref{S4.2}
  according to their similar properties. While the members of Groups 1
  and 2 can be characterized by differing physical resolution
  elements, they both have similar range and dispersion in terms of
  their $N$ values. Group 3 is the obvious outlier, hosting the only
  two galaxies with negative $N$ values. It is apparent that on
  sufficiently small scales, the K-S relation is no longer valid.}
\label{slopevalsvsresolution}
\end{figure}

\underline{AGC\,110482}: there is excellent agreement between the two
prominent \halpha\ regions and the FUV peaks. The southeastern peak of
the \ion{H}{1} ($\sim$1.7$\times$10$^{21}$ cm$^{-2}$) is exactly
co-spatial with the \halpha\ and FUV peaks in that area. There is also
a second northwestern maximum in the SF tracers which is very close to
the N$_{\rm \HI}$ $=$ 10$^{21}$ cm$^{-2}$ contour. Based on these
qualities, we expect the smooth radial profile of \ion{H}{1} surface
density and a visible bump in the \halpha\ SFR surface density profile
seen in Figure~\ref{RadProfs}. The K-S slope is steeper in Figure
\ref{FUVSFE} than in Figure~\ref{K98SFE}. The K-S index is
1.04\,$\pm$\,0.10 from the pixel-by-pixel correlation method
(Figure~\ref{SFEpixcorr}). \citet{mcquinn2015a} notes that AGC\,110482
is $\gtrsim$ 0.65 Mpc from other members of the NGC\,672 group.

\underline{AGC\,111164}: the \ion{H}{1} gas is distributed almost
circularly (but note that AGC\,111164 is one of the three galaxies
that does not have B-configuration imaging, so the resolution element
is comparatively coarse; however, because this galaxy lies at a
relatively small distance, the linear resolution is comparable to
other sources). However, the stellar component of the galaxy is
elongated in the HST and B-band, and although the \textit{Spitzer}
image shows similar overall structure, there is a line of bright
infrared emission perpendicular to the major axis of the galaxy which
is not highlighted in the other panels. The FUV and \halpha\ maxima
are co-spatial with B-band and \ion{H}{1} maxima. Additionally, the
FUV emission is somewhat extended but the \halpha\ morphology is
consistent with a single \ion{H}{2} region; the radial profiles shown
in Figure~\ref{RadProfs} highlight this structure. The peak \ion{H}{1}
column density is among the lowest in the sample (again, the coarse
resolution element likely affects this), yet massive SF is
occurring. The slopes of the FUV and the \halpha\ surface density
profiles differ. Using the FUV images, we determine $N =$
1.96\,$\pm$\,0.34 on a pixel-by-pixel basis. \citet{mcquinn2015a}
notes that this galaxy lies only $\sim$0.14 Mpc from NGC\,784 and is
part of a linear structure of galaxies associated with this larger
dwarf starburst galaxy.

\underline{AGC\,111946}: the \ion{H}{1} column density does not reach
the 10$^{21}$ cm$^{-2}$ level. However, the \ion{H}{1} gas is
generally co-spatial with the stellar component in the HST and WIYN
optical images, and the two \halpha\ and FUV peaks lie precisely
within the \ion{H}{1} peaks. There are observable bumps in the radial
profiles (Figure~\ref{RadProfs}) due to the morphology of the SF
tracers and the \ion{H}{1}.  The slope derived in
Figure~\ref{SFEpixcorr} is 1.57\,$\pm$\,0.19. \citet{mcquinn2015a}
finds that this galaxy is also located in the NGC\,784 group, but is
$\sim$1 Mpc away from its nearest neighbor, suggesting that the recent
SF activity is not driven by gravitational interactions.

\underline{AGC\,111977}: this is one of the most enigmatic SHIELD
galaxies. The peak \ion{H}{1} column density is low (although this
source was not observed in the B configuration and so the beam size is
coarse, and the relatively larger beam size yields a poorer linear
resolution). There is a significant diffuse southern \halpha\ region
that is prominent in the FUV as well. Interestingly, the \ion{H}{1}
column density maximum is on the other side of the disk from these SF
regions. Because of the large offset between \ion{H}{1} and SF peaks,
the slopes of the radial profile plots are difficult to interpret.
The pixel-by-pixel method shows a relatively shallow slope with
significant dispersion in the $\Sigma_{\rm SFR}$ dimension; the
average slope is 0.72\,$\pm$\,0.18. \citet{mcquinn2015a} finds that
AGC\,111977 is separated by $\sim$0.65 Mpc from AGC\,112521.

\underline{AGC\,112521}: the peak \ion{H}{1} column density is only
half of the canonical threshold value (although this source was not
observed in the B configuration and so the beam size is coarse,
resulting in a poorer linear resolution), and there is very little SF
currently; the FUV and \halpha\ luminosities are very low. The single
\ion{H}{2} region in the north is coincident with the FUV and B-band
maximum.  The \halpha\ and FUV radial profiles are dominated by small
number statistics.  The pixel-by-pixel method shows the smallest
scatter of any of the SHIELD galaxies in the $\Sigma_{\rm SFR}$
dimension; the average slope is
1.65\,$\pm$\,0.36. \citet{mcquinn2015a} finds that AGC\,112521 is
$\sim$0.65 Mpc from several neighbors, including AGC\,110482,
AGC\,111977, and three members of the NGC\,672 group.

\underline{AGC\,174585}: this galaxy has a small \ion{H}{1} peak that
exceeds the 10$^{21}$ cm$^{-2}$ column density threshold; this peak is
not co-spatial with the SF regions in the source, which are extended
spatially in both the \halpha\ and the FUV images.  The radial
profiles quantify what is clear from the images in
Figure~\ref{fig174585}: extended SF is associated with \ion{H}{1} gas
of a range of mass surface densities.  As expected, the slope of the
pixel-by-pixel method is among the most shallow in the sample ($N =$
0.59\,$\pm$\,0.22). \citet{mcquinn2015a} notes that AGC\,174585 is
located $\sim$0.9 Mpc away from its nearest neighbors.

\underline{AGC\,174605}: although the \ion{H}{1} column densities are
low, the \ion{H}{1} peak is tightly correlated with the \halpha\ and
FUV peaks (well within the \ion{H}{1} beam). The FUV image is the only
one in the sample at AIS depth, and the noise is high.  Similarly, the
\halpha\ image has a pronounced gradient.  The radial profiles show
somewhat steeper slopes than the pixel-by-pixel method ($N =$
0.88\,$\pm$\,0.10). \citet{mcquinn2015a} finds that AGC\,174605 is
truly isolated with no known neighbors in a 1 Mpc radius.

\underline{AGC\,182595}: this galaxy has the lowest peak \ion{H}{1}
column density of the SHIELD galaxies, and yet it is relatively
luminous in both \halpha\ and the FUV. Note that both the \halpha\ and
the FUV emission are spatially extended. There is a noticable offset
between the HI peak and the HST, B-band, and FUV peaks, and it is even
more apparent when compared to \halpha. AGC\,182595 is the only source
in the sample with a ratio of SFR$_{\rm H\alpha}$/SFR$_{\rm FUV}$ $>$
1, perhaps indicating a modest starburst episode. \citet{mcquinn2015a}
notes that it has the lowest gas-to-stellar mass ratio of the sample
and has the second-shortest gas consumption timescale.
Figures~\ref{FUVSFE}, \ref{K98SFE}, and \ref{SFEpixcorr} all show
smooth slopes consistent with trends seen in larger galaxies. Although
the $N$ values range considerably in these plots, the slopes are all
well-defined; the pixel-by-pixel average slope is
1.03\,$\pm$\,0.16. \citet{mcquinn2015a} finds that AGC\,182595 is
truly isolated with no known neighbors in a 1 Mpc radius.

\underline{AGC\,731457}: this source eclipses the 10$^{21}$ cm$^{-2}$
\ion{H}{1} column density threshold and has one of the highest SFRs in
the sample using both \halpha\ and FUV metrics. There are several
luminous \halpha\ regions; some of these are co-spatial with the
extended FUV emission, but there are regions that are \halpha\ bright
and FUV dim and vice-versa.  The \ion{H}{1} peak is slightly offset
not only from the SF regions but also from the optical counterpart of
the source as shown by the HST image.  A cursory inspection of
Figure~\ref{fig731457} shows that SF regions are associated with
\ion{H}{1} gas at a broad range of observed mass surface densities.
As in AGC\,182595, we see steeper slopes for the radial SFR surface
density plots than for the pixel-by-pixel correlation; AGC\,731457 has
the shallowest (positive) slope of any of the SHIELD galaxies via this
metric (N $=$ 0.47\,$\pm$\,0.13). \citet{mcquinn2015a} notes that
AGC\,731457 is located $\sim$0.9 Mpc away from its nearest neighbors.

\underline{AGC\,748778}: this source has a dramatic \ion{H}{1}
morphology that is highly extended relative to the stellar population,
extending more than 1 kpc to the south. Two \ion{H}{1} maxima are
observed, only one of which overlaps spatially with the stars in the
galaxy. The \halpha\ emission is very weak in this source, so much so
that it was a non-detection in our observations. Interestingly, the
source is detected with significance in the FUV image.  The recent SF
traced by this FUV flux is associated with a range of \ion{H}{1} mass
surface densities, and some FUV flux contains no associated \ion{H}{1}
gas at our current level of sensitivity. \citet{mcquinn2015a} finds
that AGC\,748778 is truly isolated with no known neighbors in a 1 Mpc
radius, suggesting that it is unlikely that gravitational interactions
have driven the recent SF activity.

\underline{AGC\,749237}: this is the largest galaxy in the sample by
mass, optical and \ion{H}{1} diameter, and it has the highest SFR (in
both FUV and \halpha). It has the second highest peak \ion{H}{1}
column density in the sample. The most intriguing aspect of the galaxy
is that two of the \halpha\ and FUV knots are co-spatial with two of
the \ion{H}{1} peaks, but not the largest and densest
one. Figure~\ref{fig749237} reveals that the highest \ion{H}{1} mass
surface densities are associated with the outer disk of the system and
contain no \halpha\ and weak FUV emission (compare the \halpha-based
SFR density profile in Figure~\ref{K98SFE} with the FUV-based profile
in Figure~\ref{FUVSFE}). In Figure~\ref{SFEpixcorr}, we see an
intriguing effect: the large spread in FUV points at the peak
\ion{H}{1} values occurs because there are both very high and very low
FUV values associated with the peak \ion{H}{1} pixels; this trend is
seen clearly in Figure~\ref{fig749237}. Nonetheless, AGC\,749237
stands out in the pixel-by-pixel diagram as having the steepest K-S
index ($N =$ 2.04\,$\pm$\,0.21). \citet{mcquinn2015a} finds that this
galaxy is truly isolated with no known neighbors in a 1 Mpc radius, so
it is likely that its recent SF activity is internally regulated.

\underline{AGC\,749241}: like AGC\,748778, this source has a highly
unusual \ion{H}{1} morphology, where the bulk of the \ion{H}{1} gas is
not co-spatial with the stellar component.  The source is a
non-detection in both the \halpha\ and infrared images. Like
AGC\,748778, it is detected with confidence in the FUV, although it
does harbor the lowest SFR$_{\rm FUV}$ in the sample, and shows an
obvious offset in the \ion{H}{1} gas and FUV emission. The
$\Sigma_{\rm SFR}$ radial profile and the pixel-by-pixel diagnostic
both suggest a negative K-S index for this source.  As for
AGC\,748778, the offset of \ion{H}{1} and FUV is the cause.  For a
galaxy to have such inefficient SF combined with a highly offset
\ion{H}{1} component is certainly an interesting scenario. The
question of whether some kind of tidal interaction has separated the
\ion{H}{1} so visibly from the stars has been addressed:
\citet{mcquinn2015a} investigates the environment surrounding this
source and finds that AGC\,749241 lies in populated region with 9
galaxies which form a linear structure 1.6 Mpc from end to end. While
it is possible that gravitational interactions among the systems in
this structure have contributed to the offset of gas and stars visible
in AGC\,749241, the low recent SF activity found in the galaxy
indicates that any gravitational interaction has not had a dramatic
impact on the star-forming properties over the past 200 Myr.

%-----------------------------------------------------------------------------%
\subsection{The Star Formation Process in Extremely Low-Mass Galaxies}
\label{S4.2}
%-----------------------------------------------------------------------------%

%-----------------------------------------------------------------------------%
\subsubsection{The Kennicutt-Schmidt Relation}
\label{S4.2.1}
%-----------------------------------------------------------------------------%

The discussion above highlights the difficulty of determining a SF
``law'' in extremely low-mass galaxies. Using \halpha\ images presents
challenges; in a given galaxy, \ion{H}{2} regions are few in number
(at most 3 well-defined clumps in a single SHIELD source), faint (low
total luminosity), and sometimes offset from the optical centers of
the sources. While the use of FUV images has certain advantages
(longer SF timescales, more uniform spatial coverage), the physical
sizes of the galaxies themselves (of order 1 kpc or smaller) present
fundamental limitations that are not encountered when undertaking
similar studies of more massive galaxy disks. Further, the sources are
relatively faint.  Neither of these problems would exist for more
massive galaxies at the same distances.

Despite these challenges, we produce three sets of plots in order to
decipher the K-S relation for each of the SHIELD galaxies
(Figures~\ref{FUVSFE}, \ref{K98SFE}, and \ref{SFEpixcorr}). Because
two different methodologies were used and two different SF tracers
were involved, the slope values derived in the plots vary
substantially. For these three figures, we derive average slopes of N
$\approx$ 1.50$\pm$0.02, N $\approx$ 0.34$\pm$0.01, and N $\approx$
0.68$\pm$0.04, respectively.  Note that the inclusion of AGC\,748778
and AGC\,749241, both of which have low or negative slopes in the
FUV-based figures and are absent from the \halpha-based figure, may
skew the average slopes to low values.  Further, the images used to
create all three figures are smoothed to a small degree, effectively
smearing out the flux from the peak and possibly making the slopes
less dramatic.

Comparing the modes of analysis noted above, we conclude that the
pixel-by-pixel correlation method (Figure~\ref{SFEpixcorr}) provides
the most holistic and reliable representation of the comparative
qualities of \ion{H}{1} and recent SF in the SHIELD galaxies. The
uncertainties for the slopes are lower for the pixel-by-pixel method
than for the radial profile methods (Figures~\ref{FUVSFE} and
\ref{K98SFE}), and the power-law slopes from the pixel-by-pixel method
more closely match what would be expected from visual inspection of
Figures~\ref{fig110482} - \ref{fig749241}. We also note that the inner
annuli used to derive the surface densities and produce
Figures~\ref{FUVSFE} and \ref{K98SFE} are sensitive to small changes
in the precise positioning of the ellipse centers; moving in a given
direction by a single pixel can include or exclude significant
\halpha\ or UV flux (moving the aperture by an entire \ion{H}{1} beam
size can result in flux changes of 50\% or more), thus changing the
shape of the resulting power-law slope. Using Figure~\ref{SFEpixcorr}
as our metric, the K-S slope values of the 12 galaxies break down as
follows: 2 sources have $N < 0.0$, 6 have $0.0 < N \leq 1.0$, and 4
have $N > 1.0$.

We must note here that our technique for analyzing the K-S relation on
sub-kpc scales differs in an important way from other recent
studies. Because the galaxies in a given sample all reside at
different distances, the physical size scale for each source will be
different. The native resolution of the HI beam will also contribute
directly to this size scale: a beam of finer resolution will yield a
smaller physical resolution. These other studies (such as Roychowdhury
\etal\ 2015\nocite{roychowdhury2015}, Roychowdhury
\etal\ 2009\nocite{roychowdhury2009}, and Bigiel
\etal\ 2008\nocite{bigiel2008}) smooth the data for each galaxy in
their sample to a common physical resolution (e.g., $\sim$400 pc or 1
kpc) in order to analyze the relationship of gas and stars on the same
physical scale. We chose not to perform this smoothing, instead
keeping the SHIELD galaxies at their native physical resolutions (see
Table~\ref{t2}). There are a number of reasons for this decision:
first, the minimum common resolution we would have had to smooth to
would have been $\sim$600 - 650 pc (limited unfortunately by those
sources which lack VLA B-configuration HI data). While this does not
pose a serious problem for the THINGS galaxies (larger spiral
galaxies) or the FIGGS galaxies (dwarf galaxies with a mean sample
distance of 4.7 Mpc), the SHIELD survey works with galaxies which are
of similar size to FIGGS but span a factor of two in distance. Our
sources are at a mean distance of 8.1 Mpc with none closer than 5.1
Mpc. This range in distances means that we can fit fewer resolution
elements across a galaxy$-$for example, even AGC\,749241 (at a
relatively small distance of 5.62 Mpc) is only $\sim$2100 pc across in
its HI map, meaning we could fit less than four smoothed resolution
elements across the galaxy, effectively smearing out any useful
spatial information. Second, the high-resolution data gives us insight
into the exact locations of the HI peaks and also lets us see the
highest column density HI gas; if we smoothed all galaxies to a
coarser common resolution, our analysis of the HI column density
threshold would be less secure. Third, our analysis of the SHIELD
galaxies on a variety of fine resolutions demonstrates the
breakdown of the K-S relation on small scales, as found in previous
studies, such as \citet{schruba2010}.

Situations such as AGC\,111977 and AGC\,749241 reveal certain
advantages of not fixing the physical resolution: both of these sources have
interesting morphologies, with HI peaks that are distinctly offset
from the SF tracers. Both of these sources are at a distance of
$\sim$6 Mpc, but AGC\,111977 lacks B-configuration data and thus has a
resolution about four times as coarse as AGC\,749241. Unfortunately,
this makes comparing their K-S relations challenging, and it also begs
the question: if higher-resolution data was available for AGC\,111977,
what would we see? Instead of having a slope of $N \approx$
0.72$\pm$0.18 when the physical resolution is $\sim$690 pc, it might
shift to a flat or negative slope, as is the case with AGC\,749241 at
a resolution of $\sim$170 pc. Likewise, smoothing the data for
AGC\,749241 to a coarse resolution might reduce the strength of the
anti-correlation we currently see in Figure~\ref{SFEpixcorr}.

%-----------------------------------------------------------------------------%
\subsubsection{Grouping the Galaxies}
\label{S4.2.2}
%-----------------------------------------------------------------------------%

Based on the analysis in Section~\ref{S4.1} and the pixel-by-pixel
metric (Figure~\ref{SFEpixcorr}), we divide our sample into three
broad categories that share similar characteristics: 1) mainly
co-spatial \ion{H}{1} and SF regions, found in systems with highest
peak \ion{H}{1} column densities and highest total \ion{H}{1} masses;
2) systems which show a range of correlation strengths between
\ion{H}{1} and SF regions, and which also span a range of peak
\ion{H}{1} column densities; and 3) obvious offsets between \ion{H}{1}
and SF peaks, found in systems with the lowest total \ion{H}{1}
masses.  These groupings are heterogeneous but useful for comparison
purposes.

Group 1 contains the systems that most closely resemble the
``expectations'' of active SF in regions of high \ion{H}{1} column
density: AGC\,110482, AGC\,111946, AGC\,174585, AGC\,731457, and
AGC\,749237. These sources demonstrate the strongest observed
correlations between the \ion{H}{1} and SF peaks, and each system
reaches a relatively high peak \ion{H}{1} column density ($\gtrsim$
9$\times$10$^{20}$ cm$^{-2}$). It is important to note that the
comparative properties of \ion{H}{1} and SF vary among and within
these sources: the pixel-by-pixel $N$ indices of these sources vary
between 0.59 and 2.04, and AGC\,110482 shows compact \ion{H}{2}
regions while AGC\,731457 shows extended \halpha\ emission and widely
extended FUV emission. These comparatively ``well-behaved'' sources
are generally at the higher-mass end of the SHIELD sample. We include
AGC\,749237 in this first group, but note that while some of its
ongoing SF is strongly correlated with high mass surface density
neutral gas, the highest \ion{H}{1} columns are devoid of SF
altogether.

Group 2 contains a variety of SHIELD sources which span a range of
properties: AGC\,111164, AGC\,111977, AGC\,112521, AGC\,174605, and
AGC\,182595. They have moderate peak \ion{H}{1} column densities
($\lesssim$ 5$\times$10$^{20}$ cm$^{-2}$) and display a range of
correlation strengths between these peaks and the regions of SF. We
must also note that the three sources which are lacking
B-configuration data are included in this group; it is likely that
their low sensitivity to high column density HI gas contributes to the
low HI column densities of those sources which have been grouped
here. Visual inspection of the images of these galaxies illustrates
some similar traits among these sources. In general, higher than
average \ion{H}{1} columns are nearly co-spatial with SF tracers
(e.g., AGC\,111164). Again, notable issues exist amongst and within
the members of this group: AGC\,111977, for example, harbors extended
SF, but the \ion{H}{1} peak is not co-spatial with the highest surface
brightness emission. With the exception of AGC\,111164, these systems
also represent the middle of the mass range for the sample.

Finally, Group 3 contains two highly unusual galaxies: AGC\,748778 and
AGC\,749241. These systems stand out compared to the other SHIELD
galaxies: they harbor the lowest stellar masses \citep{mcquinn2015a},
the narrowest \ion{H}{1} linewidths (see Table~\ref{t2}), the weakest
\halpha\ emission, and among the weakest FUV emission. Further, their
morphologies are highly unusual compared to the other sample members,
in that the \ion{H}{1} gas is significantly offset from the high
surface brightness stellar population. The negative K-S indices
highlight the extreme nature of these sources; with the present data
we are unable to determine the origin of the offset of the \ion{H}{1}
and the stars, but two scenarios are possible. One possibility is
tidal effects from the local environment near the
galaxy. \citet{mcquinn2015a} examined the environment immediately
surrounding each of the SHIELD sources in 3 dimensions using the
nearest neighbor metric, and suggested that AGC\,749241 has likely
been influenced by the external gravitational perturbations from other
systems. Another possible scenario is that both of these galaxies had
strong SF events in the past, resulting in the prominent FUV but
non-existant \halpha\ emission seen today. Due to their low mass (both
stellar and HI), these episodes might have disrupted the central HI
gas and, in turn, shut down any further SF. While the details are
complex and depend on the coupling of mechanical feedback energy to
the surrounding ISM, over time, this could create an apparent offset
between the gas and the stars. While the offsets might appear more
extreme in these two systems compared to the rest of the sample, the
observed offsets in other galaxies in the sample (e.g., AGC\,111977)
could also be explained by a major SF event disrupting the nearby
gas. In this scenario, the time elapsed since a major SF event matters
greatly: \ion{H}{2} regions would only be co-spatial with the HI gas
peaks if the SF event is young or current and hasn't had time to
disrupt the gas significantly.

Figure~\ref{resolutionvscolden} plots the SHIELD galaxies in these
three distinct groups to identify trends based on the physical
resolution and peak HI column density of each source. This plot
confirms that the peak HI column observed certainly depends on the
fineness of the resolution element, indicated by the location of Group
2 in comparison to Group 1. However, Group 3 defies this trend, and
there exist members of Group 1 which have resolutions barely better
than members of Group 2 yet which have significantly higher HI
columns. We also note that Group 2 contains all three galaxies which
lack B-configuration data; this demonstrates the degree to which the
high-resolution VLA observations allow us to see higher column density
HI gas.

Figure~\ref{slopevalsvsresolution} yields a different perspective. Again,
the three groups of galaxies do not mingle very much, and it is
apparent that some correlation exists between the physical resolution
element for a galaxy and its resulting K-S slope. At smaller
resolutions, it is more likely that the HI gas will appear offset from
the SF regions, resulting in flatter or more negative $N$ values. This
trend can also be seen in the images of the galaxies
(Figures~\ref{fig110482} - \ref{fig749241}) and in
Figure~\ref{SFEpixcorr}. These plots together provide evidence for one
of our key results: at small physical scales, the K-S relation is no
longer valid.

%-----------------------------------------------------------------------------%
\subsubsection{Atomic vs. Molecular Gas as a Tracer for Star Formation}
\label{S4.2.3}
%-----------------------------------------------------------------------------%

Taken as a composite sample, the average of the pixel-by-pixel K-S
indices for the SHIELD galaxies is $N$ $\approx$ 0.68$\pm$0.04 (based on
the slopes derived in Figure~\ref{SFEpixcorr}). To put this average
value and the individual galaxies' values in perspective, we compare
to \citet{bigiel2008} and \citet{leroy2008}, who find $N\approx$
1.0$\pm$0.2 for the molecular Schmidt law (only relating $\Sigma_{\rm
  H_{2}}$ and $\Sigma_{\rm SFR}$) in their sample of spiral
galaxies. They do not ultimately base their conclusions on the
relation of $\Sigma_{\rm \HI}$ to $\Sigma_{\rm SFR}$ because most of
their galaxies show little correlation between \ion{H}{1} mass surface
density and ongoing SF tracers. Our $N$ values are quite similar and
our method intentionally parallels theirs; however, it is important to
remain mindful that this comparison is not one-to-one, since we use
\ion{H}{1} data exclusively and the studies of \citet{bigiel2008} and
\citet{leroy2008} account for the molecular phase. Those works
conclude that, even for the spirals which are qualitatively different
from the SHIELD galaxies in fundamental ways, the SFE varies
considerably across their sample and within their individual
galaxies. The conclusion of both studies is that the SFE is set by
local environmental factors (i.e., SF is internally regulated; see
McQuinn et al. 2015a\nocite{mcquinn2015a}).

Other recent work has extended the K-S analysis to other types of
galaxies; Figure~\ref{globalsigmas} presents this comparison. A study
of low surface brightness (LSB) galaxies by \citet{wyder2009} finds
that these sources tend to lie below the canonical composite
(\ion{H}{1}+\ion{H}{2}) K-S relation from \citet{kennicutt1998a}. An
extrapolation of the slope for these sources would appear even steeper
than for the larger and higher surface brightness galaxies. However,
the analysis by \citet{roychowdhury2014} of FIGGS dwarf irregulars
yields an $N\approx$ 0.91$\pm$0.24, which is in excellent agreement
with our results. Both of these slopes are clearly shallower than the
composite slope from \citet{kennicutt1998a}, indicating a deviation
from the canonical K-S law found for higher mass galaxies but in a
different manner than found in \citet{wyder2009}. Each of these
low-mass galaxy studies$-$SHIELD, the LSB galaxy study, and
FIGGS$-$assume the molecular component of the gas is negligible. The
authors of the FIGGS study suggest that their results favor a model of
SF where thermal and pressure equilibrium in the ISM regulate the rate
at which SF occurs, where the thermal pressure in turn is set by
supernova feedback.

Our analysis leads us to a similar conclusion as those presented in
these other works: SF in extremely low-mass galaxies is dominated by
stochasticity and random fluctuations in their ISM. Using the
qualities of \ion{H}{1} gas alone, it is very challenging to predict
where a given system will show signs of recent SF. Similarly, knowing
the distribution of \halpha\ and/or FUV emission offers little insight
into where we might expect to see high \ion{H}{1} mass surface
density.

We do not see strong evidence for a threshold \ion{H}{1} column
density above which we see signs of recent SF and below which we do
not. The SHIELD sample contains multiple examples of both
``super-critical'' and ``sub-critical'' gas associated with
\halpha\ and with FUV emission. As Figure~\ref{coldentrend}
demonstrates, the column density of an \ion{H}{1} region correlates
only loosely with its SFR in our sample. While the differing HI beam
dimensions and resulting physical resolutions certainly play a role in
determining the highest column density gas detected, a sample-wide
trend would not necessarily be borne out by improved resolution for
the low column density sources; AGC\,748778 and AGC\,749241 have only
moderate HI column densities despite their fine physical resolutions.

Overall, our data suggest that SF does not only occur in regions
exceeding the suggested critical \ion{H}{1} column density. In fact,
there does not even seem to be a lower threshold: members of Group 2
(see above) all have relatively co-spatial ongoing SF and HI peaks,
yet they have column densities $\lesssim$ 5$\times$10$^{20}$ cm$^{-2}$
where the SF is occurring; members of Group 3 have significant FUV
emission in regions that are in fact devoid of \ion{H}{1} entirely
down to the limits of our observations. It is apparent from our study
of these low-mass dwarf galaxies that \ion{H}{1} is not a useful
tracer of SF when used in the K-S relation; numerous studies have come
to similar conclusions \citep{filho2016, roychowdhury2015}.

Additionally, recent studies by \citet{michalowksi2015} and
\citet{krumholz2013} have found that low-mass SF galaxies, despite
having very low molecular gas fractions, are still able to form
massive stars. This idea is supportive of our results in that
high-mass SF has occurred in the metal-poor SHIELD galaxies; the
stochastic nature of SF is certainly a contributing factor to the
variation in K-S relations seen in the sample.

%-----------------------------------------------------------------------------%
\subsubsection{Gas Consumption Timescales}
\label{S4.2.4}
%-----------------------------------------------------------------------------%

We close by commenting on the characteristic gas consumption
timescales (GCTs) of the SHIELD galaxies. There are two methods of
calculating the GCT (the time it would take for the galaxy, at its
present-day SFR, to completely use up its neutral hydrogen gas
reservoir); the first requires taking the inverse of the total SFE. In
this way, we find an average GCT for the sample of
$\sim$2$\times$10$^{9}$ yr from the global comparison of 
$\Sigma_{\rm SFR}$ to $\Sigma_{\rm HI}$. This is nearly an order of magnitude
lower than those found in \citet{roychowdhury2014} for the FIGGS
galaxies, and roughly two orders of magnitude below the timescales
derived in the outer regions of spiral galaxies.  \citet{leroy2008} and
\citet{bigiel2008} find that molecular GCTs for the THINGS galaxies
are of order $\sim$2 $\times$ 10$^{9}$ yr. The authors also suggest a
mechanism that we conclude is likely at play in our sample as well:
microphysics in the interstellar medium below the scales which we can
observe (in the form of random gas motions and stellar feedback) which
govern the formation of molecular gas from \ion{H}{1}.

Employing the second method, in which we simply divide the total HI
gas mass by the total SFR of the galaxy, we get slightly different
results. This straightforward approach yields an average GCT for the
SHIELD galaxies of $\sim$10$^{10}$ yr; if we ignore the most extreme
outlier in the sample (AGC\,112521, with a derived GCT of
$\sim$45$\times$10$^{9}$ yr), then the average drops to
$\sim$7$\times$10$^{9}$ yr. These higher average values show
consistency with results for a sample of ALFALFA dwarfs, in which the
authors found a GCT of 8.9$\times$10$^{9}$ yr
\citep{huang2012}. Consumption times derived via both methods are
included in Table~\ref{t4}. Note that the uncertainties for both modes
of calculation are significant.

%-----------------------------------------------------------------------------%
\section{Conclusions}
\label{S5}
%-----------------------------------------------------------------------------%

SHIELD is a systematic investigation of a sample of extremely low-mass
dwarf galaxies outside the Local Group. Despite the low \ion{H}{1}
column densities observed in many systems, each SHIELD galaxy has a
significant FUV luminosity, and we detect \halpha\ emission in all but
two of them. The ability to compare multi-configuration VLA \ion{H}{1}
data with SF tracers in other wavelengths allows us to examine, on a
local and global scale, the SF ``law'' in these systems. We calculate
$\Sigma_{\rm \HI}$ and $\Sigma_{\rm SFR}$ for all sample members and
find the SFEs. We derive an index for the Kennicutt-Schmidt relation
via several different methodologies. The ensemble average index using
the pixel correlation mehtod gives $N \approx$ 0.68$\pm$0.04; this is in
good agreement with other studies of low-mass dwarf galaxies, but is
shallower than the canonical K-S relation from \citet{kennicutt1998b}.
By comparing the SHIELD results to those from other major nearby
galaxy surveys, we find that HI mass and UV-based SFR are strongly
correlated over five orders of magnitude.

We stress that any one galaxy in the sample is not representative, and
that the values of $N$ vary considerably from system to
system. Instead, we focus on the narratives of the individual galaxies
and their distribution of gaseous and stellar components, which are
complex and occasionally puzzling. The average consumption time for
the sample of $\sim$2 - 10 Gyr suggests that they will consume their
gas reservoir over timescales similar to those found in other dwarf
galaxy surveys.

At the extremely faint end of the \ion{H}{1} mass function, these
systems appear to be dominated by stochastic motions in their extreme
ISM. The local microphysics within the ISM of individual galaxies may
govern whether or not they show signs of recent SF. Based
on our data, knowledge of the \ion{H}{1} properties holds little
predictive power in terms of the resulting SF
characteristics. Similarly, we see ongoing or recent SF in unexpected
regions of many of the SHIELD galaxies. The observed offsets between
gas and stars in the galaxies could originate from tidal interactions,
or the offsets might appear based on how much time has elapsed since a
major SF event disrupted the central gas component. If causal
relationships between atomic \ion{H}{1} gas and SF exist in galaxies
in this extreme mass range (6.6 $<$ log(M$_{\rm \HI}$) $<$ 7.8), these
relationships remain elusive with current data.

In addition to the analysis presented here, a companion paper by
\citet{mcnichols16} studies the HI gas kinematics and dynamics of the
SHIELD galaxies. Two- and three-dimensional analyses are used to
constrain rotational velocities.  It is argued that the SHIELD
galaxies span an important mass range where galaxies transition from
rotational to pressure support.  The sources are contextualized on the 
baryonic Tully-Fisher realtion.

The now-complete ALFALFA catalog contains dozens of galaxies with
\ion{H}{1} and stellar properties comparable to those of the 12 SHIELD
galaxies studied in this work. Observations similar to the ones
presented here are underway to characterize the SF properties of this
statistically robust sample. 

%-----------------------------------------------------------------------------%
\clearpage
\acknowledgements
%-----------------------------------------------------------------------------%

The authors acknowledge the work of the entire ALFALFA collaboration
team in observing, flagging, and extracting the catalog of galaxies
used to identify the SHIELD sample. The ALFALFA team at Cornell is
supported by NSF grants AST-0607007 and AST-1107390 to RG and MPH and
by grants from the Brinson Foundation.  Y.G.T, A.T.M., and J.M.C. are
supported by NSF grant AST-1211683. E.A.K.A. is supported by
TOP1EW.14.105, which is financed by the Netherlands Organisation for
Scientific Research (NWO).

We gratefully acknowledge Adam
Leroy for the idea of an IDL code for mask-creation, and Charlotte
Martinkus \& Ned Molter for moral support. This work would not have
been possible without the incredible support for
Python\footnote{http://www.python.org} provided by the online
community at StackOverflow.com. Plots featured in this paper have been
created with the Python package \textsc{matplotlib}.
 
Support for Hubble Space Telescope data in this work was provided by
NASA through grant GO-12658 from the Space Telescope Institute, which
is operated by Aura, Inc., under NASA contract NAS5-26555. The Arecibo
Observatory is operated by SRI International under a cooperative
agreement with the National Science Foundation (AST-1100968), and in
alliance with Ana G. M{\'e}ndez-Universidad Metropolitana, and the
Universities Space Research Association.  Galaxy Evolution Explorer
(GALEX) is a NASA Small Explorer, launched in 2003 April. We
gratefully acknowledge NASA's support for construction, operation, and
science analysis for the GALEX mission, developed in cooperation with
the Centre National d'Etudes Spatiales of France and the Korean
Ministry of Science and Technology. This research made use of NASA's
Astrophysical Data System, the NASA/IPAC Extragalactic Database which
is operated by the Jet Propulsion Laboratory, California Institute of
Technology, under contract with the National Aeronautics and Space
Administration, and Montage, funded by the NASA's Earth Science
Technology Office, Computation Technologies Project, under Cooperative
Agreement Number NCC5-626 between NASA and the California Institute of
Technology.  Montage is maintained by the NASA/IPAC Infrared Science
Archive.

\textit{Facilities: Hubble Space Telescope, Galaxy Evolution Explorer,
  \textit{Spitzer} Space Telescope, Kitt Peak National Observatory's
  Wisconsin-Indiana-Yale-NOAO Observatory 3.5m Telescope, Expanded
  Very Large Array.}

%-----------------------------------------------------------------------------%
\clearpage
\bibliographystyle{apj}                                                 

%-----------------------------------------------------------------------------%

%-----------------------------------------------------------------------------%
% Tables and Figures
%-----------------------------------------------------------------------------%
\clearpage
\thispagestyle{empty}
%\begin{landscape}
%\floattable
\begin{deluxetable}{ccccccccccc} 
%\rotate
\tabletypesize{\scriptsize}
\tablecolumns{11} 
\tablewidth{0pt} 
\tablecaption{SHIELD Galaxy Sample Properties} 
\tablehead{
\colhead{Galaxy ID} &
\colhead{R.A.} &   
\colhead{Dec.} &
\colhead{Distance} &
\colhead{$M_{B}$} &
\colhead{(B-V)} & 
\colhead{log $M_{\rm \HI}$} &
\colhead{log $M_{\star}$} &
\colhead{V$_{21}$} & 
\colhead{12+log(O/H)}& 
\colhead{A$_{\rm FUV}$} \\
\colhead{} &
\colhead{(J2000)} &
\colhead{(J2000)} &
\colhead{(Mpc)} &
\colhead{(mag)} &
\colhead{(mag)} & 
\colhead{($M_{\odot})$} &
\colhead{($M_{\odot})$} &
\colhead{(km s$^{-1}$)} &
\colhead{}&
\colhead{(mag)}
\\
\colhead{(1)} &
\colhead{(2)} &
\colhead{(3)} &
\colhead{(4)} &
\colhead{(5)} &
\colhead{(6)} & 
\colhead{(7)} &
\colhead{(8)} &
\colhead{(9)} &
\colhead{(10)}&
\colhead{(11)}
}
\startdata
AGC\,110482 & 01:42:17.4 & 26:22:00 & 7.82$\pm$0.21 & -13.02$\pm$0.13 & 0.49$\pm$0.02         & 7.28$\pm$0.05 & 7.74$^{+0.13}_{-0.18}$ & 357$\pm$1 & 7.79$\pm$0.07  & 0.75    \\
AGC\,111164 & 02:00:10.1 & 28:49:52 & 5.11$\pm$0.07 & -11.16$\pm$0.10 & 0.42$\pm$0.02         & 6.61$\pm$0.05 & 7.00$^{+0.08}_{-0.15}$ & 163$\pm$3 & 7.59$\pm$0.10  & 0.45    \\
AGC\,111946 & 01:46:42.2 & 26:48:05 & 9.02$^{+0.20}_{-0.29}$ & -11.87$\pm$0.12 & 0.30$\pm$0.02  & 7.17$\pm$0.05 & 7.23$^{+0.13}_{-0.36}$ & 367$\pm$1.5 & 7.86$\pm$0.10 & 0.68 \\
AGC\,111977 & 01:55:20.2 & 27:57:14 & 5.96$^{+0.11}_{-0.09}$ & -12.60$\pm$0.09 & 0.48$\pm$0.02  & 6.85$\pm$0.05 & 7.57$^{+0.12}_{-0.16}$ & 207$\pm$2 & 7.80$\pm$0.10& 0.58      \\
AGC\,112521 & 01:41:07.6 & 27:19:24 & 6.58$\pm$0.18 & -10.59$\pm$0.08 & 0.45$\pm$0.03         & 7.11$\pm$0.05 & 6.85$\pm$0.15 & 274$\pm$0.5 & 7.33$\pm$0.10         & 0.51  \\
AGC\,174585 & 07:36:10.3 & 09:59:11 & 7.89$^{+0.21}_{-0.17}$ & -11.32$\pm$0.13 & 0.41$\pm$0.04  & 6.90$\pm$0.05 & 6.95$^{+0.13}_{-0.20}$ & 356$\pm$3 & ...          & 0.32      \\
AGC\,174605 & 07:50:21.7 & 07:47:40 & 10.89$\pm$0.28 & -12.22$\pm$0.11 & 0.39$\pm$0.03        & 7.27$\pm$0.05 & 7.45$^{+0.17}$ & 351$\pm$1 & ...                    & 0.19   \\
AGC\,182595 & 08:51:12.1 & 27:52:48 & 9.02$\pm$0.28 & -12.70$\pm$0.13 & 0.52$\pm$0.03         & 7.00$\pm$0.05 & 7.70$^{+0.16}_{-0.34}$ & 398$\pm$2 & 7.75$\pm$0.13  & 0.34    \\
AGC\,731457 & 10:31:55.8 & 28:01:33 & 11.13$^{+0.20}_{-0.16}$ & -13.73$\pm$0.11 & 0.38$\pm$0.03 & 7.26$\pm$0.05 & 7.81$^{+0.20}_{-0.78}$ & 454$\pm$3 & 8.00$\pm$0.10& 0.23      \\
AGC\,748778 & 00:06:34.3 & 15:30:39 & 6.46$^{+0.14}_{-0.17}$ & -10.34$\pm$0.08 & 0.22$\pm$0.03  & 6.67$\pm$0.05 & 6.48$^{+0.12}_{-0.18}$ & 258$\pm$1.5 & ...        & 0.52      \\
AGC\,749237 & 12:26:23.4 & 27:44:44 & 11.62$^{+0.20}_{-0.16}$ & -14.12$\pm$0.12 & 0.45$\pm$0.03 & 7.76$\pm$0.05 & 7.72$^{+0.19}$ & 372$\pm$1 & 7.95$\pm$0.06        & 0.16     \\
AGC\,749241 & 12:40:01.7 & 26:19:19 & 5.62$^{+0.17}_{-0.14}$ & -10.25$\pm$0.19 & 0.14$\pm$0.03  & 6.75$\pm$0.05 & 6.60$^{+0.1}_{-0.30}$ & 451$\pm$1 & ...           & 0.12      \\
\enddata\clearpage
\tablecomments{Column 1 - Galaxy name. Columns 2 and 3 - Coordinates
  of galaxy in J2000. Column 4 - TRGB-derived distances from
  \citet{mcquinn2014}. Columns 5 and 6 - Optical B-band magnitude and
  (B$-$V) color from WIYN 3.5m observations
  \citep{haurberg2015}. Column 7 - Galaxy \ion{H}{1} mass from ALFALFA
  \citep{haynes2011}. Column 8 - Galaxy stellar mass from HST
  observations (see \citet{mcquinn2015a} for details). Note that
  stellar masses have also been derived from \textit{Spitzer}
  observations and are included in \citet{haurberg2015}. Column 9 -
  Systemic velocity of galaxy from ALFALFA \citep{haynes2011}. Errors
  are adopted as half of the error on the W$_{50}$ contained in Table
  \ref{t2}. Column 10 - Metallicity of galaxy from spectroscopic
  analysis in \citet{haurberg2015}.  Column 11 - Extinction in
  magnitudes calculated from Equation 6 using the E(B$-$V) values
  found in the \citet{schlegel1998} dust maps. }
\label{t1}
\end{deluxetable}
\clearpage

%%%%%%%
%\begin{landscape}
\begin{deluxetable}{ccccccc} 
\tablecolumns{7} 
\tablewidth{0pt} 
\tablecaption{SHIELD \ion{H}{1} Data} 
\tablehead{
\colhead{Gal. ID} &
\colhead{Per channel rms noise} &
\colhead{Peak \ion{H}{1} Col. Den.} &
\colhead{Beam Size} &
\colhead{Resolution} &
\colhead{W$_{\rm 50}$} &
\colhead{S$_{\rm \HI}$} \\
\colhead{} &
\colhead{(mJy\,Bm$^{-1}$)} & 
\colhead{($\times$10$^{20}$ atoms cm$^{-2}$)} &
\colhead{(\arcsec\ $\times$ \arcsec)} &
\colhead{(pc)} &
\colhead{(km s$^{-1}$)} &
\colhead{(Jy\,km\,s$^{-1}$)} \\
\colhead{(1)} &
\colhead{(2)} &
\colhead{(3)} &
\colhead{(4)} &
\colhead{(5)} &
\colhead{(6)} &
\colhead{(7)}}
\startdata
AGC\,110482 & 1.0 & 17.6$\pm$1.8   & 11.98$\times$9.04  & 450 &  30$\pm$2 & 1.33$\pm$0.04 \\
AGC\,111164 & 1.4 & 3.0$\pm$0.3    & 21.56$\times$20.24 & 530 &  27$\pm$6 & 0.65$\pm$0.04 \\
AGC\,111946 & 1.1 & 9.7$\pm$1.0    & 10.30$\times$8.86   & 450 &  21$\pm$3 & 0.76$\pm$0.03 \\
AGC\,111977 & 1.5 & 2.9$\pm$0.3    & 24.01$\times$19.95 & 690 &  26$\pm$4 & 0.85$\pm$0.05 \\
AGC\,112521 & 1.3 & 4.8$\pm$0.5    & 22.03$\times$19.51 & 700 &  26$\pm$1 & 0.69$\pm$0.04 \\
AGC\,174585 & 0.88 & 10.7$\pm$1.1   & 6.19$\times$5.52   & 240 &  21$\pm$6 & 0.54$\pm$0.04 \\
AGC\,174605 & 0.51 & 4.5$\pm$0.5    & 11.81$\times$9.99  & 620 &  24$\pm$2 & 0.66$\pm$0.04 \\
AGC\,182595 & 0.69 & 2.0$\pm$0.2    & 10.05$\times$9.93  & 440 &  20$\pm$4 & 0.42$\pm$0.03 \\
AGC\,731457 & 0.88 & 11.3$\pm$1.1   & 6.04$\times$5.53   & 330 &  36$\pm$6 & 0.62$\pm$0.04 \\
AGC\,748778 & 0.93 & 5.7$\pm$0.6    & 5.91$\times$5.23   & 190 &  16$\pm$3 & 0.46$\pm$0.04 \\
AGC\,749237 & 0.79 & 15.8$\pm$1.6   & 6.21$\times$5.59   & 350 &  65$\pm$2 & 1.80$\pm$0.05 \\
AGC\,749241 & 0.79 & 5.5$\pm$0.6    & 6.06$\times$5.82   & 170 &  18$\pm$2 & 0.76$\pm$0.03 \\
\enddata
\tablecomments{Column 1 - Galaxy name. Column 2 - Single-channel rms
  noise in the robust-weighted (briggs=0.5), full spectral resolution
  data cube. Column 3 - Peak \ion{H}{1} column density of the robust-weighted
  (briggs=0.5), spectrally-smoothed (by a factor of 3) moment-0
  map. Column 4 - Dimensions in arcseconds of the restoring beam
  (FWHM) for the robust-weighted cube. Column 5 - Resolution element
  in physical units (pc) derived from the effective radius of the beam
  described in Column 4. Column 6 - Velocity width of the galaxy 21 cm
  line profile from ALFALFA \citep{haynes2011}. Column 7 - Flux
  integral from ALFALFA \citep{haynes2011}.}
\label{t2}
\end{deluxetable}
\clearpage

%%%%%%%%%%%%%
\begin{deluxetable}{ccccccc} 
\tablecolumns{7} 
\tablewidth{0pt} 
\tablecaption{SHIELD Star Formation Properties} 
\tablehead{
\colhead{Galaxy ID} &
\colhead{log L$_{\rm FUV}$} &
\colhead{log L$_{\rm H\alpha}$} &
\colhead{log SFR$_{\rm FUV}$} &
\colhead{log SFR$_{\rm H\alpha}$} &
\colhead{$\frac{\rm SFR_{\rm H\alpha}}{\rm SFR_{\rm FUV}}$} &
\colhead{log SFR$_{\rm 200\ Myr}$}
%\colhead{log(SFR_{Lifetime})} \\
\\
\colhead{} &
\colhead{(erg s$^{-1}$ Hz$^{-1}$)} &
\colhead{(erg s$^{-1}$)} &
\colhead{(\msun\ yr$^{-1}$)} &
\colhead{(\msun\ yr$^{-1}$)} &
\colhead{} &
\colhead{(\msun\ yr$^{-1}$)}
%\colhead{(yr)} \\
\\
\colhead{(1)} &
\colhead{(2)} &
\colhead{(3)} &
\colhead{(4)} &
\colhead{(5)} &
\colhead{(6)} &
\colhead{(7)}}
%\colhead{(10)}}
\startdata
AGC\,110482 & 25.15$\pm$0.04 &                  38.43$\pm$0.04  & -2.54$\pm$0.15 & -2.67$\pm$0.04                  & 0.74$\pm$0.10   & -2.27$^{+0.15}_{-0.23}$\\
AGC\,111164 & 24.40$\pm$0.05 &                  37.70$\pm$0.05  & -3.29$\pm$0.15 & -3.40$\pm$0.05                  & 0.77$\pm$0.13   & -3.28$^{+0.16}_{-0.49}$\\
AGC\,111946 & 24.91$\pm$0.06 &                  38.00$\pm$0.07  & -2.78$\pm$0.15 & -3.10$\pm$0.07                  & 0.48$\pm$0.16   & -2.46$^{+0.17}_{-0.40}$\\
AGC\,111977 & 24.94$\pm$0.04 &                  38.04$\pm$0.08  & -2.75$\pm$0.15 & -3.06$\pm$0.08                  & 0.49$\pm$0.09   & -2.59$^{+0.18}_{-0.24}$\\
AGC\,112521 & 24.15$\pm$0.09 &                  37.00$\pm$0.08  & -3.54$\pm$0.16 & -4.10$\pm$0.08                  & 0.28$\pm$0.07   & -3.44$^{+0.41}$\\
AGC\,174585 & 24.78$\pm$0.05 &                  37.89$\pm$0.05  & -2.91$\pm$0.15 & -3.21$\pm$0.05                  & 0.50$\pm$0.08   & -2.64$^{+0.18}_{-0.36}$\\
AGC\,174605 & 25.06$\pm$0.14 &                  37.91$\pm$0.04  & -2.63$\pm$0.19 & -3.19$\pm$0.04                  & 0.28$\pm$0.11   & -2.37$^{+0.14}_{-0.38}$\\
AGC\,182595 & 24.98$\pm$0.05 &                  38.45$\pm$0.07  & -2.71$\pm$0.15 & -2.65$\pm$0.07                  & 1.15$\pm$0.22   & -2.32$^{+0.21}_{-0.12}$\\
AGC\,731457 & 25.60$\pm$0.03 &                  38.58$\pm$0.07  & -2.09$\pm$0.15 & -2.52$\pm$0.07                  & 0.37$\pm$0.06   & -1.85$^{+0.37}_{-0.37}$\\
AGC\,748778 & 24.43$\pm$0.07 &        $<$36.18\tablenotemark{b}    & -3.26$\pm$0.15 & $<$-4.92\tablenotemark{b}          & 0.02$\pm$0.003   & -3.17$^{+0.14}_{-0.25}$\\
AGC\,749237 & 25.65\tablenotemark{a}$\pm$0.03 & 38.76$\pm$0.05  & -2.05$\pm$0.15 & -2.34$\pm$0.05                  & 0.51$\pm$0.07   & -1.89$^{+0.19}_{-0.16}$\\
AGC\,749241 & 24.33$\pm$0.06 &        $<$36.10\tablenotemark{b}    & -3.36$\pm$0.15 & $<$-5.00\tablenotemark{b}          & 0.02$\pm$0.003   & -3.49$^{+0.10}$\\
\enddata
\tablecomments{Column 1 - Galaxy name. Columns 2 - FUV
  luminosity. Column 3 - \halpha\ luminosity from
  \citet{haurberg2013}. Column 4 - FUV SFR derived
  using Equation 4. Column 5 - \halpha\ SFR derived
  using Equation 5. Column 6 - Ratio of SFR$_{\rm H\alpha}$ to
  SFR$_{\rm FUV}$. Column 7 - SFR over the last
  $\sim$200 Myr derived from HST CMDs in \citet{mcquinn2015a}.}
\label{t3}
\tablenotetext{a}{estimated from NUV luminosity}
\tablenotetext{b}{upper limit}
\end{deluxetable}\clearpage

%%%%%%%%%%%%%%
\begin{deluxetable}{ccccccc} 
\tablecolumns{6} 
\tablewidth{0pt} 
\tablecaption{SHIELD Star Formation Properties (Continued)} 
\tablehead{
\colhead{Galaxy ID} &
\colhead{log($\Sigma_{\rm SFR\ FUV}$)} &
\colhead{log($\Sigma_{\rm \HI}$)} &
\colhead{log(SFE)} &
\colhead{GCT (Method 1)} &
\colhead{GCT (Method 2)} 
\\
\colhead{} &
\colhead{(\msun\ yr$^{-1}$ kpc$^{-2}$)} &
\colhead{(\msun\ pc$^{-2}$)} &
\colhead{(yr$^{-1}$)} &
\colhead{($\times$10$^{9}$ yr)} &
\colhead{($\times$10$^{9}$ yr)} 
\\
\colhead{(1)} &
\colhead{(2)} &
\colhead{(3)} &
\colhead{(4)} &
\colhead{(5)} &
\colhead{(6)}}
\startdata
AGC\,110482 & -2.75$\pm$0.16   &  0.65$\pm$0.07  & -9.40$\pm$0.33    & 3$\pm$1 & 6.6$\pm$3.5 \\
AGC\,111164 & -3.13$\pm$0.16   &  0.21$\pm$0.07  & -9.33$\pm$0.33    & 2$\pm$1 & 7.9$\pm$4.2 \\
AGC\,111946 & -2.99$\pm$0.16   &  0.50$\pm$0.07  & -9.49$\pm$0.31    & 3$\pm$1 & 8.9$\pm$4.7 \\
AGC\,111977 & -2.61$\pm$0.16   &  0.43$\pm$0.07  & -9.03$\pm$0.33    & 1$\pm$1 & 4.0$\pm$2.1 \\
AGC\,112521 & -3.48$\pm$0.17   &  0.67$\pm$0.07 & -10.14$\pm$0.30    & 1$\pm$7 & 45$\pm$12 \\
AGC\,174585 & -2.97$\pm$0.16   &  0.19$\pm$0.07  & -9.16$\pm$0.32    & 1$\pm$1 & 6.5$\pm$3.4 \\
AGC\,174605 & -3.27$\pm$0.20   &  0.32$\pm$0.07  & -9.59$\pm$0.22    & 4$\pm$2 & 7.9$\pm$5.3 \\
AGC\,182595 & -2.91$\pm$0.16   &  0.37$\pm$0.13  & -9.28$\pm$0.70    & 2$\pm$1 & 5.1$\pm$2.7 \\
AGC\,731457 & -2.54$\pm$0.16   &  0.50$\pm$0.07  & -9.04$\pm$0.26    & 1$\pm$1 & 2.2$\pm$1.2 \\
AGC\,748778 & -3.28$\pm$0.16   & -0.10$\pm$0.07  & -9.18$\pm$0.31    & 2$\pm$1 & 8.5$\pm$4.5 \\ 
AGC\,749237 & -2.76$\pm$0.16   &  0.74$\pm$0.07  & -9.50$\pm$0.33    & 3$\pm$1 & 6.5$\pm$3.4 \\
AGC\,749241 & -3.31$\pm$0.16   &  0.13$\pm$0.07  & -9.44$\pm$0.32    & 3$\pm$1 & 13$\pm$6.9 \\ 
\enddata
\tablecomments{Column 1 - Galaxy name. Column 2 - SFR$_{\rm FUV}$
  surface density. Column 3 - \ion{H}{1} surface density. Column 4 -
  SFE found by dividing the $\Sigma_{\rm SFR\ FUV}$ by the
  $\Sigma_{\rm \HI}$ (and accounting for the change from kpc$^{2}$ to
  pc$^{2}$ for the $\Sigma_{\rm SFR\ FUV}$). Column 5 - Gas
  consumption time (GCT) derived from the inverse of the SFE. The
  weighted average for the sample is $\sim$2 $\times$ 10$^{9}$
  yr. Column 6 - Gas consumption time (GCT) derived by dividing the
  total HI mass of the galaxy by its total SFR$_{\rm FUV}$ (these
  values can be found in Table~\ref{t1}, Column 7 and Table~\ref{t3},
  Column 4, respectively). The weighted average for the sample is
  $\sim$10.2 $\times$ 10$^{9}$ yr.}
\label{t4}
\end{deluxetable}\clearpage

%%%%%%%%%%%%%%%%%%%
\begin{deluxetable}{ccccc} 
\tablecolumns{5} 
\tablewidth{0pt} 
\tablecaption{Elliptical Annuli Parameters}
\tablehead{
\colhead{Galaxy ID} &
\colhead{Position Angle} &
\colhead{Ellipticity} &
\colhead{Semi-major Axis} &
\colhead{Inclination}
\\
\colhead{} &
\colhead{($^\circ$)} &
\colhead{($1-b/a$)} &
\colhead{($\arcsec$)} &
\colhead{($^\circ$)} 
\\
\colhead{(1)} &
\colhead{(2)} &
\colhead{(3)} &
\colhead{(4)} &
\colhead{(5)}}
\startdata
AGC\,110482 & 125  & 0.42  & 16 & 55   \\
AGC\,111164 & 162  & 0.36  & 21 & 50   \\
AGC\,111946 & 175  & 0.52  & 17 & 62   \\
AGC\,111977 & 207  & 0.47  & 29 & 59   \\
AGC\,112521 & 189  & 0.42  & 15 & 55   \\
AGC\,174585 & 348  & 0.26  & 11 & 42   \\
AGC\,174605 & 49   & 0.06  & 11 & 19   \\
AGC\,182595 & 106  & 0.21  & 10 & 39   \\
AGC\,731457 & 53   & 0.23  & 12 & 40   \\
AGC\,748778 & 43   & 0.23  & 10 & 40   \\
AGC\,749237 & 85   & 0.40  & 16 & 54   \\
AGC\,749241 & 119  & 0.29  & 13 & 45   \\
\enddata
\tablecomments{Column 1 - Galaxy name. Column 2 - Position angle of
  the elliptical annuli, measured counter-clockwise from North. Column
  3 - Ellipticity, where $a$ is the semi-major axis length and $b$ is the
  semi-minor axis length. Column 4 - Semi-major axis length (a) in
  arcseconds. Column 5 - Inclination derived from the axial ratio
  using the prescription of \citet{haurberg2013}.}
\label{t5}
\end{deluxetable}
\clearpage

%%%%%%%%%%%%%%%%%%%%%%%%%%%
\begin{deluxetable}{cccc} 
\tablecolumns{4} 
\tablewidth{0pt} 
\tablecaption{Star Formation Power-Law Parameters}
\tablehead{
\colhead{} &
\colhead{FUV Rad. Prof.} &
\colhead{H$\alpha$ Rad. Prof. (K98 Conv.)} &
\colhead{FUV Pix-to-Pix}
\\
\colhead{Galaxy ID} &
\colhead{Index (N)} &
\colhead{Index (N)} &
\colhead{Index (N)} 
\\
\colhead{(1)} &
\colhead{(2)} &
\colhead{(3)} &
\colhead{(4)}}
\startdata
AGC\,110482 & 1.54$\pm$0.02  &  0.35$\pm$0.02  &  1.04$\pm$0.10   \\
AGC\,111164 & 4.96$\pm$0.79  &  $-$0.89$\pm$0.52  &  1.96$\pm$0.34   \\
AGC\,111946 & 1.18$\pm$0.06  &  $-$0.04$\pm$0.01  &  1.57$\pm$0.19   \\
AGC\,111977 & 6.24$\pm$21.5  &  0.96$\pm$1.07  &  0.72$\pm$0.18   \\
AGC\,112521 & $-$0.08$\pm$1.07  & 0.71$\pm$0.30 &  1.65$\pm$0.36   \\
AGC\,174585 & 1.91$\pm$0.52  &  1.46$\pm$0.17  &  0.59$\pm$0.22  \\
AGC\,174605 & 2.84$\pm$1.29  &  1.30$\pm$0.03  &  0.88$\pm$0.10  \\
AGC\,182595 & 2.87$\pm$0.09  &  1.51$\pm$0.07  &  1.03$\pm$0.16   \\
AGC\,731457 & 2.25$\pm$0.04  &  1.19$\pm$0.04  &  0.47$\pm$0.13  \\
AGC\,748778 & 1.56$\pm$0.44  &     ...       & $-$0.28$\pm$0.15  \\
AGC\,749237 & 2.80$\pm$0.08  &  0.52$\pm$0.01  &  2.04$\pm$0.21   \\
AGC\,749241 & $-$1.79$\pm$0.06 &     ...       & $-$0.59$\pm$0.12  \\
Average     & 1.50$\pm$0.02  &  0.34$\pm$0.01  &  0.68$\pm$0.04  \\
\enddata
\tablecomments{Column 1 - Galaxy name. Columns 2, 3, and 4 - The
  slopes of the lines in Figures \ref{FUVSFE}, \ref{K98SFE}, and
  \ref{SFEpixcorr} which is the $N$ value in the Kennicutt-Schmidt
  relation.}
\label{t6}
\end{deluxetable}

\clearpage

\clearpage

\clearpage

\clearpage

\clearpage

\clearpage

\clearpage

\clearpage

\clearpage

\clearpage

\clearpage

\clearpage

\clearpage

\clearpage

\clearpage

\clearpage

\clearpage

\clearpage

\clearpage

\clearpage

%\clearpage
%\begin{figure}
%\epsscale{1.0}
%\centering
%% old location = \plotone{/Users/research/Desktop/PLOTSFORPAPER/fullL09SFEplotsall12.eps}
%\plotone{fullL09SFEplotsall12.eps}
%\caption{$\Sigma_{\rm H\alpha\ SFR}$ vs. $\Sigma_{\rm \HI}$ for each
%  of the sample members, using the SFR$_{\rm H\alpha}$ derived from
%  the \citet{lee2009} prescription. The slope of the line is the N
%  index in the Kennicutt-Schmidt relation and is included in Table
%  \ref{t6}. The white points are below our 5$\sigma$ noise threshold
%  and thus are not included in the calculation of the trendline of the
%  data. The average value for the slope is $N$ $\approx$
%  1.5$\pm$0.8. The dashed vertical line represents an \ion{H}{1}
%  column density of 10$^{21}$ atoms cm$^{-2}$.}
%\label{L09SFE}
%\end{figure}

\clearpage

\clearpage

\clearpage

\clearpage

\clearpage

\end{document}